\documentclass[aps,twocolumn,preprintnumbers]{revtex4}

\usepackage{amsthm}

\usepackage[breaklinks=true,colorlinks,citecolor=blue,linkcolor=blue,urlcolor=blue]{hyperref}

\usepackage{graphicx}
\usepackage{xcolor}
\usepackage{multirow}

\usepackage{longtable}

\usepackage{bm}        
\usepackage{amsfonts}  
\usepackage{amsmath}   
\usepackage{amssymb}   

\newtheorem{lemma}{Lemma}

\newtheorem{remark}{Remark}

\usepackage[utf8]{inputenc}
\usepackage[english]{babel}
\usepackage{physics}

\newtheorem{theorem}{Theorem}

\begin{document}

\title{Generalized Type-II Fusion of Cluster States}

\author{Noam Rimock, Khen Cohen and 
Yaron Oz}

\affiliation{School of Physics and Astronomy, Tel Aviv University, Ramat Aviv 69978, Israel}


\begin{abstract}
Measurement based quantum computation is a quantum computing paradigm that employs single-qubit measurements performed on an entangled resource state in the form of a cluster state. A basic ingredient in the construction of the resource state is the type-II fusion procedure, which  probabilistically merges two separate photonic cluster states by a quantum measurement.
We generalize the type-II fusion procedure by generalizing the measurement setup, and classify the resulting final states, which also include cluster states up to single-qubit rotations. 
We prove that the probability for the success of the generalized type-II fusion is bounded by fifty percent, and classify all the possibilities to saturate the bound.
We analyze the enhancement of the fusion success probability above the fifty percent bound, by the reduction of the entanglement entropy of the resulting state. We prove that the only states that can be obtained with a hundred percent probability of success, are product states.

\end{abstract}


\maketitle

\section{Introduction}

\begin{figure}
    \centering
    \includegraphics[width=0.8\linewidth]{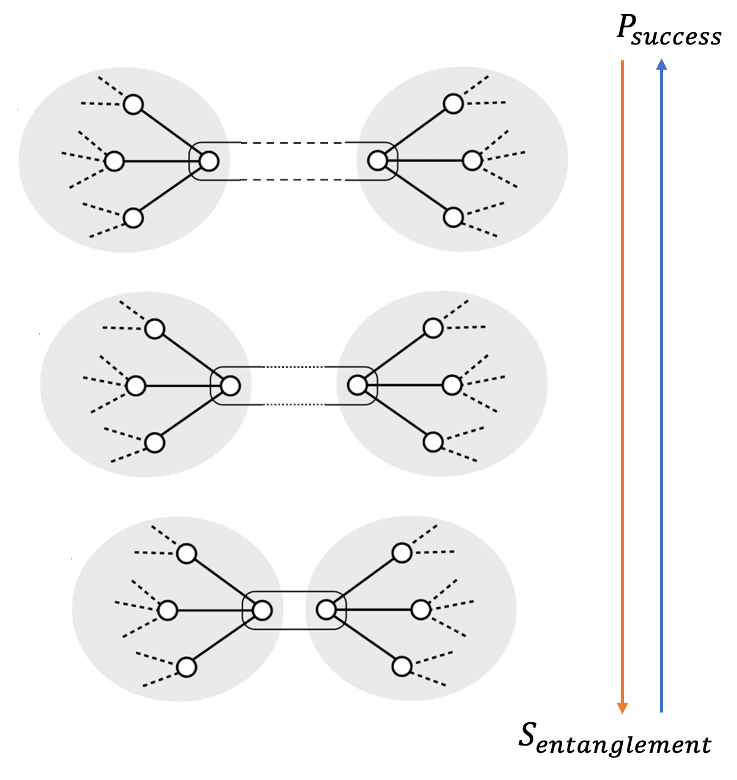} 
    \caption{Generalized type-II Fusion of cluster states: while the type-II fusion process merges two separate cluster states to a larger one by a quantum measurement, we implement a rather general measurement setup which generalizes the resulting final states. The lower the entanglement entropy between the merged clusters, the higher the probability for the success of the fusion process. The precise definitions of the entanglement entropy, the success probability and their relationship is presented in section \ref{sec:prob_def}.}
    
    \label{fig:FusionTradeoffDiagram}
\end{figure}

Measurement based quantum computation (MBQC) is a quantum computing paradigm harnessing entanglement and single-qubit measurements to perform general quantum computational tasks \cite{MBQCfirstArticle,MBQC2009}. MBQC is an alternative framework to the quantum circuit computation model, which is realized naturally in photonic platforms. At its core, MBQC employs single-qubit measurements performed on an entangled resource state \cite{MBQConClusterStatesFirstArticle,MBQC2021}, often a cluster state \cite{ClusterStatesNielsen,ClusterStatesOriginalPaper} where neighboring qubits, each in a superposition state, are entangled by a CZ gate (for an experimental realization see e.g.  \cite{RealizatonOfMBQC0,RealizatonOfMBQC1,RealizatonOfMBQC2,RealizatonOfMBQC3}). MBQC on cluster states is a universal quantum computation model that is isomorphic to the quantum circuit computation one.


The construction of cluster states is the first step to applying MBQC, and a fundamental ingredient for this task is the fusion gate, which is used to merge two smaller resource states. Fusion gates were introduced first via the KLM protocol \cite{KLMprotocolNature,UsingAncillae0}, the Nielsen protocol \cite{NielsenProtocolFirstArticle} and the Yoran-Reznik protocol \cite{YoranReznikProtocalFirstArticle}. Two types of fusion protocols, I and II, were proposed in \cite{BrowneRudolph}. 
Type-II fusion is based on qubit measurements (and so is type-I), hence it is a probabilistic procedure including projection onto a two-qubit state.
Given two separated cluster states that one wishes to merge, the type-II fusion is basically a measurement of two qubits, one from each cluster. There are two outcomes depending on the measurement result:  (i) success, in which the size of the resulting cluster state is the sum of the sizes of the two cluster states minus the two measured qubit, (ii) failure, in which the two cluster states remain separated, each losing one qubit.  
The probability for the success of type-II fusion, i.e. result (i), is 
fifty percent \cite{BrowneRudolph}.

Type-II fusion is fundamental for the construction of a two-dimensional cluster state that takes the form of an $L$-shape \cite{BrowneRudolph}, from one-dimensional ones. This is necessary in order to have a quantum advantage, since MBQC performed on one-dimensional clusters is as effective as a classical probabilistic computational model.
The construction of an $L$-shape cluster state requires on average no more than $52$ bell pairs \cite{BrowneRudolph}.
Type-II fusion is based on Bell states measurement (BSM) \cite{FusionBasedQC,ThreePhoton}, which has been proven to have a maximal value of fifty percent probability of success, when using only linear elements. i.e. that can be realized via linear optics  \cite{MaxEfficiency,BellMeasurements}. This success probability can be raised by adding ancilla resources \cite{UsingAncillae0,UsingAncillae1,UsingAncillae2,UsingAncillae3,UsingAncillae4}.
Increasing the probability of success for the fusion procedure is of utmost importance, since it reduces the resources required to build the desired graph. For example, it reduces the mean number of attempted fusions in order to succeed --- if the probability of success is $p$, then the mean number of attempted fusions until success is $\frac{1}{p}$ (that is, the mean of a geometric distribution).
The proof of the $0.5$ probability bound \cite{MaxEfficiency} is relevant for fusion type-II when it is required that the projection to a two-qubit state is to a Bell state, which in particular means projecting onto a maximally entangled two-qubit state. It has been shown that Bell states measurement isn't the only way to realize fusion type-II, and that by more general procedure the known bound of the success probability can be raised in the case where ancilla qubits are being used \cite{bartolucci2021creation}. This raises the question, whether the known $0.5$ success probability bound can be raised without using ancilla qubits, and what is the result of a projection onto a more general two-qubit state.



In this work we suggest a new framework, which generalizes the type-II fusion procedure, by generalizing the fusion matrix defined by the unitary transformation that acts on the two measured qubits before the measurement (see definition in (\ref{Unitary transformation})). We then classify the resulting final states as cluster states, stabilizer states, cluster states up to single-qubit rotations, and weighted graph states, based on the two-qubit state on which we project (so this part is relevant for every fusion protocol that resulting in a projection onto a two-qubit state).  
Obtaining cluster states up to single-qubit rotations is equivalent to obtaining cluster states since single-qubit rotations are standard operations in MBQC. Hence, our classification of such states reveals that the set of two-qubit states that projecting onto them creates fused clusters, is much larger than the set of the four Bell states. This allows more general protocols than the bell-projections-based regular type-II fusion, such as the protocol used in \cite{bartolucci2021creation}, in order to create the desired fused cluster state (note that although the protocol in\cite{bartolucci2021creation} uses ancilla qubits, this part is still relevant because we only assumed a fusion that results in two-qubit state projection).

We generalize the proof of \cite{MaxEfficiency}, by relaxing the condition on the two-qubit state on which we project, and allow having a general maximally entangled two-qubit state.
We prove that the success probability of the generalized type-II fusion is bounded by fifty percent, and provide a precise mathematical description of the ways to saturate the bound.
We analyze the cases of projecting on non-maximally entangled two-qubit states, and perform a numerical optimization analysis of fusion gates, allowing lower entanglement entropy. This enlarges the set of two-qubit states on which we project, and as a result we obtain higher probability of fusion as in  Fig. \ref{fig:FusionTradeoffDiagram}.


Throughout, we also consider weighted graph states that are a continuous generalization of cluster states (see exact definition in subsection \ref{Subsection:Generalized Graph States}), with a potential trade-off between entanglement strength and generation success probability. Some of the weighted graph states have been shown to constitute a universal resource for quantum computing \cite{gross2007measurement}.
In our setting (no ancilla qubits), we show that gates creating weighted graph states have at least $0.5$ probability of success as well (by providing explicit examples for such gates). Further study is needed in order to find, whether weighted graph states can be realized with probabilities above $0.5$ for suitable generalized type-II fusion.

The paper is organized as follows: In section \ref{sec:Fusion_type2} we briefly review cluster states and type-II fusion. In Section \ref{sec:Generalized Type II Fusion and the resulting state} we generalize type-II fusion, and classify the final states that arise from this fusion (or from a general projection onto two-qubit state). 
In section \ref{sec:prob_def} we study the possibility of reducing the entanglement entropy of the two-qubit state on which we project.
We define the fusion problem as a mathematical optimization problem, and correlate the fusion success probabilities and the values of the entanglement entropies. 
We lay the framework to show in section \ref{sec:numerical_res} that a reduction in the entanglement entropy of the two-qubit state on which we project, allows for an increase in the fusion success probability.
In Section \ref{sec:anal_bound} we conduct an analytical analysis and prove the fifty percent probability success bound for a maximal entanglement entropy as well as deriving several other results. 
In Section \ref{sec:numerical_res} we outline our numerical optimization method, followed by a demonstration of the fusion probability success versus the entanglement entropy.
Section \ref{sec:discussion} is devoted to a discussion and an outlook.
Detailed proofs of the theorems are given in the appendix.


\section{Cluster States and Type-II Fusion}
\label{sec:Fusion_type2}

In this section we will briefly review the definitions of stabilizer states, graph states and cluster states and the process of type-II fusion. Detailed reviews of optical quantum computing can be found in \cite{ReviewOfOpticalQuantumComputing0,ReviewOfOpticalQuantumComputing1,5lectures}.

\subsection{Stabilizer States}
\label{Subsection:Stabilizer States}

A state $\ket{\phi}$ of $n$ qubits is called a stabilizer state, if there exists a subgroup $S$ of the Pauli group with $n$ generators ($|S|=2^n$) such that for any operator $A\in S$, $A\ket{\phi}=\ket{\phi}$. If 
$g_1, g_2,...g_n$ are the generators of $S$ then an equivalent requirement 
is $g_i\ket{\phi}=\ket{\phi}$ for every $1\leq i\leq n$, and $g_i$ are called stabilizers. The set of stabilizer states is the set of all the states that can be obtained by acting on $\ket{0}^{\otimes n}$ with elements of the Clifford algebra \cite{GeneralizedGraphStatesAndStabilizerStates}.

For instance, the two-qubit state ($a$ and $b$ denote the qubits):
\begin{gather}
    \frac{1}{\sqrt{2}}(\ket{0}_a\ket{+}_b+\ket{1}_a\ket{-}_b) \ ,
    \label{example to stabilizer state}
\end{gather}
is a stabilizer state, with the generators:
\begin{gather}
    g_1=X_a Z_b,~~
    g_2=X_b Z_a \ ,
\end{gather}
where $X,Z$ are the Pauli matrices.

\subsection{Weighted Graph States}
\label{Subsection:Generalized Graph States}

\begin{figure}
    \centering
    \includegraphics[width=1.0\linewidth]{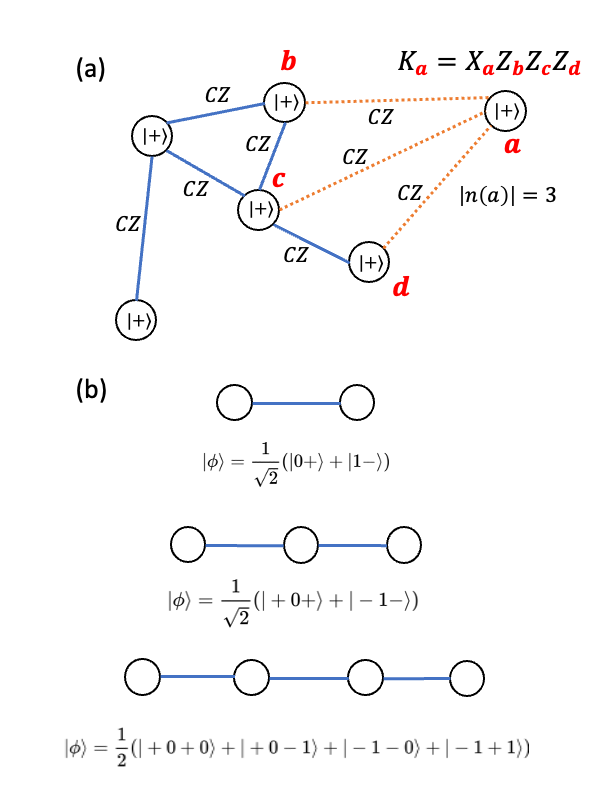}
    \caption{Graph states: (a) A graph state is generated by assigning to each node a
    single qubit in the state $\ket{+}$, and applying a $CZ$ gate to two neighboring (connected by an edge) vertices $i,j$. This construction can be described recursively as follows. Choose a  vertex $a$ in the graph,
    set the qubits without $a$ and its edges in the graph state  $\ket{\phi}_{V_a\setminus \{a\}}$, and apply $CZ$ gates to $a$ and its neighbors.
    The resulting graph state is as in Eq. (\ref{Graph state recursive definition}). We denote by $K_a$ the stabilizer of qubit $a$, and the graph wave function is an eigen-state of this stabilizer with eigenvalue $1$. (b) Examples of one-dimensional graph states of $n=2,3,4$ qubits. The corresponding wave functions are evaluated recursively using Eq. (\ref{Graph state recursive definition}).}
    \label{fig:graph_state_diagram}
\end{figure}

Given a graph \cite{GraphTheoryBook} with a set of vertices $V$ and a set of edges $E$, one associates to it a graph state as follows. We assign to each vertex a qubit in the superposition state $\ket{+}=\frac{1}{\sqrt{2}}(\ket{0}+\ket{1})$,
and apply a $CZ$ gate
to every two qubits $i,j$ that are connected by an edge:

\begin{gather}
    \ket{\phi}=\prod_{(i,j)\in E} CZ^{i,j} \prod_{k} \ket{+}_k \ .
    \label{Graph state product of CZ definition}
\end{gather}

Since the $CZ$ gates commute with each other, the order of the application them is irrelevant.
The resulting state  is called a graph state.  
For instance, the graph state of two qubits connected by an edge reads:
\begin{gather}
    CZ\ket{+}\ket{+}=\frac{1}{\sqrt{2}}(\ket{0}\ket{+}+\ket{1}\ket{-}) \ ,
\end{gather}
and is the stabilizer state (\ref{example to stabilizer state}).

Graph states have an equivalent definition as a stabilizer states of the subgroup $S$ of the Pauli group generated by $K_{a}$:
\begin{eqnarray}
    K_{a}\ket{\phi}&=&\ket{\phi},
    \nonumber\\
    K_{a}&=&X_{a} \otimes 
    \prod _{b\in n(a)} Z_{b}  \ ,
    \label{Graph state eigenvalue equation}
\end{eqnarray}
where $X_{a}$ and $Z_{b}$ are Pauli operators acting on the qubits labeled by
$a$ and $b$, respectively, and $n(a)$ denotes the vertices connected to $a$. The mutually commuting generators $K_{a} (K_{a}^2 = I_d$)
are the stabilizers of the state $\ket{\phi}$.

The construction of graph states can be defined recursively as follows.
Choose any vertex $a$ in the graph, denote
the graph state wave function without this vertex and without the edges that are connected to it as
$\ket{\phi}_{V_a\setminus \{a\}}$. Then, add the vertex $a$ to the graph and apply $CZ$ gates to
 $a$ and its neighbors $n(a)$. The resulting graph state wave function reads:
\begin{gather}
    \ket{\phi}=\frac{1}{\sqrt{2}}\left(\ket{0}_a\ket{\phi}_{V_a\setminus \{a\}}+\ket{1}_a \prod_{b\in n(a)} Z_b \ket{\phi}_{V_a\setminus \{a\}}\right) \ .
    \label{Graph state recursive definition}
\end{gather}

In order to construct the desired graph state (\ref{Graph state product of CZ definition}) recursively, we start in the first step with a single qubit in the state $\ket{+}$. Then, at each step we add a qubit and the edges connecting this qubit to the already existing qubits, and update the state by the recursive relation (\ref{Graph state recursive definition}). For simplicity, in the following we will omit the $\frac{1}{\sqrt{2}}$ normalization factor and absorb it in $\ket{\phi}_{V_a\setminus \{a\}}$.
For completeness, we include the proof of the equivalence of (\ref{Graph state eigenvalue equation})
and (\ref{Graph state recursive definition}) 
in appendix \ref{sec:Proof of equivalence}. 
It is important to note that $\ket{\phi}_{V_a\setminus \{a\}}$ and $\prod_{b\in n(a)} Z_b \ket{\phi}_{V_a\setminus \{a\}}$ are orthogonal to each other, and both have norm 1 \cite{GeneralizedGraphStatesAndStabilizerStates}.


One can generalize the definition of the graph states as follows. Consider a graph with $V$ vertices and $E$ edges.
Assign a qubit in the state $\ket{+}$ to each vertex and apply to each pair of qubits connected by an edge
a two-qubit gate, which is not the global phase
gate and not necessarily $CZ$. The set of two-qubit gates must be commuting so that the order of their application becomes irrelevant.
Also, all the gates must differ from the global phase gate, otherwise they will trivially commute with all the other gates and generate a product state of two subgraphs.
The resulting state is called a weighted graph state \cite{hein2004multiparty,dur2005entanglement}.

By applying Pauli $Z$ rotations on qubits in a weighted graph state, one can recast the two-qubit gates acting on qubits $a$ and $b$ in the form \cite{GeneralizedGraphStatesAndStabilizerStates}:
\begin{gather}
        U_{ab}=e^{i\chi_{ab}\ket{1_a,1_b}\bra{1_a,1_b}}
        =\ket{0_a,0_b}\bra{0_a,0_b}+\ket{0_a,1_b}\bra{0_a,1_b}\nonumber \\
        +\ket{1_a,0_b}\bra{1_a,0_b}+e^{i\chi_{ab}}\ket{1_a,1_b}\bra{1_a,1_b} \ .
        \label{The 2-qubits gates for generalized graph state after z-rotations}
\end{gather}

As will be shown later, the requirement for maximal entanglement between two connected qubits implies that  $\chi_{ab}=\pi$ \cite{GeneralizedGraphStatesAndStabilizerStates}.

\subsection{Cluster States}

A cluster state is a specific form of a graph state, where the graph is a subset of a lattice. A cluster state can be specified by the stabilizers:
\begin{eqnarray}
    K_{a}\ket{\phi}&=&(-1)^{k_{a}}\ket{\phi},~~~k_{a} \in \{0, 1\},
    \nonumber\\
    K_{a}&=&X_{a} \otimes 
    \prod _{b\in n(a)} Z_{b}  \ .
    \label{Clusters eigenvalue equation}
\end{eqnarray}

The difference between the stabilizers definition of cluster states (\ref{Clusters eigenvalue equation}) and graph states (\ref{Graph state eigenvalue equation}) are $\{k_{a}\}$, which are zero for a graph state. The operator $Z_a$ commutes with $K_b$ when $b\neq a$ and anti-commutes with $K_{a}$. Thus, one can construct any cluster state from a corresponding graph state by acting with $Z_{a}$ on each vertex $a$ where $k_{a}=1$.




As with graph states, the cluster state wave function can be constructed recursively. Choose a vertex $a$ in the graph, and denote the wave function without this vertex and without the edges that are connected to it as $\ket{\phi}_{V_a\setminus \{a\}}$. Next, we add the vertex $a$ to the graph and if $k_a=0$ we act with  $CZ$ gates on the qubit $a$ and its neighbors $n(a)$, which results in the recursive relation (\ref{Graph state recursive definition}). If $k_a=1$ then we also act on $a$ with the $Z_a$ gate.

The terminology  "cluster state" is does not have a universal meaning in the literature. Sometimes every graph state is called a cluster state, and sometimes only graph states that are lattice-based are called  cluster states. In this paper, in the type-II fusion procedure we start with two one-dimensional cluster states, and end with a graph state in the shape "+", which we will refer to as the "required fused cluster state" --- even though it is sometimes called only a graph state and not a cluster state.

\subsection{Type-II Fusion}


We will work within the linear optics framework, where a qubit is realized by a photon with a fixed frequency, and the quantum state is encoded in the photon's polarization, where $|0\rangle$ and $|1\rangle$ correspond to horizontal and vertical polarizations, respectively \cite{PhotonicQubits}.
The photon creation operators $a^\dag=(a^\dag_H,a^\dag_V)$ are defined, such that $a^\dag_H$ or $a^\dag_V$ acting on the vacuum state $\ket{vac}$, create a photon with a horizontal or vertical polarization, respectively.
A wave function $|\Psi\rangle = {\cal O}\ket{vac}$ will be denoted by the operator ${\cal O}$ made of the photon creation operators.
For instance, a two-qubit Bell state reads in this notation: 
\begin{equation}
|Bell\rangle : \frac{1}{\sqrt{2}}\left(a^{\dag}_H b^{\dag}_H + a^{\dag}_V b^{\dag}_V\right) \ .
\end{equation}
The entangling gates  are realized by linear optics elements and measurements, and are therefor probabilistic.
One constructs a larger cluster state from smaller ones, by employing
probabilistic fusion gates. In the following, we will focus on type-II fusion gates.
\begin{figure}
    \centering
    \includegraphics[width=1.0\linewidth]{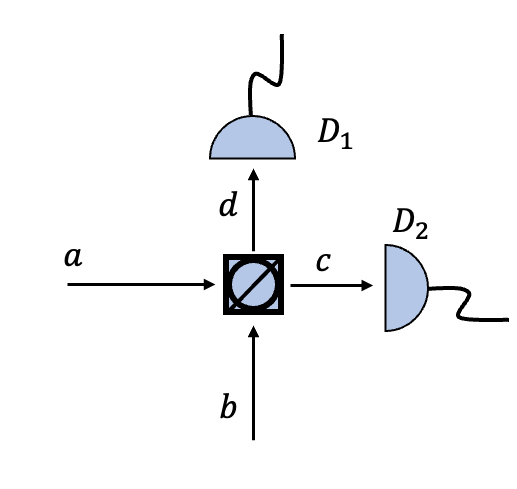}
    \caption{Type-II fusion optical setup, where $a$ and $b$ are the incoming photon modes that pass a diagonal polarization beam splitter  and exit as  $c$ and $d$ modes which are measured by the detectors $D_2$ and $D_1$.}
    \label{fig:typeII_optical_implementation}
\end{figure}
Consider two cluster states, whose wave functions are denoted 
by the operators  $\hat{C_1}$ and $\hat{C_2}$, and we mark one photon in each one of them, $a$ and $b$ respectively: 
\begin{gather}
    \hat{C_1}=\frac{1}{\sqrt{2}} (f_1 a^\dag_H + f_2 a^\dag_V),
    \hspace{0.5cm}
    \hat{C_2} =\frac{1}{\sqrt{2}} (f_3 b^\dag_H + f_4 b^\dag_V) \ .  
    \label{C_1 and C_2}
\end{gather}
$a^\dag$ and $b^\dag$ are the creation operators of the marked photons and $f_i,i=1,2,3,4$ denote the other parts of the two cluster states. Note that $f_1,f_2$ are orthogonal wave functions and the same holds for $f_3,f_4$. Thus, one can consider the wave functions of the two clusters as two effectively Bell pairs. The product state of these two cluster states reads:
\begin{gather}
    \hat{C_1}\hat{C_2} =\frac{1}{2} \left( f_1 a^\dag_H + f_2 a^\dag_V \right) \left( f_3 b^\dag_H + f_4 b^\dag_V \right) \nonumber\\
    =\frac{1}{2} (f_1 f_3 a^\dag_H b^\dag_H  + f_2 f_3 a^\dag_V b^\dag_H + f_1 f_4 a^\dag_H b^\dag_V + f_2 f_4 a^\dag_V b^\dag_V) \ .
    \label{C_1*C_2}
\end{gather}
In the optical Type-II fusion setup we pass two photons through a diagonal polarization beam splitter (PBS2), as in Figure \ref{fig:typeII_optical_implementation}, and the output modes read:
\begin{eqnarray}
    c_H^{\dag} &=&\frac{1}{2}\left(a_H^{\dag}+a_V^{\dag}+b_H^{\dag}-b_V^{\dag}\right),
    \nonumber\\
    c_V^{\dag} &=&\frac{1}{2}\left(a_H^{\dag}+a_V^{\dag}-b_H^{\dag}+b_V^{\dag}\right),\nonumber\\
    d_H^{\dag}&=&\frac{1}{2}\left(a_H^{\dag}-a_V^{\dag}+b_H^{\dag}+b_V^{\dag}\right),
    \nonumber\\
    d_V^{\dag} &=&\frac{1}{2}\left(-a_H^{\dag}+a_V^{\dag}+b_H^{\dag}+b_V^{\dag}\right) \ .
\label{c,d by a,b in standart fusion type 2}
\end{eqnarray}
Substituting (\ref{c,d by a,b in standart fusion type 2}) in (\ref{C_1*C_2}) gives the state of the two clusters in terms of $c,d$ (up to normalization):
\begin{gather}
    (f_1+f_2)(f_3-f_4)({c_H^{\dag}}^2-{c_V^{\dag}}^2)\nonumber \\+(f_1-f_2)(f_3+f_4)({d_H^{\dag}}^2-{d_V^{\dag}}^2)\nonumber \\+2(f_1f_3+f_2f_4)c_H^{\dag}d_H^{\dag}+2(f_1f_4+f_2f_3)c_H^{\dag}d_V^{\dag}\nonumber\\+2(f_1f_4+f_2f_3)c_V^{\dag}d_H^{\dag}+2(f_1f_3+f_2f_4)c_V^{\dag}d_V^{\dag} \ .
\end{gather}
When measuring the output photons (the order of the measurement does not matter because these outputs are orthogonal), there is a probability $\frac{1}{2}$ to measure two photons at the same output port, $c$ or $d$, that have the same polarization $H$ or $V$. In such a case, the resulting wave functions are:
\begin{eqnarray}
\frac{1}{2}(f_1+f_2)(f_3-f_4)~or~
\frac{1}{2}(f_1-f_2)(f_3+f_4) \ ,
\end{eqnarray}
and they are separable states.
 There is also a $\frac{1}{2}$  probability to measure one output photon in each port, one in $c$ and one in $d$, where they can be in an H or V state. The resulting wave functions read:
 \begin{eqnarray}
 \frac{1}{\sqrt{2}}(f_1f_3+f_2f_4)~or~\frac{1}{\sqrt{2}}(f_1f_4+f_2f_3) \ ,
\end{eqnarray}
and are the required Bell type entanglement between the two clusters. Thus, this procedure yields a maximal entanglement with a $\frac{1}{2}$ probability of success \footnote{For further reading see \cite{5lectures,BrowneRudolph,stanisic2015universal}.
 We use the same notations as in \cite{5lectures}.}. It's important to note that one should use feed-forward as part of the fusion, so in the case the resulting state is $\frac{1}{\sqrt{2}}(f_1f_4+f_2f_3)$, a series of single-qubit rotations will transfer this state to the state $\frac{1}{\sqrt{2}}(f_1f_3+f_2f_4)$ --- see equation (\ref{Cluster state after measuring a,b and getting 01+10}),(\ref{Multipication of all the Zd where d in n(b)}).

\begin{figure} [h!]
    \centering
    \includegraphics[width=0.9\linewidth]{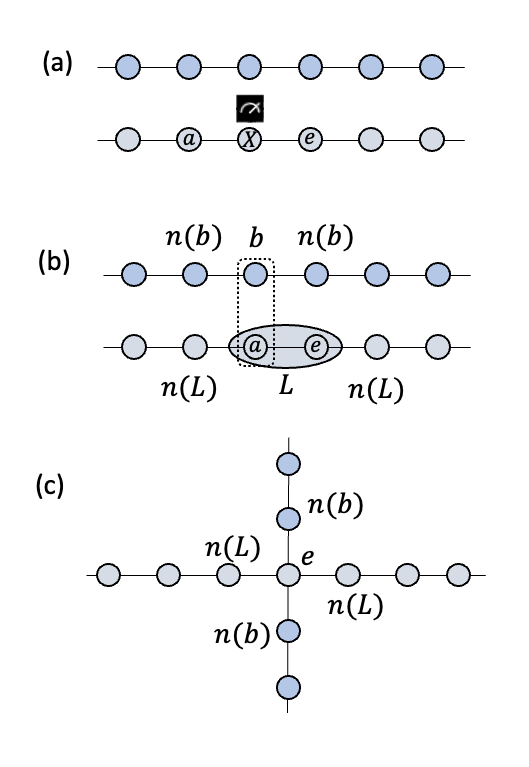}
    \caption{Type-II Fusion: (a) Two one-dimensional clusters. An $X$ measurement is performed on a qubit that belongs to one of the clusters, erases it and connects its two neighbors $a$ and $e$ to one logical qubit $L$. (b) Type-II fusion is performed
    on qubit $a$ in $L$ and qubit $b$ from the second cluster. (c) The result of the fusion: qubits $a,b$ are deleted, and qubit $e$ is connected to the previous neighbors $n(L)$ of the logical qubit $L$, and to the previous neighbors $n(b)$ of the qubit $b$.}
    \label{fig:FusionType2}
\end{figure}
The Type-II fusion plotted in Figure \ref{fig:FusionType2} is used for the construction of an $L$-shape cluster state in order to build a two-dimensional cluster from one-dimensional ones. 
Step (a): one begins with two one-dimensional clusters, and performs an $X$ measurement on one of the qubits of one of the two clusters. This erases the qubit from the cluster, and combines its two neighbors $a$ and $e$. 
Step (b): one combines $a,e$ qubits to one logical qubit $L$:
\begin{gather}
    \ket{0}_L=\ket{0}_a\ket{0}_e
    \hspace{0.5cm}
    \ket{1}_L=\ket{1}_a\ket{1}_e \ ,
\label{Logic qubit of a,e}
\end{gather}
and denotes by $b$ a qubit from the second cluster. Using the recursive relation in (\ref{Graph state recursive definition}), the total wave function takes the form:
\begin{align}
    \ket{\phi}=\left(\ket{0}_L\ket{\phi_L}_{V_L\setminus \{L\}}+\ket{1}_L \prod_{c\in n(L)} Z_c \ket{\phi_L}_{V_L\setminus \{L\}}\right) \nonumber \\
    \left(\ket{0}_b\ket{\phi_b}_{V_b\setminus \{b\}}+\ket{1}_b \prod_{d\in n(b)} Z_d \ket{\phi_b}_{V_b\setminus \{b\}} \right) \ ,
    \label{Cluster state before measuring a,b}
\end{align}
where $n(L),n(b)$ denote the neighbors of qubits $L$ and $b$, and we use for simplicity of notation
the same $V$ to denote the set of all the qubits in the clusters corresponding to $L$ and $b$. Note, that we assumed that $k_a=k_b=0$, but this can be modified by acting  with $Z_a$ and $Z_b$.

A successful type-II fusion on the qubits $a$ and $b$ is defined as:
operating with:
\begin{equation}
{\rm Type}~{\rm II}:~ \bra{0}_a\bra{0}_b+\bra{1}_a\bra{1}_b~ or~ \bra{0}_a\bra{1}_b+\bra{1}_a\bra{0}_b \ .
\end{equation}
Operating with $\bra{0}_a\bra{0}_b+\bra{1}_a\bra{1}_b$ yields the state:
\begin{gather}
    \ket{0}_e\ket{\phi_L}_{V_L\setminus \{L\}}\ket{\phi_b}_{V_b\setminus \{b\}} +
    \nonumber\\
    +\ket{1}_e\prod_{c\in n(L)} Z_c \ket{\phi_L}_{V_L\setminus \{L\}}\prod_{d\in n(b)} Z_d \ket{\phi_b}_{V_b\setminus \{b\}}
\label{Cluster state after measuring a,b and getting 00+11} \ ,
\end{gather}
which is the cluster state (c), as in the recursive definition (\ref{Graph state recursive definition}). In the new cluster, $e$ is connected both to the previous neighboring qubits
of the logical qubit $L$, as well as to the neighboring qubits of $b$, and have $k_e=0$ while it remains unchanged for all the other qubits.

Operating with $\bra{0}_a\bra{1}_b+\bra{1}_a\bra{0}_b$ yields the state:
\begin{gather}
    \ket{0}_e\ket{\phi}_{V_L\setminus \{L\}}\prod_{d\in n(b)} Z_d\ket{\phi}_{V_b\setminus \{b\}} +
    \nonumber\\
    +\ket{1}_e\prod_{c\in n(L)} Z_c \ket{\phi}_{V_L\setminus \{L\}} \ket{\phi}_{V_b\setminus \{b\}}
\label{Cluster state after measuring a,b and getting 01+10} \ , 
\end{gather}
which is a cluster state, having the same structure as in Figure \ref{fig:FusionType2} (c), but with different eigenvalues of the stabilizers in (\ref{Clusters eigenvalue equation}) compared to (\ref{Cluster state after measuring a,b and getting 00+11}): for all the qubits $d \in n(b)$, $(-1)^{k_d}$ change their sign, because this state is obtained from (\ref{Cluster state after measuring a,b and getting 00+11}) by operating with $Z_d$ on all $d\in n(b)$.
Thus, if one connects two graph states and aims at a resulting
graph state, an action with $Z_d$ on all $d\in n(b)$:
\begin{equation}
\label{Multipication of all the Zd where d in n(b)}
\ket{\phi'}_{V_b\setminus \{b\}}=\prod_{d\in n(b)} Z_d\ket{\phi}_{V_b\setminus \{b\}} \ ,
\end{equation}
is needed.
Substituting this in (\ref{Cluster state after measuring a,b and getting 01+10}) using $Z^2=I$ gives a state of the form of (\ref{Cluster state after measuring a,b and getting 00+11}).

\section{Generalized Type-II Fusion and the Final States}
\label{sec:Generalized Type II Fusion and the resulting state}

\begin{figure}
    \centering
    \includegraphics[width=1.0\linewidth]{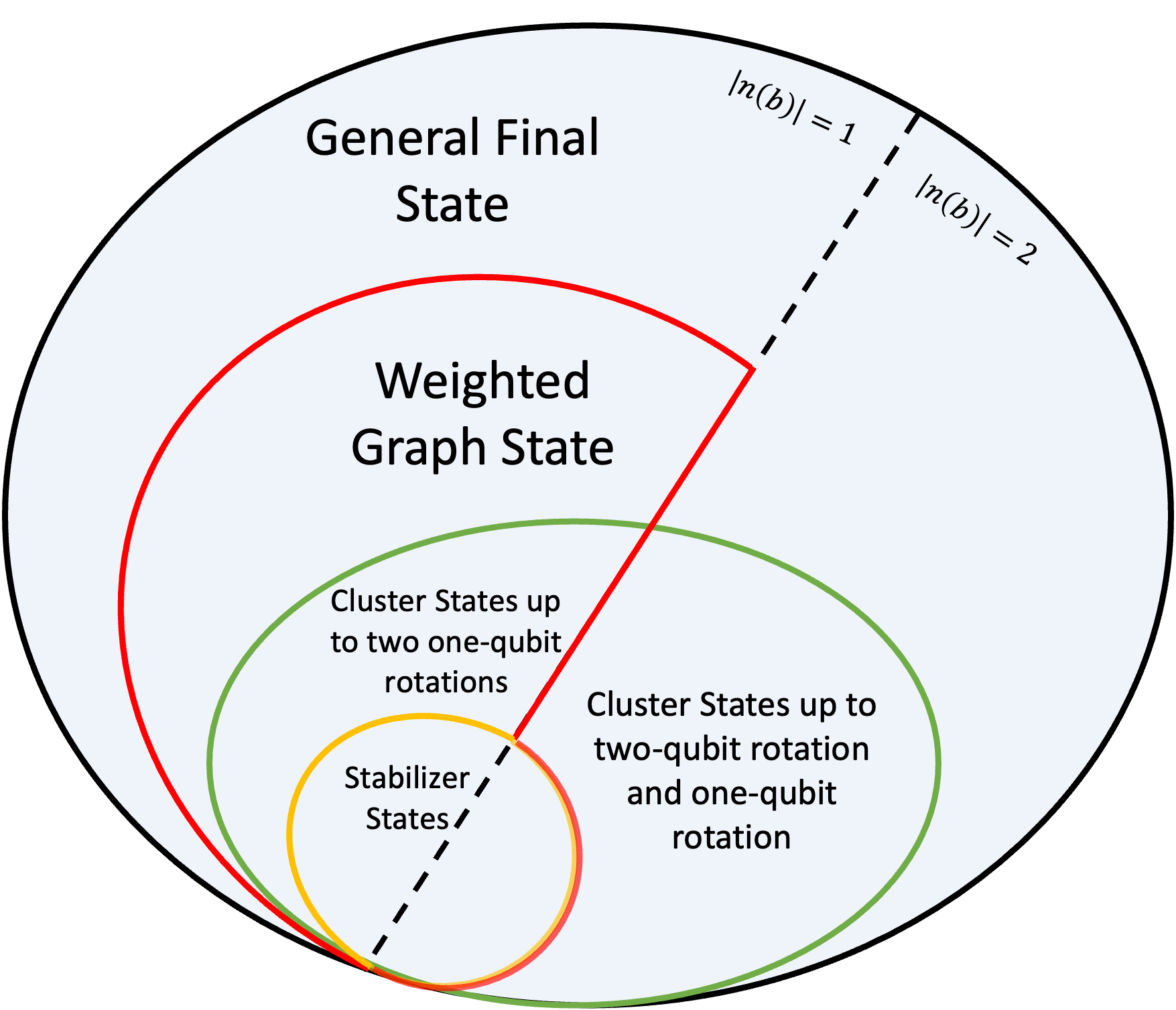}
    \caption{All the possible final states after performing a generalized type-II fusion.  The black outer set is the set of all the final states. The yellow set is the set of all the final states that are stabilizer states. The red set is the set of all the final states that are weighted graph states. The green set is the set of all the final states that are cluster states up to one two-qubit rotation when $|n(b)|=2$, and single-qubit rotations when $|n(b)|=1$.}
    \label{fig:graphs_heirarchy}
\end{figure}

In the following we introduce a generalization of type-II fusion, and the result of a general projection onto a two-qubit state, and prove three theorems on the conditions needed to obtain a stabilizer state, a weighted graph state or a cluster state up to one two-qubit rotation. The results of the theorems are summarized in Figure \ref{fig:graphs_heirarchy}, and by the explicit formulas in Figure \ref{fig:SetsHeirarchy} in section \ref{sec:Summary of Generalized Type II Fusion} in the appendix.

Consider Figure \ref{fig:graphs_heirarchy}. There is the set of all the possible final states of the two fused clusters, which are not necessarily cluster states. Within this general set we define three classes of final states. 
The first class consists of stabilizer states, which we consider in theorem \ref{Theorem - when it's cluster state up to 1-qubit gate operation}, where we can operate with a phase shift gate on qubit $e$ to generate a cluster state. The set of stabilizer states is marked by yellow in Figure \ref{fig:graphs_heirarchy}. 

The second class consists of weighted graph states having the same structure as the required fused cluster states. There are two cases, which we consider in theorem \ref{Theorem - when it's graph state}: (i) $|n(b)|=2$, which means that the qubit $b$ that we fuse is from the middle of its cluster. Then, every weighted graph state with the same structure is also a stabilizer state. (ii) $|n(b)|=1$, which means that the qubit $b$ that we fuse is from the edge of its cluster. Then, the final states are also cluster states up to one two-qubit rotation. Note, however, that such a cluster state can be generated also by applying one $CZ$ gate. The set of Weighted graph states is marked by red in Figure \ref{fig:graphs_heirarchy}. 

The third class consists of cluster states up to one two-qubit rotation and one single-qubit rotation, when $|n(b)|=2$, and two single-qubit rotations, when $|n(b)|=1$. This is a set of states that can take the form of the cluster state by operating with two-qubit gate on $n(b)$ and a single qubit gate on $e$ in the case $|n(b)|=2$, and single qubit gates on $n(b)$ and  a single qubit gate on $e$ when $|n(b)|=1$. The case $|n(b)|=1$ is very useful. The case $|n(b)|=2$ is less useful, as it requires the application of a 2-qubit gate, which however is
still less than the three $CZ$ gates, that need to be applied in order to construct the required fused cluster state from the original one-dimensional clusters without fusion. In theorem \ref{Theorem - when it can be made cluster state by operating with 2 and 3 qubits gates} we identify these states and prove that if $|n(b)|=1$, then they are weighted graph states. In theorem \ref{Theorem - maximally entangled graph state} we prove that when $|n(b)|=1$, it includes all the weighted graph states with maximal entanglement entropy. This set of states that are cluster states up to one two-qubit rotation (and one single-qubit rotation) or single-qubit rotations is marked by green in Figure \ref{fig:graphs_heirarchy}.

Figure \ref{fig:graphs_heirarchy} shows the hierarchy of the sets of final states after fusion. The outer set is divided by a line to two subsets for the cases $|n(b)|=1$ and $|n(b)|=2$, corresponding to the cases where the qubit $b$ is or is not at the end of its one-dimensional cluster before the fusion, respectively. When $|n(b)|=1$, the yellow set of stabilizer states is a subset of the green set, that includes cluster states up to single-qubit rotations, which is a subset of the red set of weighted graph states. When $|n(b)|=2$, 
the yellow set of stabilizers states is identified with the red set of weighted graph states, and is as subset of the green set of cluster states up to a one two-qubit rotation.

It is important to note that all the results from equation (\ref{Coefficients A,B,C,D}) and forward, including all the theorems of this part, are relevant for any projection onto a two-qubit state with general parameters $A,B,C,D$ that are defined later in equation (\ref{projectionOperatorForGeneralU}).

\subsection{Generalized Type-II Fusion}

Consider two incoming photons corresponding to creation operators $a^\dag_H,a^\dag_V,b^\dag_H,b^\dag_V$. Using linear optics devices, the transformation of the creation operators is unitary $U$ \cite{CreatingU0,CreatingU1,CreatingU2}:
\begin{equation}
    \begin{bmatrix}
        a^\dag_H \\
        a^\dag_V \\
        b^\dag_H \\
        b^\dag_V
    \end{bmatrix} = \begin{bmatrix}
     U_{11} & U_{12} & U_{13} & U_{14} \\
     U_{21} & U_{22} & U_{23} & U_{24} \\
     U_{31} & U_{32} & U_{33} & U_{34} \\
     U_{41} & U_{42} & U_{43} & U_{44}
    \end{bmatrix}
    \begin{bmatrix}
        c^\dag_H \\
        c^\dag_V \\
        d^\dag_H \\
        d^\dag_V
    \end{bmatrix} \ .
    \label{Unitary transformation}
\end{equation}
For instance, the PBS2 discussed in the previous section corresponds to
\begin{gather}
    U=\frac{1}{2}\begin{bmatrix}
    1 & 1 & 1 & -1 \\
     1 & 1 & -1 & 1 \\
     1 & -1 & 1 & 1 \\
     -1 & 1 & 1 & 1
\end{bmatrix} \ .
\label{PBS2 matrix U}
\end{gather}
We will refer to the matrix $U$ (\ref{Unitary transformation}) as the "fusion matrix" further on.

Any unitary matrix $U$ can be realized \cite{CreatingU0,CreatingU1,CreatingU2} by using beam splitters and phase shifters, that act as a general $2\times 2$ unitary matrices on two channels.

\begin{remark}
    In general, $U$ can be $4$ on $n\geq 4$ isometry --- which is equivalent to adding $n-4$ vacuum ancilla qubits to the two incoming photons and then doing $n\times n$ unitary. In this work we assume no such vacuum ancilla qubits.
\end{remark}

As in the previous section, we express (\ref{C_1*C_2}) in terms of $c^\dag$ and $d^\dag$ creation operators using the matrix elements of $U$ and measure $(c,d)$. We get
the final cluster state written in terms of  of $f_1, f_2, f_3, f_4$ as:
\begin{gather}
    Af_1f_3+Bf_1f_4+Cf_2f_3+Df_2f_4 \ ,
\label{Coefficients A,B,C,D} 
\end{gather}
for some $A,B,C,D$ coefficients that depend on the elements of $U$ and the results of the measurements of $(c,d)$ (see (\ref{The coefficients of the wave function}) and remark \ref{Remark: ABCD coeffients} for the calculation of these coefficients). In the language of cluster state construction, we assume that the logical qubit $L$ is composed of physical qubits $a$ and $e$ and is part of a one-dimensional cluster, while qubit $b$ belongs to a second cluster. Then we operate on the state (\ref{Cluster state before measuring a,b}) with the fusion operator (up to normalization):
\begin{gather}
    \label{projectionOperatorForGeneralU}
    A\bra{0}_a\bra{0}_b+B\bra{0}_a\bra{1}_b+C\bra{1}_a\bra{0}_b+D\bra{1}_a\bra{1}_b \ .
\end{gather}
This yields a state:
\begin{gather}
    A\ket{0}_e\ket{\phi}_{V_L\setminus \{L\}}\ket{\phi}_{V_b\setminus \{b\}}
    \nonumber\\
    +B\ket{0}_e\ket{\phi}_{V_L\setminus \{L\}}\prod_{d\in n(b)} Z_d\ket{\phi}_{V_b\setminus \{b\}}
    \nonumber\\
    +C\ket{1}_e\prod_{c\in n(L)} Z_c \ket{\phi}_{V_L\setminus \{L\}} \ket{\phi}_{V_b\setminus \{b\}}
    \nonumber\\
    +D\ket{1}_e\prod_{c\in n(L)} Z_c \ket{\phi}_{V_L\setminus \{L\}}\prod_{d\in n(b)} Z_d \ket{\phi}_{V_b\setminus \{b\}} \ .
    \label{Cluster state after measuring a,b for general unitary transformation}
\end{gather}

From this point on, we assume that $A,B,C,D$ are describing a general projection operator in (\ref{projectionOperatorForGeneralU}) and not necessarily one that arises from the fusion matrix $U$ and the results of a measurement of $c,d$ as in (\ref{The coefficients of the wave function}). For example, a projection onto a two-qubit state of qubits $a,b$ using also ancilla qubits, as in \cite{bartolucci2021creation}. Given the state (\ref{Cluster state after measuring a,b for general unitary transformation}), one can ask if it is a stabilizer state, and whether one can define its stabilizers as in (\ref{Clusters eigenvalue equation}).
In theorem \ref{Theorem - when it's cluster state up to 1-qubit gate operation} we will consider these questions, provide a description of these states and how
they can be recast in the cluster state form, by applying a phase-shift gate to the qubit $e$. This already allows
for more final states than the four Bell states in the proof of \cite{MaxEfficiency}.
One can also ask, whether the final states can be weighted graph states, and whether they can be transformed
to cluster states by rotations.
In theorem \ref{Theorem - when it's graph state}, we will provide a complete analysis of the final states that are weighted graph states. We will show that the difference between them and the cluster states is one edge. In theorem \ref{Theorem - when it can be made cluster state by operating with 2 and 3 qubits gates}, we will describe the states that can be recast as cluster states by one two-qubit rotation. We will prove in theorem \ref{Theorem - maximally entangled graph state}, that when $|n(b)|=1$ these states are weighted graph states with maximal entanglement entropy, and can be recast as cluster states by single-qubit rotations.

\subsection{Final Stabilizer States}
\label{Subsection:Final Stabilizer State}

In this subsection we will characterize final states (\ref{Cluster state after measuring a,b for general unitary transformation}) that are stabilizer states, as defined in subsection \ref{Subsection:Stabilizer States}.
We will retain the structure of the stabilizers as in (\ref{Clusters eigenvalue equation}), and allow the stabilizers to be composed of single qubit operators that are not necessarily Pauli matrices. We will consider the cases were the stabilizer operators differ from the stabilizer operators of the required fused cluster state (\ref{Clusters eigenvalue equation}) only by the operators acting on qubit $e$. In particular, every final state that is the required fused cluster state up to a single-qubit rotation on qubit $e$, is also a stabilizer state because applying such a rotation on the operators acting on $e$, on the stabilizers of the required fused cluster state yields the set of stabilizers for the appropriate final state. In theorem \ref{Theorem - when it's cluster state up to 1-qubit gate operation} we show that the converse also holds - the final states that are stabilizer states are also the required fused final state up to a single-qubit rotation acting on $e$, specifically a $Z$ rotation. The results are summarized in Figure \ref{fig:stab_condition}.

\begin{theorem}
    [Stabilizer state] The state (\ref{Cluster state after measuring a,b for general unitary transformation}) is a stabilizer state iff $A=D=0$ and $|B|=|C|$, or $B=C=0$ and $|A|=|D|$. In these two cases, the cluster state is obtained
    by applying a single qubit rotation gate on the qubit $e$. This theorem holds for any value of $|n(b)|$ (even $|n(b)|>2$).
    \label{Theorem - when it's cluster state up to 1-qubit gate operation}
\end{theorem}

\begin{proof}
    For $a'$ that is different from $e$ and not in $n(L)\cup n(b)$, the stabilizers remain the same as those of the wanted cluster as in equation (\ref{Clusters eigenvalue equation}). This holds also for $a'\in n(L)$, because then the $A$ and $C$ terms in (\ref{Cluster state after measuring a,b for general unitary transformation}) are multiplied by $(-1)^{k_{a'}}$, and this also applies to the $B$ and $D$ terms (this does not hold for $a'\in n(b)$).

For $a'=e$ we would like to replace $X_e$ from equation (\ref{Clusters eigenvalue equation}) with a new operator, and for $a'\in n(b)$ we would like to replace the $Z_e$ operator in the product in equation (\ref{Clusters eigenvalue equation}) with a new operator. 

\begin{lemma}
\label{Lemma - Conditions for stabilizer in e and the stabilizer}
    If $a'=e$ then the conditions for having a proper stabilizer $K_e$ are $|A|=|D|$, $|B|=|C|$ and $AC=BD$, which means that there exists a real number $\Phi$ such that $D=e^{i\Phi}A$ and $C=e^{i\Phi}B$. In such a case, the stabilizer
    reads:

    \begin{gather}
    \label{New stabilizer for e}
        K_{e}=T_e\prod_{c\in n(L)\cup n(b)} Z_c \ ,
    \end{gather}
 where we defined the operator that replaces $X_e$ in (\ref{Clusters eigenvalue equation}):
 \begin{gather}
    \label{Operator that replacing Xe}
    T_e=
    \begin{bmatrix}
        0&e^{i\Phi}\\
        e^{-i\Phi}&0
    \end{bmatrix} \ ,
    \end{gather}
which commutes with all the other stabilizers as required. 

\begin{figure}
    \centering
    \includegraphics[width=1.0\linewidth]{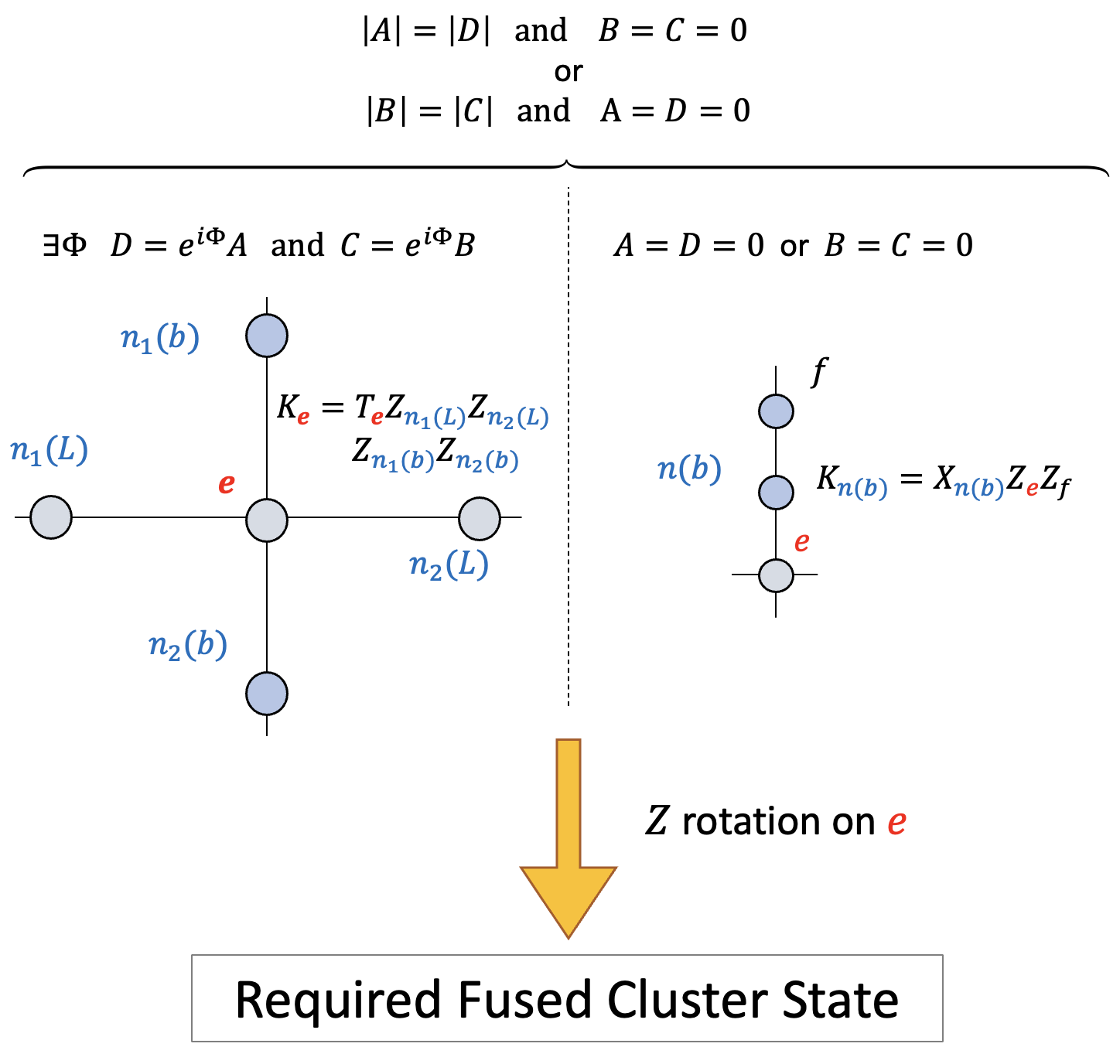}
    \caption{Stabilizer states - For all the qubits that differ from $e$ and are not in $n(b)$, the  appropriate stabilizers exist and are the same as the stabilizers (\ref{Clusters eigenvalue equation}) of the required fused cluster state. For $e$ and $n(b)$ such stabilizers do not necessarily exist. The left part of the figure displays the conditions on the coefficients $A,B,C,D$ of the final state (\ref{Cluster state after measuring a,b for general unitary transformation}), such that there exists a stabilizer for $e$, and the appropriate stabilizers, where the operator $T_e$ is given in (\ref{Operator that replacing Xe}). The right part of this figure displays the conditions on the coefficients $A,B,C,D$, such that there exist stabilizers for the qubits in $n(b)$, with the appropriate stabilizers being the same as the stabilizers (\ref{Clusters eigenvalue equation}) of the required fused cluster state. The combination of these two conditions yields the conditions on $A,B,C,D$, in order for the final state to be a stabilizer state (as displayed at the top of the figure). One can transform this state to the required fused cluster state by applying a $Z$ rotation on $e$ (as displayed at the bottom).}
    \label{fig:stab_condition}
\end{figure}

\end{lemma}

The conditions of Lemma \ref{Lemma - Conditions for stabilizer in e and the stabilizer} and the appropriate stabilizer (\ref{New stabilizer for e}) are displayed in the left side of Figure (\ref{fig:stab_condition}).

\begin{proof}
    See proof \ref{Proof appendix - Lemma - Conditions for stabilizer in e and the stabilizer} in proof appendix.
\end{proof}

\begin{lemma}
    \label{Lemma - Conditions for stabilizer in n(b) and the stabilizer}
    If $a'\in n(b)$ then in order to have a proper stabilizer $K'_{a'}$ we must have $A=D=0$ or $B=C=0$, and the stabilizer is the stabilizer as in (\ref{Clusters eigenvalue equation}), which commutes with all the other stabilizers, including (\ref{New stabilizer for e}).
\end{lemma}

The conditions of Lemma \ref{Lemma - Conditions for stabilizer in n(b) and the stabilizer} are displayed in the right side of Figure (\ref{fig:stab_condition}).

\begin{proof}
    See proof \ref{Proof appendix - Lemma - Conditions for stabilizer in n(b) and the stabilizer} in proof appendix.
\end{proof}

To conclude, in order to have the stabilizers for $e$ and $n(b)$, we must have $A=D=0$ or $B=C=0$, and $|A|=|D|$ or $|B|=|C|$, which ends the proof. 

In both cases, we have a state state of the form (\ref{Cluster state after measuring a,b and getting 00+11}) or (\ref{Cluster state after measuring a,b and getting 01+10}), where $\ket{0}_e$, $\ket{1}_e$ are multiplied by phases $\delta_0$ and $\delta_1$, which we can correct by operating on $e$ with single qubit gates of the form:

\begin{gather}
    \begin{bmatrix}
        e^{-i\delta_0}&0\\
        0&e^{-i\delta_1} 
    \end{bmatrix} \ . 
\end{gather}
 Up to a total phase, this is a $Z$ rotation. After this correction, the final state is the standard cluster state.

\end{proof}

\begin{remark}
    If one requires the stabilizers to be elements of the Pauli group, then  $\Phi\in\{0,\pm\frac{\pi}{2},\pi\}$ in (\ref{Operator that replacing Xe}). In this case the coefficients $A,B,C,D$ (\ref{Cluster state after measuring a,b for general unitary transformation}) satisfy $D\in\{\pm A,\pm iA\}$ and $B=C=0$, or 
    $C\in\{\pm B,\pm iB\}$ and $A=D=0$. When $D=\pm A$ and $B=C=0$ or $C= \pm B$ and $A=D=0$
    we will get that the projection operator (\ref{projectionOperatorForGeneralU}) is on a Bell state and the final state  (\ref{Cluster state after measuring a,b for general unitary transformation}) is the required fused cluster state. For example, when $A=D$ and $B=C=0$ the final state is as in (\ref{Cluster state after measuring a,b and getting 00+11}), and when $B=C$ and $A=D=0$ the final state is as in (\ref{Cluster state after measuring a,b and getting 01+10}).
\end{remark}

 \begin{remark}
 \label{remark:proof is not complete}
    If one wishes to identify all the final states (\ref{Cluster state after measuring a,b for general unitary transformation}) that are stabilizer states with the regular definition (i.e. the stabilizers belongs to the Pauli group), without requiring the stabilizers to have the same structure as in (\ref{Clusters eigenvalue equation}), then this is not a complete proof.  
\end{remark}

\begin{remark}
    As a continuation to remark \ref{remark:proof is not complete}, even without the requirement that the new stabilizers differ from those of the required fused cluster state we will have more states. In the case $|n(b)|=1$ there are possible final states that are the required fused cluster state up to two single-qubit rotations on $e$ and $n(b)$ (shown in subsection \ref{Subsection:Cluster State Up To two-Qubit Rotations}), and these states will have proper stabilizers where the operators (in the stabilizers) of $e$ and the qubit in $n(b)$ will be rotated accordingly.
\end{remark}

\subsection{Final Weighted Graph States}
\label{Subsection:Final Generalized Graph State}
\begin{figure}
    \centering
    \includegraphics[width=1.0\linewidth]{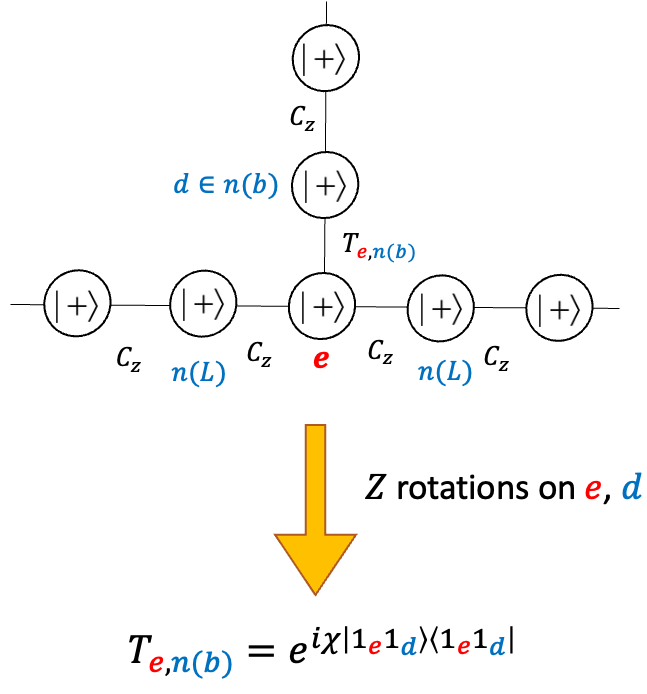}
    \caption{Weighted graph states - by theorem \ref{Theorem - when it's graph state}, when $|n(b)|=1$ then the final state (\ref{Cluster state after measuring a,b for general unitary transformation}) is a weighted graph state having the same structure as the required fused cluster state, iff the coefficients $A,B,C,D$ satisfy (\ref{Conditions on A,B,C,D for graph state}). In this case, the final state (\ref{Cluster state after measuring a,b for general unitary transformation}) can be described by starting with all the qubits in a $\ket{+}$ state, and applying $CZ$ gates for every two qubits that are connected by an edge, as in the construction of the regular graph state (\ref{Graph state product of CZ definition}), except for the gate between $e$ and the single qubit in $n(b)$, which by $Z$ rotations on these qubits can be transform to the form (\ref{The 2-qubits gate between e and n(b) for graph state after z-rotations}). When $|n(b)|=2$, then the final state (\ref{Cluster state after measuring a,b for general unitary transformation}) is a weighted graph state having the same structure as the required fused cluster state, iff it is a stabilizer state as in subsection \ref{Subsection:Stabilizer States}.}
    \label{fig:generalized_gs}
\end{figure}

Although the result of the generalized type-II fusion is almost always not a stabilizer state, one can still ask whether the state  (\ref{Cluster state after measuring a,b for general unitary transformation}) is a weighted graph state as defined in subsection \ref{Subsection:Generalized Graph States}. The following theorem answers this question and is summarized in Figure \ref{fig:generalized_gs}.

\begin{theorem}[Weighted graph states]
\label{Theorem - when it's graph state}

    If the constants $A,B,C,D$ in (\ref{Cluster state after measuring a,b for general unitary transformation}) satisfy:
    \begin{gather}
        |A|^2+|B|^2=|C|^2+|D|^2=\frac{1}{2} \nonumber \\
        \Re{AB^*}=\Re{CD^*}=0 \ , 
        \label{Conditions on A,B,C,D for graph state}
    \end{gather}
    and $n(b)$ contain only one qubit, then it is  a weighted graph state of the form (\ref{Graph state product of CZ definition}), but with a different two-qubit gate than $CZ^{e,n(b)}$, acting on $e$ and the $a$ qubit from $n(b)$. It is:
    \begin{gather}
        T_{e,n(b)}=A\ket{0}_e\bra{0}_e+B\ket{0}_e\bra{0}_eZ_d
        \nonumber \\+C\ket{1}_e\bra{1}_e+D\ket{1}_e\bra{1}_eZ_d \ .
        \label{The 2-qubits gate between e and n(b) for graph state}
    \end{gather}
So the wave function (\ref{Cluster state after measuring a,b for general unitary transformation}) takes the form:
\begin{gather}
        \ket{\phi}=T_{e,n(b)}\prod_{(i,j)\in E\setminus \{(e,n(b))\}} CZ^{i,j} \prod_{k} \ket{+}_k \ .
        \label{The generalized graph state as a product of the CZs and T}
    \end{gather}
    
\end{theorem}

\begin{proof}
    See proof \ref{Proof appendix - Theorem - when it's graph state} in proof appendix.
\end{proof}

\begin{remark}{}
By $Z$ rotations on the qubits $e,d$ (which can be performed, since these rotations commute with all the two-qubit gates that construct the weighted graph state in equation (\ref{The generalized graph state as a product of the CZs and T})), we can transform (\ref{The 2-qubits gate between e and n(b) for graph state}) to the form (\ref{The 2-qubits gates for generalized graph state after z-rotations}):

    \begin{gather}
        T_{e,n(b)}=e^{i\chi\ket{1_e,1_d}\bra{1_e,1_d}}
        =\ket{0_e,0_d}\bra{0_e,0_d}+\ket{0_e,1_d}\bra{0_e,1_e}\nonumber \\
        +\ket{1_e,0_d}\bra{1_e,0_e}+e^{i\chi}\ket{1_e,1_d}\bra{1_e,1_e} \ .
        \label{The 2-qubits gate between e and n(b) for graph state after z-rotations}
    \end{gather}
\end{remark}

\begin{remark}
    When $A,B,C,D$ do not satisfy these conditions, one can operate with a two-qubit gate on $d,e$, but only after operating with all the $CZ$ gates on all the nearest-neighbor qubits. These operations do not commute, and the wave function is no longer a weighted graph state.
\end{remark}

\begin{remark}
    If $|n(b)|=2$ (or larger), then it is a weighted graph state (with the same structure as the cluster state that arises via the regular type-II fusion), iff the state is a stabilizer state - satisfying the conditions of theorem \ref{Theorem - when it's cluster state up to 1-qubit gate operation}.
\end{remark}

\begin{remark}
    The conditions of (\ref{Conditions on A,B,C,D for graph state}) are equivalent to the existence of real numbers $\theta_1$, $\phi_1$, $\theta_2$, $\phi_2$ such that:
    \begin{gather}
        (A,B)=\frac{e^{i\theta_1}}{\sqrt{2}}(\cos{\phi_1}, i\sin{\phi_1}), \nonumber \\
        (C,D)=\frac{e^{i\theta_2}}{\sqrt{2}}(i\sin{\phi_2},\cos{\phi_2}) \ .
        \label{Parameterization of A,B,C,D in the case of graph state}
    \end{gather}
 In this form, the angle $\chi$ (\ref{The 2-qubits gate between e and n(b) for graph state after z-rotations}) is:
       \begin{gather}
        \chi=2(\phi_1-\phi_2)+\pi \ .
        \label{chi by phi1,2}
    \end{gather} 
\end{remark}

\begin{proof}
    For remarks for theorem \ref{Theorem - when it's graph state} see also proof \ref{Proof appendix - Theorem - when it's graph state} in proof appendix.
\end{proof}
 
Note, that when $|n(b)|=1$, they are also cluster states up to one two-qubit rotation, since we can apply two-qubit gate as in (\ref{The 2-qubits gate between e and n(b) for graph state after z-rotations}), that transform $\chi$ to $\pi$ yielding the cluster state (because $T_{e,n(b)}$  becomes $CZ$). This does reduce two-qubit gate resources, since instead of applying this two-qubit gate, we can instead of the fusion apply $CZ$ gate at the beginning between a qubit from one of the original one-dimensional clusters, and another qubit from the edge of the other one-dimensional cluster,
and obtain the required fused cluster state (still with $|n(b)|=1$). 

The natural question that arises is whether a projection operator (\ref{projectionOperatorForGeneralU}) that satisfy the conditions (\ref{Conditions on A,B,C,D for graph state}) for weighted graph state can arise from the generalized fusion we defined (\ref{Unitary transformation}) --- the answer for that is affirmative; see subsection \ref{Subsection:Example for generating weighted graph state}.

\subsection{Schmidt Decomposition}
\label{Subsection:Schmidt Decomposition}
Theorem \ref{Theorem - when it's graph state} does not guarantee that the final state (\ref{Cluster state after measuring a,b for general unitary transformation})
is the required fused cluster state. It will be useful to recast the final state via a Schmidt decomposition of the normalized state:
\begin{gather}
    Af_1f_3+Bf_1f_4+Cf_2f_3+Df_2f_4=\alpha g_1 g_3 +\beta g_2 g_4    \ .
    \label{Schmidt decomposition f's language}
\end{gather}
Here, $\{g_1,g_2\}$ and $\{g_3,g_4\}$ are orthonormal bases of $span\{f_1,f_2\}$ and $span\{f_3,f_4\}$, respectively. $\alpha, \beta$ are non-negative real numbers that satisfy
 $|\alpha |^2+|\beta |^2=1$. 
The wave function (\ref{Cluster state after measuring a,b for general unitary transformation}) takes the 
Schmidt decomposition form:
\begin{gather}
\label{Schmidt decomposition final state language}
    \alpha \ket{0'}_e\ket{\phi'_L}_{V_L\setminus \{L\}} \ket{\phi'_b}_{V_b\setminus \{b\}}
    +\beta \ket{1'}_e\ket{\phi''_L}_{V_L\setminus \{L\}} \ket{\phi''_b}_{V_b\setminus \{b\}} \ , 
\end{gather}
where, the two bases are:
\begin{gather}
    \{\ket{0'}_e\ket{\phi'_L}_{V_L\setminus \{L\}},\ket{1'}_e\ket{\phi''_L}_{V_L\setminus \{L\}}\},
    \nonumber\\
    \{\ket{\phi'_b}_{V_b\setminus \{b\}},\ket{\phi''_b}_{V_b\setminus \{b\}}\} \ .
\end{gather}
These two bases can be reached by performing unitary transformations on
\begin{equation}
\{\ket{0}_e\ket{\phi_L}_{V_L\setminus \{L\}},\ket{1}_e\prod_{c\in n(L)} Z_c \ket{\phi_L}_{V_L\setminus \{L\}}\} \ ,     
\end{equation}
and on
\begin{equation}
\{\ket{\phi_b}_{V_b\setminus \{b\}},\prod_{d\in n(b)} Z_d \ket{\phi_b}_{V_b\setminus \{b\}}\} \ .
\end{equation}
These transformations can be performed by 
\begin{gather}
    T_{11}\ket{0}_e\bra{0}_e\prod_{c\in n(L)} I_c+T_{12}\ket{1}_e\bra{0}_e\prod_{c\in n(L)} Z_c
    \nonumber \\
    +T_{21}\ket{0}_e\bra{1}_e\prod_{c\in n(L)} Z_c + T_{22}\ket{1}_e\bra{1}_e\prod_{c\in n(L)} I_c  \ ,
    \nonumber \\
    S_{11}\ket{\phi_b}_{V_b\setminus \{b\}}\bra{\phi_b}_{V_b\setminus \{b\}}  +S_{12} \prod_{d\in n(b)} Z_d \ket{\phi_b}_{V_b\setminus \{b\}}\bra{\phi_b}_{V_b\setminus \{b\}}
    \nonumber \\
    +S_{21}\ket{\phi_b}_{V_b\setminus \{b\}}\bra{\phi_b}_{V_b\setminus \{b\}}\prod_{d\in n(b)} Z_d
    \nonumber \\ +S_{22} \prod_{d\in n(b)} Z_d \ket{\phi_b}_{V_b\setminus \{b\}}\bra{\phi_b}_{V_b\setminus \{b\}}\prod_{d\in n(b)} Z_d \ ,
    \label{Hard transformations}
\end{gather}
where 
\begin{gather}
    \begin{bmatrix}
            T_{11}&T_{12} \\
            T_{21}&T_{22}
    \end{bmatrix} \ ,
    \begin{bmatrix}
            S_{11}&S_{12} \\
            S_{21}&S_{22}
    \end{bmatrix} \ ,
\end{gather}
are unitary matrices. Note, that (\ref{Hard transformations}) are hard transformations in general, that may require $\abs{V_b\setminus \{b\}}$ and three-qubit operations (if $L$ at the edge of it's original cluster then two-qubit operation is required instead of three-qubit operation). In special cases, that will be discussed
in the next subsection, we will be able to arrive at the Schmidt decomposition (\ref{Schmidt decomposition final state language})
by a simpler set of operations.


\subsection{Cluster States Up To Two-Qubit Rotation}
\label{Subsection:Cluster State Up To two-Qubit Rotations}

\begin{figure}
    \centering
    \includegraphics[width=1.0\linewidth]{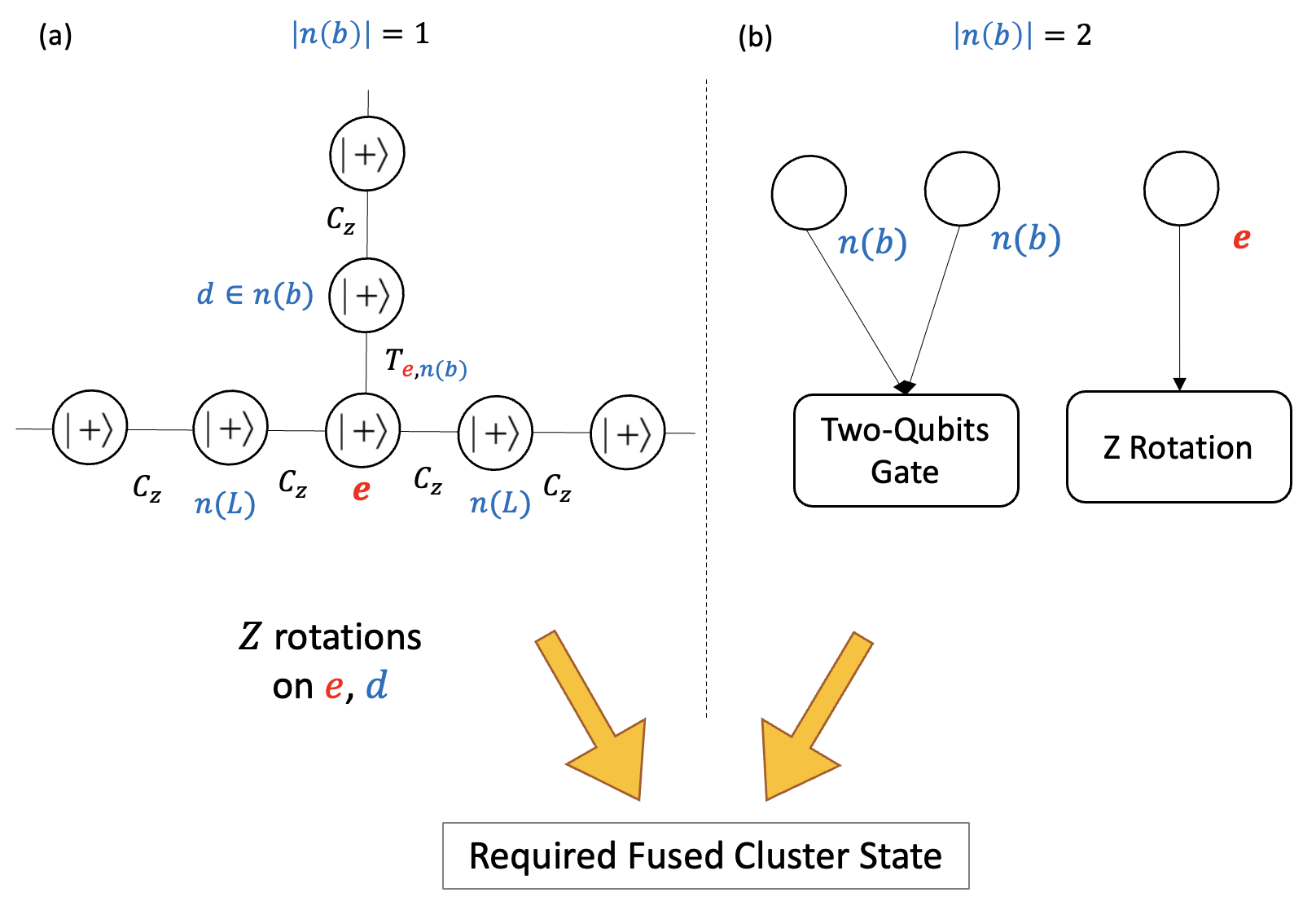}
    \caption{Cluster State Up To Two-Qubit Rotation: as proven in theorem \ref{Theorem - when it can be made cluster state by operating with 2 and 3 qubits gates}, the set of final states (\ref{Cluster state after measuring a,b for general unitary transformation}) that satisfy conditions (\ref{Conditions to get cluster state after hard transformations}) depends on $|n(b)|$. a) When $|n(b)|=1$ their are weighted graph states with $\chi$ (\ref{chi by phi1,2}) equal to zero. Hence, the required fused Cluster state can be achieved from those states by $Z$ rotations on $e$ and the qubit $d$ in $n(b)$. b) When $|n(b)|=2$ we can apply a two-qubit rotation on $n(b)$ and $Z$ rotation on $e$ that will transfer those states to the required fused Cluster state.}
    \label{fig:cluster_state_up_to_2qubit_rot}
\end{figure}

After performing the transformation (\ref{Hard transformations}) on the final state (\ref{Cluster state after measuring a,b for general unitary transformation}), we arrived at a Schmidt decomposition form (\ref{Schmidt decomposition final state language}). 
In the following theorem we will answer the question, when is the state (\ref{Schmidt decomposition final state language}) a cluster state, summarized in figure \ref{fig:cluster_state_up_to_2qubit_rot}. That is, when does it satisfy the conditions: 
\begin{eqnarray}
    \ket{\phi''_L}_{V_L\setminus \{L\}}&=&\prod_{c\in n(L)} Z_c\ket{\phi'_L}_{V_L\setminus \{L\}},
    \nonumber \\
    \ket{\phi''_b}_{V_b\setminus \{b\}}&=&\prod_{d\in n(b)} Z_d\ket{\phi'_b}_{V_b\setminus \{b\}} \ ,\nonumber\\
    \alpha &=& \beta \ .
    \label{Conditions to get cluster state after hard transformations}
\end{eqnarray}
We will also derive transformations that require less resources than (\ref{Hard transformations}), which when acting on (\ref{Cluster state after measuring a,b for general unitary transformation}), result in a cluster state.

\begin{theorem}[Cluster States Up To Two-Qubit Rotation]
    \label{Theorem - when it can be made cluster state by operating with 2 and 3 qubits gates}
    
    (i) Conditions (\ref{Conditions to get cluster state after hard transformations})  are satisfied iff there exist real numbers $\theta_1$, $\theta_2$ and $\phi$ such that:
\begin{gather}
        \label{General form of A,B,C,D in order to get cluster state after hard transformations}
        A,D=\frac{e^{i\theta_{1,2}}}{\sqrt{2}}\cos{\phi}  \hspace{0.5cm}B,C=\frac{ie^{i\theta_{1,2}}}{\sqrt{2}}\sin{\phi} 
        \ .
    \end{gather}
    
    (ii) If conditions (\ref{Conditions to get cluster state after hard transformations})  are satisfied then the final state (\ref{Cluster state after measuring a,b for general unitary transformation}) is a cluster state up to a two-qubit rotation on $n(b)$ and a Z rotation on $e$.

    We will refer to such states as 'cluster states up to a two-qubit rotation'.
\end{theorem}

\begin{remark}
    If $|n(b)|=1$ then the final state (\ref{Cluster state after measuring a,b for general unitary transformation}) is a cluster state up to Z rotations on $n(b)$ and $e$. For a general value of $|n(b)|$ the final state (\ref{Cluster state after measuring a,b for general unitary transformation}) is a cluster state up to $|n(b)|$-qubit rotation on $n(b)$ and Z rotation on $e$.
\end{remark}

\begin{proof}
    See proof \ref{Proof appendix - Theorem - when it can be made cluster state by operating with 2 and 3 qubits gates} in proof appendix.
\end{proof}

When $|n(b)|=1$, then the states that satisfy theorem \ref{Theorem - when it can be made cluster state by operating with 2 and 3 qubits gates} are cluster states up to two single-qubit rotations, in comparison to the weighted graph states, that are cluster states up to a two-qubit rotation. Also, when $|n(b)|=1$, the states that satisfy theorem \ref{Theorem - when it can be made cluster state by operating with 2 and 3 qubits gates} are weighted graph states, since they satisfy the conditions of theorem \ref{Theorem - when it's graph state}, with $\phi_1=\phi_2$ in the parametrization (\ref{Parameterization of A,B,C,D in the case of graph state}) (which is equivalent to $\chi$ (\ref{chi by phi1,2}) being zero).
 Furthermore, in theorem \ref{Theorem - maximally entangled graph state} in the next section, we will prove that these states consist of the set of weighted graph states with maximal entanglement entropy.
 

When $|n(b)|=2$, the requirement to use a two-qubit gate in theorem \ref{Theorem - when it can be made cluster state by operating with 2 and 3 qubits gates}, also requires less resources since the alternative to the fusion process is to operate with three $CZ$ gates on the original one-dimensional clusters: we disconnect two adjacent qubits in one cluster by applying $CZ$ (since $(CZ)^2$ is the identity), and connect both to a third qubit from the second cluster by another two $CZ$ operations (one can even use only two $CZ$ gates, if there are three one-dimensional clusters).


\section{The Entanglement of Fusion}
\label{sec:prob_def}

In this section we compute (and define how because there are a some different ways to do so) the entanglement entropy of type-II fusion, which we will refer to as the "entanglement entropy of the fusion link" further on. We pose a question about the relationship
between the value of the entanglement entropy of the fusion, and its success probability. We prove certain theorems: Given a unitary transformation $U$, what are all the possible final states and their entanglement entropies, and for each state, what is the probability to obtain it.

\begin{figure}
    \centering
    \includegraphics[width=1.0\linewidth]{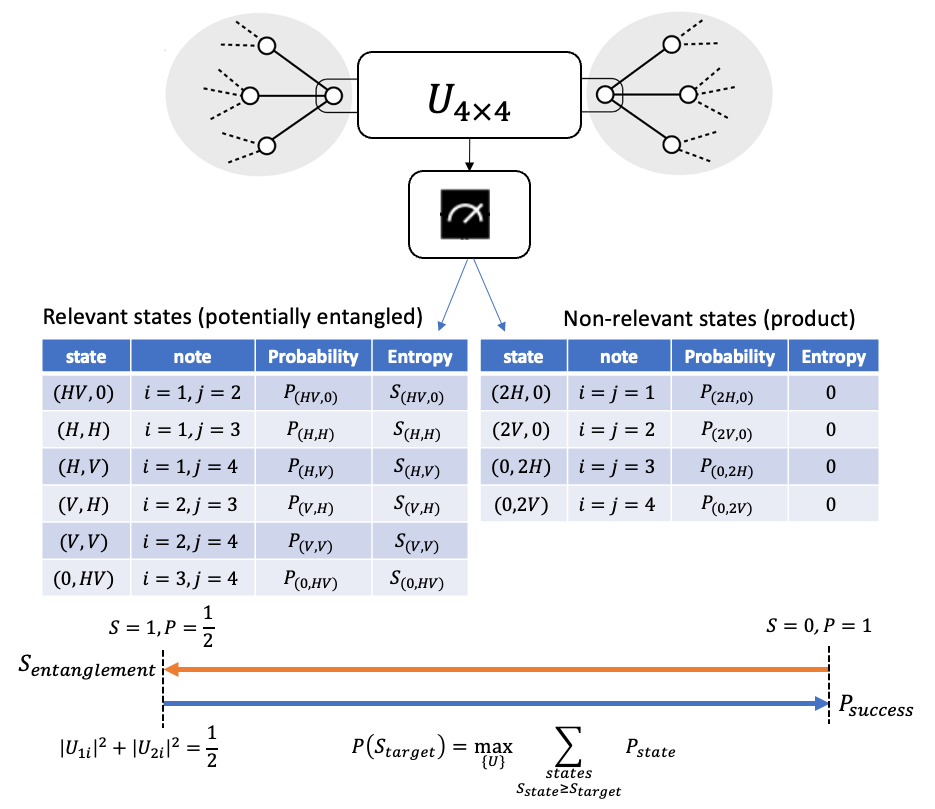}
    \caption{Reduction of the entanglement entropy and an increase of fusion success probability: given an entanglement entropy and a fusion matrix U, we calculate the probability of the fusion to result in a final state with 
    at least this entanglement. Given a value of the entanglement entropy, we define the success probability as the maximal probability over all the unitary matrices $U$, to have this entanglement entropy or higher. For the maximal entropy $S=1$, we prove analytically in theorems \ref{Theorem - the 0.5 bound} and \ref{Theorem - reaching the 0.5 bound} that $P=\frac{1}{2}$, which means that for a general fusion matrix $U$ the probability to have a maximally entangled state is less or equal to $\frac{1}{2}$, and that the matrix $U$ saturates this bound iff $|U_{1i}|^2+|U_{2i}|^2=\frac{1}{2}, i=1,2,3,4$.  For probability $P=1$, we prove analytically in theorem \ref{Theorem - we can't have probability 1} that $S=0$, which means that there is no entanglement entropy greater than zero that can be obtained in the generalized type-II fusion process with probability one.}
    \label{fig:FusionDiagram}
\end{figure}

We specifically compute the Von Neumann entropy \cite{EntanglementEntropyReview} of the new created link in the final state (\ref{Cluster state after measuring a,b for general unitary transformation}), as follows.  


We partition the set of $N$ qubits to two subsets $N_a$ and $N_b$, which we naturally choose as the two one-dimensional clusters (minus the measured qubits $a,b$). This induces a partition of the Hilbert space
${\cal H} = {\cal H}_{N_a}\times {\cal H}_{N_b}$. The entanglement entropy is the Von Neumann entropy of the reduced density matrix of ${N_a}$ obtained by tracing out ${N_b}$, $\rho_{N_a}= Tr_{N_b} \left(\rho_{{N_a}{N_b}}\right)$:
\begin{equation}
S = -Tr \left(\rho_{N_a} \log_2 \rho_{N_a} \right) \ .
\label{Formal definition of entanglement entropy}
\end{equation}

In the fusion process we obtained new links between $e$ and the qubits from $n(L)$ and $n(b)$, where the links between $e$ and $n(L)$ replaced the links between $L$ and $n(L)$ in the original one-dimensional cluster, and retain the same
structure, i.e  comparing (\ref{Graph state recursive definition}) and (\ref{Cluster state after measuring a,b for general unitary transformation}) one sees that in (\ref{Cluster state after measuring a,b for general unitary transformation})
$\ket{0}_e$ multiplies $\ket{\phi_L}_{V_L\setminus \{L\}}$ and $\ket{1}_e$ multiplies $\prod_{c\in n(L)} Z_c \ket{\phi_L}_{V_L\setminus \{L\}}$ as in the recursion relation (\ref{Graph state recursive definition}). 
One can also see this by the fact that the stabilizers of $n(L)$ retain their form as shown in the proof of theorem \ref{Theorem - when it's cluster state up to 1-qubit gate operation}. 
The link between $e$ and the two qubits $n(b)$ is symmetric under the exchange of the latter two, which supports the above partition of all the qubits, e.g. $N_a$ and $N_b$ are the initial one-dimensional clusters except the measured $a$ and  $b$  qubits.

We choose an orthonormal basis of $H_{N_a}$ that includes the states $\ket{0}_e\ket{\phi_L}_{V_L\setminus \{L\}}$ and $\ket{1}_e\prod_{c\in n(L)} Z_c \ket{\phi_L}_{V_L\setminus \{L\}}$, and an orthonormal basis of $H_{N_b}$ that includes $\ket{\phi_b}_{V_b\setminus \{b\}}$ and $\prod_{d\in n(b)} Z_d \ket{\phi_b}_{V_b\setminus \{b\}}$. In these bases, the decomposition of the Hilbert space and the corresponding entanglement entropy ({\ref{Formal definition of entanglement entropy}}) are analogous to having in (\ref{Coefficients A,B,C,D}), $f_1,f_2$ as $k^\dag_H,k^\dag_V$ and similarly $f_3,f_4$ as $p^\dag_H,p^\dag_V$, where before the fusion we effectively consider two Bell pairs. The fusion process erases one qubit from each pair, and we end up with a two-qubit state whose entanglement entropy (Von Neumann entropy) reads:
\begin{gather}
    S=-|\alpha|^2\log_{2}{|\alpha|^2}-|\beta|^2\log_{2}{|\beta|^2} \ ,
    \label{Entanglement entropy}
\end{gather}
where $\alpha,\beta$ are the coefficients from (\ref{Schmidt decomposition f's language}). Note that this is also the entanglement entropy of the two-qubit state $A\ket{0}_a\ket{0}_b+B\ket{0}_a\ket{1}_b+C\ket{1}_a\ket{0}_b+D\ket{1}_a\ket{1}_b$ on which we project (\ref{projectionOperatorForGeneralU}).

To summarize, when measuring the qubits $(c,d)$, there are several possible final states that can be obtained with certain probabilities, and associated entanglement entropies (\ref{Entanglement entropy}), defined by the Von Neumann entropy of the reduced density matrix after tracing out $f_3$,$f_4$ from (\ref{Coefficients A,B,C,D}). 
Since there are several possible final states, there are various ways to define the entanglement entropy of the type-II fusion process, such as the mean entanglement entropy.
We will consider the following question: Given an entanglement entropy $S$, what is the highest probability to obtain a state, whose entanglement entropy is at least $S$.
We will refer to such $S$, as the target entropy of the fusion.

Given a target entropy $S_{target}$, we define the probability to reach this target entropy, as the sum of the probabilities over all final states, whose entanglement entropy is at least $S_{target}$. Our goal will be to maximize the probability to reach $S_{target}$ in the fusion process, thus we will calculate:
\begin{equation}
    P(S_{target}) = \max_{\{U\}} \sum_{
        state ,\ S_{state} \geq S_{target}
    } p_{state} \ ,
    \label{eq:P_S_target}
\end{equation}
where the maximum is over all the possible fusion matrices $U$ (\ref{Unitary transformation}).

Figure \ref{fig:FusionDiagram} displays (\ref{eq:P_S_target}): we are given two cluster states that
we fuse. In each cluster we choose one qubit, and measure the two qubits.
After the measurement, there are ten possible final states, and as we will show, four of them are product states having zero entanglement entropy (non-relevant states), while the other six states are potentially entangled states (relevant states). For each final state we calculate the probability to obtain it and its entanglement entropy,
and evaluate the probability of the target entropy $P(S_{target})$.
By optimization over $U$ we find (\ref{eq:P_S_target}). By decreasing of the target entanglement entropy, we increase the probability and vice versa, as shown by the orange and blue arrows. 

In the following two sections we will prove analytically that $P(1)=\frac{1}{2}$, and this is satisfied for $U$ iff $|U_{1i}|^2+|U_{2i}|^2=1$ for every $i$. Also, $P(0)=1$ meaning that if $S_{target}>0$ then $P(S_{target})<1$, as featured also in Figure \ref{fig:FusionDiagram}. Then, in section \ref{sec:numerical_res} we plot numerically the graph of $P(S)$, as derived by numerical optimization.
In table \ref{tab:ComputationsSummarized} in section \ref{sec:Summary of usefull computations} in the appendix, we summarize the calculations and results of this section and the next one that do not appear in Figure \ref{fig:FusionDiagram}.

\subsection{Notations}

For simplicity, we will denote in the following the measurement channels $c_{H}, c_{V}, d_{H}, d_{V}$ by the indices $1,2,3,4$, respectively. Referring to the $(i,j)$ state will mean referring to the state outcome, after measuring a single photon in each of the $i,j$ states (if $i=j$ then we measure two photons in $i$ state). For instance, the state $(1,2)$ is the state that we obtain after measuring one photon in $c_H$ and one photon in $c_V$.
We define the variables:
\begin{gather}
    m_i=|U_{1i}|^2+|U_{2i}|^2
    \hspace{1cm}
    n_i=\frac{1}{2}-m_i \nonumber\\
    t_i=U_{1i}U_{2i}^*
    \hspace{1cm}
    k_i=|U_{1i}|^2-|U_{2i}|^2 \ .
    \label{m,n,t,k definitions}
\end{gather}

Note, that since U is unitary, one has the following relations (see proof \ref{Proof appendix - m,n,t,k relations} in proofs appendix for the full derivation):
\begin{gather}
    m_1+m_2+m_3+m_4=2 \nonumber\\
    n_1+n_2+n_3+n_4=0 \nonumber\\
    t_1+t_2+t_3+t_4=0 \nonumber\\
    k_1+k_2+k_3+k_4=0 \ .
    \label{m,n,t,k relations}
\end{gather}

\subsection{The Wave Function}

After measuring (\ref{C_1*C_2}) in the basis of $(c,d)$, we get a new wave function of the joined of two clusters (\ref{Cluster state after measuring a,b for general unitary transformation}). 
\begin{lemma}
\label{Lemma - the wave function}
    
The wave function (\ref{Cluster state after measuring a,b for general unitary transformation}) after measurement (\ref{C_1*C_2}) is as follows:\\
\noindent
(i) The wave function of the state $(i,i)$ reads:
\begin{gather}
    \ket{\phi}_{ii}=\frac{1}{N_{ii}}(U_{1i}U_{3i}f_{1}f_{3}+U_{1i}U_{4i}f_{1}f_{4} \nonumber\\ +U_{2i}U_{3i}f_{2}f_{3}+U_{2i}U_{4i}f_{2}f_{4})\nonumber\\
    =\frac{1}{N_{ii}}(U_{1i}f_{1}+U_{2i}f_{2})(U_{3i}f_{3}+U_{4i}f_{4}) \ ,
    \label{Wave function of non-relevant state}
\end{gather}
where $N_{ii}$ is the normalization factor. \\
\noindent
(ii) The wave function of the state (i,j) with $i\ne j$ reads:
\begin{equation}
    \ket{\phi}_{ij}=\frac{1}{N_{ij}}(a_{ij}f_{1}f_{3}+b_{ij}f_{1}f_{4}+c_{ij}f_{2}f_{3}+d_{ij}f_{2}f_{4}) \ ,
    \label{Wave function of relevant state}
\end{equation}
where the coefficients are: 
\begin{gather}
    a_{ij}=U_{1i}U_{3j}+U_{1j}U_{3i}
    \hspace{0.5cm}
    b_{ij}=U_{1i}U_{4j}+U_{1j}U_{4i} \nonumber\\
    c_{ij}=U_{2i}U_{3j}+U_{2j}U_{3i}
    \hspace{0.5cm}
    d_{ij}=U_{2i}U_{4j}+U_{2j}U_{4i} \nonumber\\
    N_{ij}=\sqrt{|a_{ij}|^2+|b_{ij}|^2+|c_{ij}|^2+|d_{ij}|^2}=\sqrt{4p_{ij}} \ ,
    \label{The coefficients of the wave function}
\end{gather}
and $p_{ij}$ is the probability to get the state $(i,j)$ which will be discussed in the next subsection.
\end{lemma}

\begin{remark}
\label{Remark: ABCD coeffients}
    Note that for every $i,j$, the coefficients $A,B,C,D$ (\ref{Cluster state after measuring a,b for general unitary transformation}) of the resulting state, are $a_{ij},b_{ij},c_{ij},d_{ij}$ up to the normalization factor $N_{ij}$, meaning $A=\frac{a_{ij}}{N_{ij}}$, etc.
\end{remark}

\begin{proof}
    This is proven in proofs appendix, proof \ref{Proof appendix - Lemma - the wave function}.
\end{proof}

From equation (\ref{Wave function of non-relevant state}) one concludes that $(i,i)$ is a product state.
We will refer to these product states as irrelevant states, and will refer to the $i\neq j$ states as relevant states, since they are potentially entangled.
\begin{remark}
\label{Remark - when two elements of U are zero}
    For a relevant state $(i,j)$, if two of the four elements $U_{1i},U_{2i},U_{1j}, U_{2j}$ are zero, then its is a product state, except for the cases $U_{1i}=U_{2j}=0$ or $U_{1j}=U_{2i}=0$. In particular if $n_i=\frac{1}{2}$ then the state $(i,j)$ is a product state.
\end{remark}

\subsection{The States Probabilities}

\begin{lemma}
\label{Lemma - the states probabilities}
    The probabilities to reach any of the states are as follows:\\
 \noindent   
(i)    
    The probability to obtain the relevant state $(i,j)$ ($i\neq j$) is:
    \begin{gather}
        p_{ij}=\frac{1}{4}(m_{i}(1-m_{j})+m_{j}(1-m_{i})) \nonumber\\ - \frac{1}{2} |U_{1i}U_{1j}^{*}+U_{2i}U_{2j}^{*}|^2 \nonumber\\
        = \frac{1}{8} - \frac{1}{2} n_{i}n_{j} - \frac{1}{2} |U_{1i}U_{1j}^{*}+U_{2i}U_{2j}^{*}|^2 \leq\frac{1}{4} \ .
        \label{Probability of relevant state}
    \end{gather}
\noindent 
(ii)
    The probability to obtain the non-relevant state $(i,i)$ is:
    \begin{gather}
        p_{ii}=\frac{1}{2}m_i(1-m_i)=\frac{1}{8}-\frac{1}{2}n_{i}^2 \ .
        \label{Probability of non-relevant state}
    \end{gather}

\end{lemma}

\begin{proof}
 This is proven in proofs appendix, proof \ref{Proof appendix - Lemma - the states probabilities}.
\end{proof}
\begin{remark}
\label{Remark: equality pij=one over four}
    If equality holds in equation (\ref{Probability of relevant state}), meaning $p_{ij}=\frac{1}{4}$, then in particular one of $n_i,n_j$ is $\frac{1}{2}$ and the other one is $-\frac{1}{2}$, and by remark \ref{Remark - when two elements of U are zero} the resulting state $(i,j)$ (\ref{Wave function of relevant state}) is a product state.
\end{remark}
Summing over $i$ in equation (\ref{Probability of non-relevant state}), we get that the probability of obtaining one of the non-relevant states is:

\begin{gather}
    p_{non-relevant}=\frac{1}{2}(1-n_{1}^2-n_{2}^2-n_{3}^2-n_{4}^2)\leq\frac{1}{2} \ .
    \label{Probability of all non-relevant states}
\end{gather}

\subsection{Reduced Density Matrix and Entanglement Entropy}

The reduced density matrix (obtained by tracing out $f_3$,$f_4$ from $\rho_{ij}=\ket{\phi}_{ij}\bra{\phi}_{ij}$) is:
\begin{equation}
    \rho_{ij}=\frac{1}{N_{ij}^{2}}
    \left(
    \begin{bmatrix}
            |a_{ij}|^2+|b_{ij}|^2 & a_{ij}^*c_{ij}+b_{ij}^*d_{ij} \\ a_{ij}c_{ij}^*+b_{ij}d_{ij}^* & |c_{ij}|^2+|d_{ij}|^2
    \end{bmatrix}
    \right) \ .
    \label{Density matrix}
\end{equation}
Denote by $\lambda_{ij}$, $1-\lambda_{ij}$ the eigenvalues of $\rho_{ij}$, then the Von Neumann entropy reads: 
\begin{gather}
    S_{ij}=-\lambda_{ij} \log_2 \lambda_{ij} -(1-\lambda_{ij} ) \log_2 {(1-\lambda_{ij})} \ .
    \label{Entropy}
\end{gather}
The entropy increases monotonically for $\lambda \in [0,\frac{1}{2}]$ and the maximum entanglement entropy is at $\lambda=\frac{1}{2}$.
It will be useful to work with the determinant that shares the same monotonicity property:
\begin{gather}
    det(\rho_{ij})=\lambda_{ij}(1-\lambda_{ij}) \ .
    \label{Determinant}
\end{gather}
\begin{lemma}
\label{Lemma - the determinant satisfies}
    The determinant satisfies:
    \begin{gather}
    det(\rho_{ij})=\frac{|a_{ij}d_{ij}-b_{ij}c_{ij}|^2}{N_{ij}^4}\nonumber\\=\left|\frac{(U_{1i}U_{2j}-U_{1j}U_{2i})(U_{3j}U_{4i}-U_{3i}U_{4j})}{4p_{ij}}\right|^2 \ .
    \label{Determinant identities}
    \end{gather}
\end{lemma}

\begin{proof}
    See details of the computation in \ref{Proof appendix - Lemma - the determinant satisfies} in the proofs appendix.
\end{proof}
We will mostly work with the determinant since it is easy to calculate. The entanglement entropy (\ref{Entropy}) can be calculated
using:
\begin{gather}
    \lambda_{ij}=\frac{1+\sqrt{1-4\det(\rho_{ij})}}{2} \nonumber \\
    1-\lambda_{ij}=\frac{1-\sqrt{1-4\det(\rho_{ij})}}{2} \ .
    \label{eigenvalues of the reduced density matrix}
\end{gather}

\begin{remark}
    \label{Remark:iff conditions for a relevant state to be a product state}
    One can deduce from (\ref{Determinant identities}) that the relevant state $(i,j)$ is a product state iff $U_{1i}U_{2j}=U_{2i}U_{1j}$ or $U_{3j}U_{4i}=U_{3i}U_{4j}$.
\end{remark}

In relation to the sets of final states that we discussed in section \ref{sec:Generalized Type II Fusion and the resulting state} as summarized in table \ref{tab:states_det_s}, the determinant shows that any final state that is a stabilizer state (satisfying the conditions of theorem \ref{Theorem - when it's cluster state up to 1-qubit gate operation}), or a cluster state up to single-qubit gate operations and one two-qubit gate operation (satisfying the conditions of theorem \ref{Theorem - when it can be made cluster state by operating with 2 and 3 qubits gates}), is a maximal entanglement entropy state. A final state that is a weighted graph state (satisfying the conditions of theorem \ref{Theorem - when it's graph state}) is not necessarily a maximal entanglement entropy state. If $|n(b)|=2$ then by theorem \ref{Theorem - when it's graph state} it is a stabilizer state, thus a maximally entangled state. However, if $|n(b)|=1$, then we can represent $a,b,c,d$ in the parametrization (\ref{Parameterization of A,B,C,D in the case of graph state}) (for which $N=1$), and by plugging it in (\ref{Determinant identities}) we get the determinant:
 \begin{gather}
    det(\rho)=\frac{\cos^2(\phi_1-\phi_2)}{4}=\frac{1-\cos{\chi}}{8} \ ,
    \label{determinant of graph state}
    \end{gather}
where we substitute $\chi$ (\ref{The 2-qubits gate between e and n(b) for graph state after z-rotations}) from (\ref{chi by phi1,2}). The appropriate eigenvalues (\ref{eigenvalues of the reduced density matrix}) are $\frac{1 \pm \abs{\cos{\frac{\chi}{2}}}}{2}$, from which one can compute the entropy (\ref{Entropy}). This is not necessarily a maximally entangled state, as will be shown in the following theorem.

\begin{theorem}[Maximally entangled weighted graph states]
\label{Theorem - maximally entangled graph state}
    If $|n(b)|=1$, then a weighted graph state is maximally entangled iff it is a cluster state up to single-qubit gates.
\end{theorem}

\begin{table}
    \centering
    \begin{tabular}{|p{5cm}|p{3cm}|} \hline 
         \textbf{Final State}& \textbf{Det / Entropy}\\ \hline 
         General Final State (\ref{Cluster state after measuring a,b for general unitary transformation})& $0 \leq S \leq 1 \hspace{1.25cm}$   
$0 \leq det \rho \leq \frac{1}{4}$\\ \hline 
 Required Final Cluster State& $S=1$ (max S)\\ \hline 
         Stabilizer State (\ref{Subsection:Final Stabilizer State})& $S=1$ (max S)\\ \hline 
         Weighted Graph State (\ref{Subsection:Final Generalized Graph State})& $\det \rho = \frac{1-\cos \chi }{8}$ (\ref{determinant of graph state}) (max S iff $\chi=\pi$)\\ \hline 
 Cluster Up to Two-Qubit Rotation (\ref{Subsection:Cluster State Up To two-Qubit Rotations})&$S=1$ (max S)\\ \hline
    \end{tabular}
    \caption{The determinant (\ref{Determinant})/entropy (\ref{Entropy}) of the states, described in section \ref{sec:Generalized Type II Fusion and the resulting state}.}
    \label{tab:states_det_s}
\end{table}

\begin{proof}
    The state is maximally entangled iff the determinant in (\ref{determinant of graph state}) equals $\frac{1}{4}$, which is equivalent to $\phi_1=\phi_2$, and by theorem \ref{Theorem - when it can be made cluster state by operating with 2 and 3 qubits gates} is equivalent to the state being a cluster state up to single-qubit gates.
    
\end{proof} 

\begin{remark}
    The form of the cluster states up to one and two-qubit gates (\ref{General form of A,B,C,D in order to get cluster state after hard transformations}) is obtained by substituting $\theta_b=\theta_a+\frac{\pi}{2}$ and $\theta_a=\theta_1,\theta_d=\theta_2$ in the general form of two-qubit maximally entangled states:
    \begin{gather}
    A=\frac{1}{\sqrt{2}}e^{i\theta_a}\cos{\phi}\hspace{0.5cm}
    B=\frac{1}{\sqrt{2}}e^{i\theta_b}\sin{\phi}
    \nonumber \\
    C=\frac{1}{\sqrt{2}}e^{i(\theta_a+\theta_d-\theta_b)}\sin{\phi}\hspace{0.5cm}
    D=-\frac{1}{\sqrt{2}}e^{i\theta_d}\cos{\phi} \ .
    \label{General form of 2-qubits maximally entangled state}
    \end{gather}
    The form (\ref{General form of 2-qubits maximally entangled state})  is shown in proof \ref{Proof appendix - General form of 2-qubits maximally entangled state}.  
\end{remark}

\section{Analytical Probability Bounds and a Constructive Example}
\label{sec:anal_bound}
Building on the state classification in section \ref{sec:Generalized Type II Fusion and the resulting state} and the link-entropy definition in section \ref{sec:prob_def}, we derive
in subsection \ref{subsection:Endpoints values of P(S)} the values of the edges of graph of the fusion success probability as a function of the target entanglement $S_{target}$ (\ref{eq:P_S_target}). We prove a tight $0.5$ cap (no ancilla qubits) for producing maximally entangled links and identify exactly when the bound is saturated.  We also show that unit success probability is possible only for product outcomes, thereby fixing the endpoints of the trade-off curve: $P(1)=\tfrac12$ and $P(0)=1$ (while for every $S>0$, $P(S)<1$).
In addition, in subsection \ref{Subsection:Example for generating weighted graph state} we present a concrete fusion example that generates a weighted graph state with probability $0.5$ when the fused boundary has degree one (as explained in subsection \ref{Subsection:Final Generalized Graph State}): the surviving neighbors acquire a tunable controlled-phase, and the six relevant two-click outcomes split cleanly into cluster, weighted, and product classes.

\subsection{Endpoints values of $P(S_{target})$}
\label{subsection:Endpoints values of P(S)}
Our goal in theorems \ref{Theorem - the 0.5 bound} and \ref{Theorem - reaching the 0.5 bound} is to show that the highest probability to obtain a maximally entangled state is $\frac{1}{2}$, and to identify all the unitary matrices $U$ that yield this probability. Our proof will be more general than the proof in \cite{MaxEfficiency}, since we will allow any final maximally entangled state and not only one of the four bell states (up to global phase) as in \cite{MaxEfficiency}. This is important because we showed in theorems \ref{Theorem - when it's cluster state up to 1-qubit gate operation} and \ref{Theorem - when it can be made cluster state by operating with 2 and 3 qubits gates} that there are final states other than the four Bell states from which one can construct the required cluster state. Unlike the proof in \cite{MaxEfficiency} we will not allow vacuum modes, i.e. no transformation from the qubits $a,b$ to a general n-qubit set $c_1,c_2,...,c_n$ with $n>2$, which are all being measured. In theorem \ref{Theorem - we can't have probability 1} we will prove that one cannot obtain a nonzero entanglement entropy with probability one. Thus, we will characterize the two edges of the graph of $P(S)$, as illustrated in Figure \ref{fig:FusionDiagram}.

\begin{theorem}[The probability to obtain a maximally entangled state bounded by $\frac{1}{2}$] 
\label{Theorem - the 0.5 bound}
 For every fusion matrix $U$, when measuring $c$ and $d$, the probability of obtaining a maximally entangled state is bounded from above by $\frac{1}{2}$.

\end{theorem}

\begin{proof}

This theorem is proved using the following Lemmas.

\begin{lemma}
\label{Lemma - Conditions for Bell state}
    The necessary and sufficient conditions for a relevant $(i,j)$ state to be a maximally entangled state are:
    \begin{gather}
        n_it_j+n_jt_i=0 \ ,
        \label{condition 1 for Bell state}
    \end{gather}      
        and 
        \begin{gather}
        n_ik_j+n_jk_i=0 \ .
        \label{condition 2 for Bell state}
    \end{gather}
\end{lemma}

    \begin{proof}
        The reduced density matrix $\rho_{ij}$ (\ref{Density matrix}) of a maximally entangled relevant state $(i,j)$ is $\frac{1}{2} Id_{2\times2}$. Thus, the off diagonal elements of $\rho_{ij}$ and difference between the two diagonal elements are $0$.  We impose these two conditions, with a multiplication by $N_{ij}^2$. 
        The first condition reads:
        \begin{gather}
        0=a_{ij}c_{ij}^*+b_{ij}d_{ij}^*=2t_in_j+2t_jn_i \ ,
        \label{Condition 1 with details}
        \end{gather}
        (see details of computation in proof \ref{Proof appendix - Lemma - Conditions for Bell state - condition 1} in proof appendix)
        and the second condition reads:
        
        \begin{eqnarray}
        0= |a_{ij}|^2+|b_{ij}|^2-|c_{ij}|^2-|d_{ij}|^2   \nonumber\\
        =  4n_i|U_{1j}|^2+4n_j|U_{1i}|^2+4n_in_j-n_i-n_j \nonumber\\
        =  2n_ik_j+2n_jk_i \ ,
        \label{eq:Condition 2 with details}
        \end{eqnarray}
        
        (see details of computation in proof \ref{Proof appendix - Lemma - Conditions for Bell state - condition 2} in proof appendix).
    \end{proof}
    \begin{lemma}
    \label{Lemma - probability for specific Bell state}
        If $(i,j)$ is a maximally entangled relevant state, then the probability to obtain it is bounded from above by  $\frac{1}{8}$. If the bound is saturated then $n_i=n_j=0$.
    \end{lemma}
    \begin{proof}
         See proof \ref{Proof appendix - Lemma - probability for specific Bell state} in proof appendix.
    \end{proof}
   
    \begin{lemma}
    \label{Lemma - six Bell states}
        When all the relevant states are maximally entangled states then $n_1=n_2=n_3=n_4=0$,  and the probability to obtain a maximally entangled state is $\frac{1}{2}$.
    \end{lemma}
   
    \begin{proof}
        When all the relevant states are maximally entangled states, then from (\ref{condition 1 for Bell state}) one gets that either all $n$'s or all $t$'s are zero (for details see proof \ref{Proof appendix - Lemma - six Bell states - completing the proof} in proof appendix).
        If all the $t$'s vanish, then for every $i$, one of $U_{1i}$,$U_{2i}$ is zero. Thus, we can choose two indices $(i,j)$ for which either $U_{1i}=U_{2i}=0$ or $U_{2i}=U_{2j}=0$, and the state $(i,j)$ is a product state (see Remark \ref{Remark - when two elements of U are zero}), hence a contradiction.
        Therefore, all the $n$'s are zero. In this case, the probability for non-relevant states (\ref{Probability of all non-relevant states}) is $\frac{1}{2}$, which is also the probability for all the relevant states.  When $n_i=n_j=0$ then by Lemma \ref{Lemma - Conditions for Bell state} the two conditions (\ref{condition 1 for Bell state}),(\ref{condition 2 for Bell state}) for $(i,j)$ state to be a maximally entangled state are fulfilled, hence, in this case all the relevant states are maximally entangled states and we have $\frac{1}{2}$ probability to get a maximally entangled state.
    \end{proof}
    \begin{lemma}
    \label{Lemma - five Bell states}
         There cannot be exactly five relevant states that are maximally entangled states, except when all the $n$'s are zero. 
    \end{lemma}
     
    \begin{proof}
        See proof \ref{Proof appendix - Lemma - five Bell states} in proof appendix.
    \end{proof}

Now, we can complete the proof: If we have four maximally entangled states or less, then the probability
to get a maximally entangled state is bounded from above by $4\times \frac{1}{8}= \frac{1}{2}$ by Lemma \ref{Lemma - probability for specific Bell state}. Else, by Lemma \ref{Lemma - five Bell states} we have six maximally entangled states, and by Lemma \ref{Lemma - six Bell states} the probability is $\frac{1}{2}$. 

\end{proof}

\begin{theorem}[Saturating the $P= \frac{1}{2}$ bound] 
\label{Theorem - reaching the 0.5 bound}
The set of fusion matrices $U$ which lead to a saturation of the $\frac{1}{2}$ bound, is the set of all the matrices for which all the $n$'s (\ref{m,n,t,k definitions}) are zero, that is the matrices $U$ that satisfy:
\begin{equation}
    |U_{1i}|^2+|U_{2i}|^2=\frac{1}{2},~~~i=1,2,3,4 \ .
    \label{Condition for 0.5 bound} 
\end{equation}

\end{theorem}

\begin{proof}
    If all the six states are maximally entangled states, then we have already shown that all the n's are zero. We also showed, that the number of maximally entangled states cannot be five by Lemma \ref{Lemma - five Bell states}, thus we are left with the case that there are four maximally entangled states or less. If a relevant state is maximally entangled state then by Lemma \ref{Lemma - probability for specific Bell state} the probability to obtain this state is bounded from above by $\frac{1}{8}$, so, if we have four states or less, then the only option to get to a total probability $P=\frac{1}{2}$ is if to have exactly four maximally entangled states, with probability $\frac{1}{8}$ for each one. By Lemma \ref{Lemma - probability for specific Bell state}, we have $n_i=n_j=0$ for every $(i,j)$ of the four states.

    \begin{lemma}
    \label{Lemma - subsets of (1,2,3,4)}
        If there are four different subsets of the set $\{1,2,3,4\}$, each one consisting of two elements, then one can choose two of the subsets such that their union is the set $\{1,2,3,4\}$ \ .
    \end{lemma}

    \begin{proof}
        see proof \ref{Proof appendix - Lemma - subsets of (1,2,3,4)} in proofs appendix.
    \end{proof}

    By Lemma \ref{Lemma - subsets of (1,2,3,4)} one can choose two of the four states, $(i,j)$ and $(k,l)$ such that $\{i,j,k,l\}=\{1,2,3,4\}$. Hence, $n_i=n_j=n_k=n_l=0$, thus all the $n$'s are zero and we are done.
    
\end{proof}

\begin{theorem}[A Minimal number of beam splitters and phase shifters]
\label{Theorem - min number of gates}
    The minimal number of basic linear optical devices (beam splitters and phase shifters), that are needed to construct the fusion matrix $U$ (\ref{Unitary transformation}) that saturates the $P=\frac{1}{2}$ bound is two.
\end{theorem}

\begin{proof}
    If one uses only one beam splitter/phase shifter, then in the fusion matrix $U$ there will be two rows, that are the same as in the identity matrix, so it will not obey the condition (\ref{Condition for 0.5 bound}).
    One can use two basic linear optical devices to construct a fusion matrix $U$ that obeys (\ref{Condition for 0.5 bound}).
    For instance, a $45$ degree rotation for $a_H,b_H$ and a $45$ degree rotation for $a_V,b_V$ (both can be realized by beam splitters):

    \begin{gather}
        U=\begin{bmatrix}
        \frac{1}{\sqrt{2}} & 0 & \frac{1}{\sqrt{2}} & 0 \\
        0 & 1 & 0 & 0 \\
        -\frac{1}{\sqrt{2}} & 0 & \frac{1}{\sqrt{2}} & 0 \\
        0 & 0 & 0 & 1
    \end{bmatrix}
    \begin{bmatrix}
         1 & 0 & 0 & 0 \\
     0 & \frac{1}{\sqrt{2}} & 0 & \frac{1}{\sqrt{2}} \\
     0 & 0 & 1 & 0 \\
     0 & -\frac{1}{\sqrt{2}} & 0 & \frac{1}{\sqrt{2}}
    \end{bmatrix}
    \nonumber\\
    =\frac{1}{\sqrt{2}}\begin{bmatrix}
      1 & 0 & 1 & 0 \\
     0 & 1 & 0 & 1 \\
     -1 & 0 & 1 & 0 \\
     0 & -1 & 0 & 1
    \end{bmatrix} \ ,
    \label{U by two 1-qubit gates}
    \end{gather}
and this matrix indeed obeys (\ref{Condition for 0.5 bound}). 
    
\end{proof}

If the fusion of (\ref{U by two 1-qubit gates}) succeeds, then the fusion operator is $\bra{0}_a\bra{1}_a\pm \bra{0}_a\bra{1}_a$ and one arrives at the same situation as in (\ref{Cluster state after measuring a,b and getting 01+10}). This is a projection to one of the Bell states that yields the required cluster state.

It is important to note that two basic optical devices is less than the three, that are theoretically being used in the construction of the regular type-II fusion matrix as in (\ref{PBS2 matrix U}). But in practice the regular type-II fusion matrix is simply realized by PBS2.

\begin{theorem}[Probability $P(S)=1$ iff $S=0$]
\label{Theorem - we can't have probability 1}
    The success probability is one iff the target entropy is zero, i.e. the final state (\ref{Cluster state after measuring a,b for general unitary transformation}) is a product state.
\end{theorem}
\begin{proof}
    Assume that there is $P(S)=1$ probability of success to obtain a final state with nonzero entropy $S$ or higher. Then, the probability for all the non-relevant states (\ref{Probability of all non-relevant states}) must be zero, and from equations (\ref{m,n,t,k relations}) and (\ref{Probability of all non-relevant states}) one concludes that two of the $n$'s are $\frac{1}{2}$ and the other two are $-\frac{1}{2}$. 
    By remark \ref{Remark - when two elements of U are zero}, every $(i,j)$ state (\ref{Wave function of relevant state}) with $n_i=\frac{1}{2}$ or $n_j=\frac{1}{2}$ is a product state.
    Thus, one has five relevant states, that are product states, and the sixth state has $n_i=n_j=-\frac{1}{2}$, hence, its probability is zero (by equation \ref{Probability of relevant state}). Therefore, one gets a $P=1$ probability to obtain product state and $S=0$.
\end{proof}

\subsection{Weighted-Edge Fusion: A Constructive Example}
\label{Subsection:Example for generating weighted graph state}
In subsection \ref{Subsection:Final Generalized Graph State} we computed the conditions (\ref{The 2-qubits gate between e and n(b) for graph state}) for a specific relevant state to be a weighted graph state (assuming that the qubit $b$ was in the edge of its cluster, see subsection \ref{Subsection:Final Generalized Graph State}). In this subsection we provide an example of unitary $U$ (\ref{Unitary transformation}) for which two relevant states are weighted graph states --- choose an angle $\eta$ and define the unitary matrix:

\begin{equation}
    \label{Example for two states being weighted graph state and two states being graph states with one parameter}
    U=\frac{1}{\sqrt{2}}\begin{bmatrix}
     \cos{\eta} & \sin{\eta} & \cos{\eta} & \sin{\eta} \\
     i\cos{\eta} & i\sin{\eta} & -i\cos{\eta} & -i\sin{\eta} \\
     \sin{\eta} & -\cos{\eta} & \sin{\eta} & -\cos{\eta} \\
     i\sin{\eta} & -i\cos{\eta} & -i\sin{\eta} & i\cos{\eta}
    \end{bmatrix} \ ,
\end{equation}

for which, together with the non-relevant states, the states $i=1,j=2$ and $i=3,j=4$ are product states (by remark \ref{Remark:iff conditions for a relevant state to be a product state}) with probabilities (computed by (\ref{Probability of relevant state})):

\begin{gather}
    p_{12}=p_{34}=\frac{\cos^2{2\eta}}{4}  \ ,
\end{gather}

the states $i=1,j=3$ and $i=2,j=4$ are the required fused cluster state (by equations (\ref{The coefficients of the wave function}) the resulting wave function (\ref{Cluster state after measuring a,b for general unitary transformation}) is (\ref{Cluster state after measuring a,b and getting 01+10}) up to operating with $Z_e$ and a total phase) with probabilities (computed by (\ref{Probability of relevant state})):

\begin{gather}
    p_{13}=p_{24}=\frac{1-\cos^2{2\eta}}{8} \ ,
\end{gather}

and the states $i=1,j=4$ and $i=2,j=3$ are weighted graph states (the coefficients $A,B,C,D$ computed by (\ref{The coefficients of the wave function}) and remark \ref{Remark: ABCD coeffients} are satisfying the conditions (\ref{The 2-qubits gate between e and n(b) for graph state})), with probabilities (computed by (\ref{Probability of relevant state})):

\begin{gather}
    p_{14}=p_{23}=\frac{1+\cos^2{2\eta}}{8} \ ,
\end{gather}

and determinants (computed by (\ref{Determinant identities})):

\begin{gather}
    det(\rho_{14})=det(\rho_{23})=\frac{1}{4}\left(\frac{1-\cos^2{2\eta}}{1+\cos^2{2\eta}}\right)^2 \ .
\end{gather}

Notice that the total probability for a weighted graph state/required fused cluster state is $p_{13}+p_{14}+p_{23}+p_{24}=0.5$, and with probability $0.5$ we get a product state. The division of the $0.5$ probability for a weighted graph state/required fused cluster state varies by $|\cos{2\eta}|$.
For $|\cos{2\eta}|$ close to zero, the probability for the weighted graph states is close to $\frac{1}{4}$, and their determinant is close to its maximal value $\frac{1}{4}$ (and those states are in fact becoming the required fused cluster state). 
As $|\cos{2\eta}|$ increases, the probability for the weighted graph states increases and their determinant decreases, and at the edge $|\cos{2\eta}|=1$ we get a product states with the entire $0.5$ probability. 

This example sets a lower bound of $0.5$ for the optimal success probability for creating a weighted graph state with a target entropy (for any target entropy) --- the same question as in (\ref{eq:P_S_target}), where the sum is taken on all the final states with $S_{state}\ge S_{target}$ and that are also weighted graph states.

This example also showed that a specific relevant state can be a weighted graph state, with any probability less than $\frac{1}{4}$, as by remark \ref{Remark: equality pij=one over four}, $\frac{1}{4}$ is the unreachable bound for the probability of every relevant state that is not a product state. This is much higher than the maximum probability for a specific relevant state that is maximally entangled which is $\frac{1}{8}$ by lemma \ref{Lemma - probability for specific Bell state}, thus creates a possibility to potentially bypass the $0.5$ probability using three relevant states or more.

\section{Numerical Simulations}
\label{sec:numerical_res}

In order to quantitatively evaluate (\ref{eq:P_S_target}),
and to validate our analytical analysis we construct a simulator using the PyTorch package, Python and
optimize general $4\times4$ unitary matrices $U$ (\ref{Unitary transformation}), which we parameterize by Euler angles \cite{SUN,AnotherSUN}. To improve the convergence rate, we initialized the optimization parameters after taking the best result among randomly-sampled matrices, which we sample using the Haar probability measure (for details see appendix  \ref{sec:Haar measure}).
In subsection \ref{Subsection:Expectation Entropy Optimization} we will present and discuss the results of the optimization of the expectation value of the entanglement entropy, and in subsection \ref{Subsection:Optimize the Minimum Entropy}, the optimization of the minimal entanglement entropy. 
The code generating this simulations is given in \cite{simulationCode}.


\subsection{Optimization of the Mean Entanglement Entropy}
\label{Subsection:Expectation Entropy Optimization}
\begin{figure}
    \centering
    \includegraphics[width=1.0\linewidth] {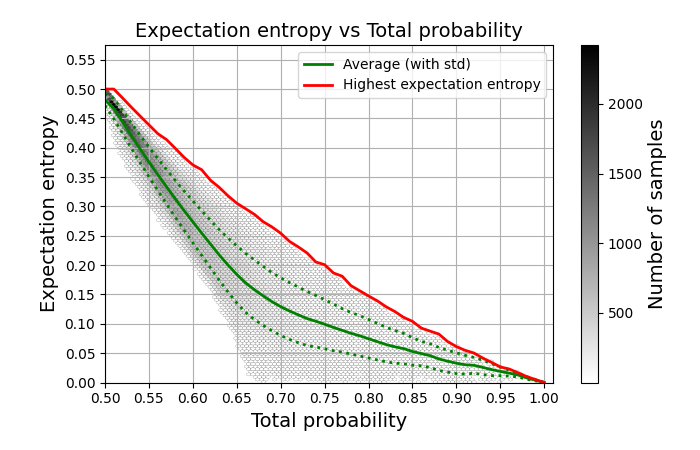}
    \caption{Plot of the expectation value of the entanglement entropy (expectation entropy) versus the total success probability. The gray area is created by randomly sampling fusion matrices. The green line represents the average expectation entropy for each total probability (plus or minus the standard deviation in the dashed line). The red line shows the maximal (red) expectation value of entropy versus the probability, after using an optimization scheme. The gray shading indicates the density of sampled matrices. }
    \label{fig:s_vs_p_exp}
\end{figure}

Define the expectation value of the entanglement entropy of the final states as
\begin{equation}
    \langle S \rangle=\sum_{(i,j) state} p_{ij}S_{ij} \ ,
    \label{Expectation value of entropy definition}
\end{equation}
where the summation is over all the final states $(i,j)$ (\ref{Wave function of relevant state}), $p_{ij}$ (\ref{Probability of relevant state})
is the probability to obtain the state $(i,j)$, and $S_{ij}$ is its entanglement 
entropy (\ref{Entropy}). Since all the non-relevant (\ref{Wave function of non-relevant state}) states are product states having zero entropy, one can replace the summation by a sum (\ref{Expectation value of entropy definition}) over the relevant states.
Define the total probability  $p_{total} \equiv p_{relevant}=1-p_{non-relevant}$ to be the sum of the probabilities of all the relevant states, which is (by (\ref{Probability of all non-relevant states})):
\begin{gather}
 p_{total} 
 =\frac{1}{2}(1+n_1^2+n_2^2+n_3^2+n_4^2) \ .
    \label{Probability for all relevant states}
\end{gather}
The goal is, given a target total probability $p$, to maximize the expectation value of the entropy
\begin{gather}
    \langle S \rangle_{max}(p) \equiv \max_{\{U | p_{total}(U)=p\}} \langle S \rangle \ .
    \label{<S>}
\end{gather}
We randomly sample fusion matrices, locate them in the $S-P$ plane, and calculate the mean and the standard deviation of the expectation value of the entanglement entropy for each probability sum. These are the gray shaded area, and the green line, respectively, in Figure  \ref{fig:s_vs_p_exp}). 
Using the optimization scheme, we sweep over all the probabilities between $0.5$ and $1.0$, (in steps of 0.01),  and for each point we optimize the target probability and maximal expectation value of the entropy. We will refer in the following to the expectation value of the entropy
and the expectation entropy. We average the results over 20 runs, while we initialize the optimization to be one of the highest expectation entropy among those found in the random steps. Each run is optimized to find a specific target probability and a maximal expectation entropy. The cost function for the optimization reads:
\begin{gather}
    \mathcal{L}_{expectaion} \equiv \alpha \left|| p\left(U\right) - p_{target} |\right|^2 - \langle S \rangle_{max}\left(U \right) \ .
\end{gather}
In practice, we add baseline to the cost function in order to avoid negative values (for numerical stability in the gradient descent). $\alpha$ is a hyperparameter which we tuned, and practically set it to 1. We use Adam Optimizer for $1,000$ epochs, with a learning rate of $0.001$. The results are shown in Figure \ref{fig:s_vs_p_exp} in the red line.

Note, that the relation between the expectation entropy and the total probability is monotonically decreasing (as can be seen in the graph), namely, during the optimization process, increasing the expectation entropy leads to a decrease in the total probability. Therefore, there is a small bias of the optimization results towards lower probability.
The graph exhibits an intriguing  close to linear structure :
\begin{gather}
    S = 1-P \ ,
\end{gather}
for which we lack an analytical derivation.
The plot reveals a critical probability, around $\frac{2}{3}$, where for any probability below  it there is at least one relevant state that is not separable, since the average entropy is always greater than zero in this range. 

The range of values of the expectation entropy is the interval $[0,1]$.
When the total probability is $0.5$, then all the relevant states are maximally entangled states,
as we proved in theorem \ref{Theorem - reaching the 0.5 bound}, and the expectation entropy is equal to $0.5\cdot 1=0.5$.
When the total probability is $1$, then as shown in theorem \ref{Theorem - we can't have probability 1}, all the relevant states are product states, with entropy zero, and the expectation entropy is therefor equal to zero.
Note, also that the variance of the expectation entropy is relatively small near the $0.5$ and $1$, and increases in between.
This implies that choosing a random fusion unitary operation is not the right strategy, when attempting to have a high probability for the fusion success, while reducing the entanglement entropy of the fusion link.


\subsection{Optimization of the Minimal Entanglement Entropy}
\label{Subsection:Optimize the Minimum Entropy}
 
\begin{figure}
    \includegraphics[width=1.\linewidth] {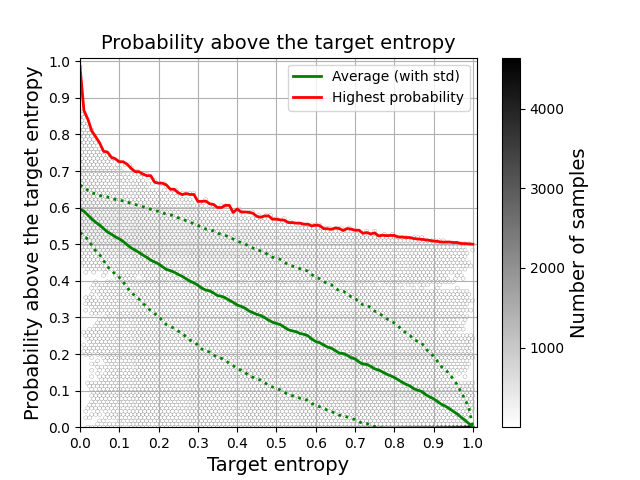}
    \caption{The probability for the success of the fusion process versus the target entropy, according to definition \ref{eq:P_S_target}. The gray area is constructed by randomly sampling fusion matrices. The red line shows the maximal  probability for each target entropy, as is obtained by the optimization. The gray shading indicates the density of sampled matrices.}
    \label{fig:PvsS_threshold}
\end{figure}

\begin{figure}
\hfill
    \includegraphics[width=1.0\linewidth]{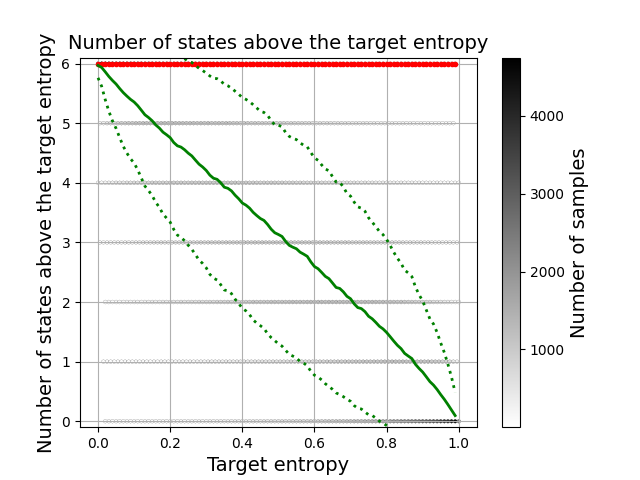}
    \caption{Number of states above the target entropy for each of the points in Figure \ref{fig:PvsS_threshold} (the gray points match, as well as the red points). The vertical axis represents the number of states among the six states, and the horizontal axis represents the target entropy. 
    It is clear from this graph that the optimal fusion matrices founded by the numerical simulation employ all the six states.}
\label{fig:NOSvsS_threshold}
\end{figure}

Our aim is to maximize the probability to reach $S_{target}$ in the fusion process (\ref{eq:P_S_target}), by optimizing the fusion matrix $U$.
The cost function for the optimization is:
\begin{gather}
    \mathcal{L} \equiv  - P(S_{target}) \ ,
\end{gather}
and the results are shown in Figure \ref{fig:PvsS_threshold}.
The number of states, whose entropy is not lower than the given target entropy $S_{target}$,
is shown in Figure \ref{fig:NOSvsS_threshold}.
Figure \ref{fig:PvsS_threshold} shows, as expected,  a monotonically decreasing graph.
The graph is convex, the decrease is faster in the left regime, and slows down with the increase
of the target entropy. In the right regime, the graph is nearly flat. 
Thus, the numerical data suggests that the left regime is relevant for decreasing the target entropy and getting a higher probability of success.
On the other hand, in the right regime, it is useful to increase the target entropy to one, since it lowers only 
slightly the probability of success.
The number of states that are being used in the optimal fusion matrix is fixed at six, which means that all the relevant states contribute for the optimal solutions for any target entropy. Therefore, the y-axis corresponding to
the red line $p(S)$ in Figure \ref{fig:PvsS_threshold} is $p_{relevant}=\frac{1}{2}(1+n_1^2+n_2^2+n_3^2+n_4^2)$. 
However, this does not mean that there are no optimal solutions with less than six states that the numerical simulation does not convergence to. For example, the regular fusion type-2 is an optimal solution (for $S=1$) with four states, with the other two relevant states having a zero success probability, but for $S=1$ the number of degrees of freedom of the optimal solutions having two relevant states with zero probability is less than the number of degrees of freedom of the general optimal solutions (for $S=1$) and hence the simulation does not converge to those solutions and only find optimal solutions that are using all the relevant states. This may be true also for other entropies.
Inspecting the mean $p(S)$ (before optimization) and the distribution in Figure \ref{fig:PvsS_threshold}, reveals
a large distance between the mean and the maximal values, which emphasizes the significance
of finding the optimal $U$'s, rather than having a random one. One also sees in Figure \ref{fig:NOSvsS_threshold}, that the average number of states that are being used, is decreasing, as expected.






\section{Discussion and Outlook}
\label{sec:discussion}


We generalized the type-II fusion procedure, and analyzed it both analytically and numerically.
We classified all the possible final states of the generalized type-II fusion, including the probability for the success of the fusion, and its correlation with the entanglement entropy of the fusion link.
In addition to having as final states maximally entangled states, we consider weighted graph states whose entanglement entropy is not maximal, while potentially having a higher success probability for the fusion --- right now we have an example of $0.5$ success probability, and we are planning further investigations in order to find the optimal probabilities to generating weighted graph states.
As noted above, it has been shown that some of such weighted graph states can be utilized as resource states for quantum computation, though the classification of all of those weighted graph states is still unknown \cite{gross2007measurement}. 

We found, that allowing maximally entangled states, which are not Bell states, as final states, does not increase the probability of the fusion success above the fifty percent bound, if no ancilla qubits are used. 
It is important to find, whether and how this probability changes with the addition of ancilla qubits, as done in the fusion process in \cite{bartolucci2021creation}.

Analytically, we proved in a rather general setup, the fifty percent bound on the fusion probability success, and classified the fusion matrices and the set of final states, which is larger than the set that has been previously considered.
We proved, that only fusion links with zero entanglement entropy can be constructed with a hundred percent probability of success.
We have not obtained, however, an analytical formula for the fusion probability success $p(S)$ for a general entanglement entropy $S$ of the fusion link, and we analyzed it numerically. It would be interesting to further explore this analytically.

Another route to take is to generalize the analysis to the cases, where one allows a unitary operation that involves more registers as in \cite{MaxEfficiency}.
Since MBQC can be defined on resource states made of qudits, it is natural to generalize our work to such cases.
Finally, it would be interesting to have experimental realizations of the various resource states that we studied.

\vspace{0.5cm}

\textbf{Acknowledgments}
We would like to thank I. Kaplan for participation in the early stages of this project, 
M. Erew for a valuable discussion and Ori Shem-Ur for comments on the manuscript. 

\textbf{Funding} - This work is supported in part by the Israeli Science Foundation Excellence Center, the US-Israel Binational Science Foundation, and the Israel Ministry of Science.


\textbf{Data availability} -
The code generating the simulations in section \ref{sec:numerical_res} is given in \cite{simulationCode}.

\textbf{The author declare no conflict of interest.}

\bibliographystyle{unsrt.bst}
\bibliography{main.bib}

@article{5lectures,
  author    = {Pieter Kok},
  title     = {Five Lectures on Optical Quantum Computing},
  journal   = {arXiv: Quantum Physics},
  volume    = {787},
  year      = {2007},
  pages     = {187-219},
  note      = {\url{
https://doi.org/10.1007/978-3-642-02871-7_7}},
}

@article{ClusterStatesNielsen,
  author    = {M. A. Nielsen},
  title     = {Cluster-state quantum computation},
  journal   = {ArXiv},
  year      = {2005},
  note      = {\url{
https://doi.org/10.1016/S0034-4877%2806%2980014-5}},
}

@article{BrowneRudolph,
  author    = {D. E. Browne and T. Rudolph},
  title     = {Resource-efficient linear optical quantum computation},
  journal   = {ArXiv},
  year      = {2004},
  note      = {\url{https://doi.org/10.1103/PhysRevLett.95.010501}},
}

@misc{stanisic2015universal,
  title={Universal quantum computation by linear optics},
  author={Stanisic, Stasja},
  year={2015},
  publisher={unpublished}
}

@article{ThreePhoton,
  author    = {Gimeno-Segovia, M. and Shadbolt, P. and Browne, D. E. and Rudolph, T.},
  title     = {From three-photon GHZ states to ballistic universal quantum computation},
  year      = {2015},
  note      = {\url{https://doi.org/10.1103/PhysRevLett.115.020502}},
}

@article{MaxEfficiency,
  author    = {Calsamiglia, J. and Lütkenhaus, N.},
  title     = {Maximum efficiency of a linear-optical Bell-state analyzer},
  year      = {2001},
  note       = {\url{https://doi.org/10.1007/s003400000484}},
}

@article{BellMeasurements,
  author    = {Lütkenhaus, N. and Calsamiglia, J. and Suominen, K-A.},
  title     = {On Bell measurements for teleportation},
  year      = {1999},
  note      = {\url{https://doi.org/10.1103/PhysRevA.59.3295}},
}

@article{FusionBasedQC,
  author    = {Bartolucci, S. and Birchall, P. and Bombín, H. and Cable, H. and Dawson, C. and Gimeno-Segovia, M. and Johnston, E. and Kieling, K. and Nickerson, N. and Pant, M. and Pastawski, F. and Rudolph, T. and Sparrow, C.},
  title     = {Fusion-based quantum computation},
  year      = {2021},
  note      = {\url{https://doi.org/10.1038/s41467-023-36493-1}},
}

@article{CreatingU0,
  author    = {Reck, M. and Zeilinger, A. and Bernstein, H. J. and Bertani, P.},
  title     = {Experimental Realization of Any Discrete Unitary Operator},
  journal   = {Arxiv},
  year      = {1994},
  note      = {\url{https://doi.org/10.1103/PhysRevLett.73.58}},
}

@article{CreatingU1,
    author = { de Guise, H. and  Sanders, B. C. and Bartlett, S. D. and  Zhang, W.},
    title = {Geometric Phase in SU(N) Interferometry},
    journal = {Arxiv},
    year = {2001},
    note={\url{https://doi.org/10.1023/A%3A1017585321693}},
}

@article{CreatingU2,
    author = {Dhand, I. and Goyal, S. K.},
    title = {Realization of arbitrary discrete unitary transformations using spatial and internal modes of light},
    journal = {Arxiv},
    year = {2015},
    note={\url{https://doi.org/10.1103/PhysRevA.92.043813}},
}

@article{SUN,
  author    = {Tilma, T. and Sudarshan, E. C.},
  title     = {Generalized Euler Angle Parameterization for SU(N)},
  journal   = {ArXiv},
  year      = {2002},
  note      = {\url{
https://doi.org/10.1088/0305-4470/35/48/316}},
}

@article{AnotherSUN,
  author    = {Tilma, T. and Byrd, M. and Sudarshan, E. C. G.},
  title     = {A Parametrization of Bipartite Systems Based on SU(4) Euler Angles},
  journal   = {ArXiv},
  year      = {2002},
  note      = {\url{
https://doi.org/10.1088/0305-4470/35/48/315}},
}

@article{GeneralizedGraphStatesAndStabilizerStates,
    author = {Hein, M. and Dur, W. and Eisert, J. and  Raussendorf, R. and Nest, M. V. D. and Briegel, H. J.},
    title = {Entanglement in Graph States and its Applications},
    journal = {Arxiv},
    year = {2006},
    note = {\url{https://doi.org/10.48550/arXiv.quant-ph/0602096}},
}

@book{HaarMeasureBook,
    author = {Meckes, E.},
    title = {The Random Matrix Theory of the
Classical Compact Groups},
    publisher = {Cambridge University Press},
    year ={2019},
    note = {\url{https://doi.org/10.1017/9781108303453}},
}

@article{MBQCfirstArticle,
    author = {Raussendorf, R. and Briegel, H.},
    title = {Computational model underlying the one-way quantum computer},
    journal = {Arxiv},
    year = {2002},
    note ={\url{https://doi.org/10.48550/arXiv.quant-ph/0108067}},
}

@article{NielsenProtocolFirstArticle,
    author = {Nielsen, M. A.},
    title = {Optical quantum computation using cluster states},
    journal = {Arxiv},
    year ={2004}, 
    note = {\url{https://doi.org/10.1103/PhysRevLett.93.040503}},
}

@article{YoranReznikProtocalFirstArticle,
    author ={Yoran, N. and Reznik, B.} ,
    title = {Deterministic linear optics quantum computation utilizing linked photon circuits},
    journal ={Arxiv} ,
    year = {2003},
    note={\url{https://doi.org/10.1103/PhysRevLett.91.037903}},
}

@article{ClusterStatesOriginalPaper,
    author = {Briegel, H. J. and Raussendorf, R.},
    title = {Persistent entanglement in arrays of interacting particles},
    journal ={Arxiv} ,
    year ={2000} ,
    note={\url{https://doi.org/10.1103/PhysRevLett.86.910}},
}

@article{MBQConClusterStatesFirstArticle,
    author = {Raussendorf, R. and Browne, D. E. and Briegel, H. J.},
    title ={Measurement-based quantum computation with cluster states} ,
    journal ={Arxiv} ,
    year = {2003},
    note={\url{https://doi.org/10.1103/PhysRevA.68.022312}},
}

@article{MBQC2009,
    author = {Briegel, H. J. and Browne, D. E. and  D{\"u}r, W. and Raussendorf, R. and Van den Nest, M.},
    title = {Measurement-based quantum computation},
    journal = {Arxiv},
    year = {2009},
    note={\url{https://doi.org/10.1038/nphys1157}}
}

@article{MBQC2021,
    author = {Shah, Swapnil Nitin},
    title = {Realizations of Measurement Based Quantum Computing},
    journal ={Arxiv} ,
    year = {2021},
    note={\url{https://doi.org/10.48550/arXiv.2112.11601}},
}

@article{UsingAncillae0,
    author = {Knill, E. and Laflamme, R. and Milburn, G.},
    title ={Efficient Linear Optics Quantum Computation} ,
    journal ={Arxiv} ,
    year = {2000},
    note={\url{https://doi.org/10.48550/arXiv.quant-ph/0006088}},
}

@article{UsingAncillae1,
    author = {Grice, W. P.},
    title = {Arbitrarily complete Bell-state measurement using only linear optical elements},
    journal ={Physical Review A} ,
    year = {2011},
    note={\url{https://doi.org/10.1103/PhysRevA.84.042331}}
}

@article{UsingAncillae2,
    author = {Ewert, F. and Loock, P. V.},
    title = {3/4-efficient Bell measurement with passive linear optics and unentangled ancillae},
    journal = {Arxiv},
    year ={2014},
    note={\url{https://doi.org/10.1103/PhysRevLett.113.140403}},
}

@article{UsingAncillae3,
    author ={Kilmer, T. and  Guha, S.} ,
    title = {Boosting Linear-Optical Bell Measurement Success Probability with Pre-Detection Squeezing and Imperfect Photon-Number-Resolving Detectors},
    journal = {Arvix},
    year ={2018},
    note={\url{https://doi.org/10.1103/PhysRevA.99.032302}},
}

@article{UsingAncillae4,
    author = {Bayerbach, M. J. and  D’Aurelio, S. E. and Loock, P. V. and  Barz, S.},
    title = {Bell-state measurement exceeding 50% success probability with linear optics},
    journal = {Arxiv},
    year = {2022},
    note={\url{https://doi.org/10.1126/sciadv.adf4080}}
}

@article{KLMprotocolNature,
    author = {Knill, E. and Laflamme, R. and Milburn, G.},
    title ={A scheme for efficient quantum computation with linear optics} ,
    journal ={Nature} ,
    year = {2001},
    note={\url{https://doi.org/10.1038/35051009}}
}

@article{RealizatonOfMBQC0,
    author = {Walther, P. and Resch, K. J. and Rudolph, T. and Schenck, E. and Weinfurter, H. and Vedral, V. and Aspelmeyer, M. and Zeilinger, A.},
    title = {Experimental One-Way Quantum Computing},
    journal = {Arxiv},
    year = {2005},
    note={\url{https://doi.org/10.1038/nature03347}},
}

@article{RealizatonOfMBQC1,
    author ={Chen, Kai and Li, Che-Ming and Zhang, Qiang and Chen, Yu-Ao and Goebel, Alexander and Chen, Shuai and Mair, Alois and Pan, Jian-Wei} ,
    title = {Experimental Realization of One-Way Quantum Computing with Two-Photon Four-Qubit Cluster States},
    journal = {Arxiv},
    year = {2007},
    note={\url{https://doi.org/10.1103/PhysRevLett.99.120503}},
}

@article{RealizatonOfMBQC2,
    author = {Vallone, Giuseppe and Pomarico, Enrico and  Martini, Francesco De and Mataloni, Paolo},
    title ={Active one-way quantum computation with 2-photon 4-qubit cluster states} ,
    journal ={Arxiv} ,
    year = {2007},
    note={\url{https://doi.org/10.1103/PhysRevLett.100.160502}},
}

@article{RealizatonOfMBQC3,
    author = {Tame, M. S. and Bell, B. A. and Di Franco, C. and Wadsworth, W. J. and  Rarity, J. G.},
    title ={Experimental Realization of a One-way Quantum Computer Algorithm Solving Simon's Problem} ,
    journal = {Arxiv},
    year ={2014},
    note={\url{https://doi.org/10.1103/PhysRevLett.113.200501}},
}

@article{PhotonicQubits,
    author = {Adami, C. and Cerf, N. J.},
    title = {Quantum computation with linear optics},
    journal = {Arxiv},
    year = {1998},
    note={\url{https://doi.org/10.48550/arXiv.quant-ph/9806048}}
}

@article{ReviewOfOpticalQuantumComputing0,
    author = {Kok, P. and Munro, W. J. and Nemoto, K. and Ralph, T. C. and Dowling, J. P. and Milburn, G. J.},
    title = {Linear optical quantum computing with photonic qubits},
    journal = {Reviews of Modern Physics},
    year = {2007},
    note={\url{https://doi.org/10.1103/RevModPhys.79.135}},
}

@article{ReviewOfOpticalQuantumComputing1,
    author = {Ralph, T. C. and  Pryde, G. J.},
    title = {Optical Quantum Computation},
    journal = {Progress in Optics},
    year = {2011},
    note={\url{https://doi.org/10.1016/S0079-6638%2810%2905409-0}},
}

@article{EntanglementEntropyReview,
    author ={Zyczkowski, Karol and Bengtsson, Ingemar} ,
    title = {An Introduction to Quantum Entanglement: a Geometric Approach},
    journal = {Arxiv},
    year = {2006},
    note={\url{https://doi.org/10.48550/arXiv.quant-ph/0606228}},
}

@book{GraphTheoryBook,
    author = {West, Douglas Brent and others},
    title = {Introduction to graph theory},
    publisher = {Prentice hall Upper Saddle River},
    year = {2001}
}

@article{gross2007measurement,
  title={Measurement-based quantum computation beyond the one-way model},
  author={Gross, Daivd and Eisert, Jens and Schuch, Norbert and Perez-Garcia, David},
  journal={Physical Review A},
  volume={76},
  number={5},
  pages={052315},
  year={2007},
  publisher={APS},
  note = {\url{https://doi.org/10.1103/PhysRevA.76.052315}},
}

@article{hein2004multiparty,
  title={Multiparty entanglement in graph states},
  author={Hein, Marc and Eisert, Jens and Briegel, Hans J},
  journal={Physical Review A},
  volume={69},
  number={6},
  pages={062311},
  year={2004},
  publisher={APS},
  note = {\url{https://doi.org/10.1103/PhysRevA.69.062311}},
}

@article{dur2005entanglement,
  title={Entanglement in spin chains and lattices with long-range ising-type interactions},
  author={D{\"u}r, W and Hartmann, L and Hein, M and Lewenstein, M and Briegel, H-J},
  journal={Physical review letters},
  volume={94},
  number={9},
  pages={097203},
  year={2005},
  publisher={APS},
  note = {\url{https://doi.org/10.1103/PhysRevLett.94.097203}},
}

@article{bartolucci2021creation,
  title={Creation of entangled photonic states using linear optics},
  author={Bartolucci, Sara and Birchall, Patrick M and Gimeno-Segovia, Mercedes and Johnston, Eric and Kieling, Konrad and Pant, Mihir and Rudolph, Terry and Smith, Jake and Sparrow, Chris and Vidrighin, Mihai D},
  journal={arXiv preprint arXiv:2106.13825},
  year={2021},
  note = {\url{https://doi.org/10.48550/arXiv.2106.13825}},
}

@article{simulationCode,
    title={The simulations code can be found in this link},
    note={\url{https://doi.org/10.5281/zenodo.17370302}}
}

\appendix

\section{Detailed Proofs}
\label{sec:proof_appendix}

\subsection{Proof of Lemma \ref{Lemma - Conditions for stabilizer in e and the stabilizer}}
    \label{Proof appendix - Lemma - Conditions for stabilizer in e and the stabilizer}

    If $a'=e$, then acting on (\ref{Cluster state after measuring a,b for general unitary transformation}) with $\prod_{c\in n(L)\cup n(b)} Z_c$ yields the state:
 \begin{gather}
        A\ket{0}_e\prod_{c\in n(L)} Z_c \ket{\phi_L}_{V_L\setminus \{L\}}\prod_{d\in n(b)} Z_d \ket{\phi_b}_{V_b\setminus \{b\}}
        \nonumber\\
        +B\ket{0}_e\prod_{c\in n(L)} Z_c\ket{\phi_L}_{V_L\setminus \{L\}}\ket{\phi_b}_{V_b\setminus \{b\}}
        \nonumber\\
        +C\ket{1}_e \ket{\phi_L}_{V_L\setminus \{L\}} \prod_{d\in n(b)} Z_d\ket{\phi_b}_{V_b\setminus \{b\}}
        \nonumber\\
        +D\ket{1}_e \ket{\phi_L}_{V_L\setminus \{L\}} \ket{\phi_b}_{V_b\setminus \{b\}} \ .
        \label{Acting on the cluster state with multiplication of Zc when c neighbor of e}
    \end{gather}
    In order to be able to have an hermitian  (since it commutes with all the $Z_c$ operators and their multiplication should be hermitian) operator, that acts on $span\{\ket{0}_e,\ket{1}_e\}$, such that the state (\ref{Acting on the cluster state with multiplication of Zc when c neighbor of e}) becomes (\ref{Cluster state after measuring a,b for general unitary transformation}), we have to exchange the $A$ and $D$ terms 
    as well as the $B$ and $C$ terms. Thus, we need to have $AC=BD$, and if this holds then the corresponding operator takes the form  $T_a=
        \begin{bmatrix}
            0&t\\
            \frac{1}{t}&0
        \end{bmatrix}$ when $D=tA$ and $C=tB$, where $t=e^{i\Phi}$ to ensure herimiticity.
        Hence, we get (\ref{Operator that replacing Xe}), and therefore the stabilizer (\ref{New stabilizer for e}).
    Since $t=e^{i\Phi}$ we require $|A|=|D|$, $|B|=|C|$ and $AC=BD$.
    Note, that in this case, the resulting stabilizer (\ref{New stabilizer for e}) commutes with all the other stabilizers $K_{a'}$ when $a'$ in neither $e$ nor in $n(b)$. If $a'$ in not in $n(L)$ then this is trivial, and if it is, then $Z_{a'}$ from $K'_{e}$ and $X_{a'}$ from $K_{a'}$ are anti-commuting, but so do $T_a$ from $K'_{e}$ and $Z_e$ from $K_{a'}$.
    
\subsection{Proof of Lemma \ref{Lemma - Conditions for stabilizer in n(b) and the stabilizer}}
    \label{Proof appendix - Lemma - Conditions for stabilizer in n(b) and the stabilizer}

    When $a'\in n(b)$, then acting on (\ref{Cluster state after measuring a,b for general unitary transformation}) with the operator $X_{a'}\prod_{b'\in n(a')\setminus \{e\}} Z_{b'}$ yields the state:
 \begin{gather}
        (-1)^{k'_{a'}}(A\ket{0}_e\ket{\phi_L}_{V_L\setminus \{L\}}\ket{\phi_b}_{V_b\setminus \{b\}}
        \nonumber\\
        -B\ket{0}_e\ket{\phi_L}_{V_L\setminus \{L\}}\prod_{d\in n(b)} Z_d\ket{\phi_b}_{V_b\setminus \{b\}}
        \nonumber\\
        +C\ket{1}_e\prod_{c\in n(L)} Z_c \ket{\phi_L}_{V_L\setminus \{L\}} \ket{\phi_b}_{V_b\setminus \{b\}}
        \nonumber\\
        -D\ket{1}_e\prod_{c\in n(L)} Z_c \ket{\phi_L}_{V_L\setminus \{L\}}\prod_{d\in n(b)} Z_d \ket{\phi_b}_{V_b\setminus \{b\}}
        ) \ .
        \label{Acting on the cluster state with multiplication of Zb' when b' neighbor of a', multiple by Xa'}
    \end{gather}
    In order to be able to act on $span{\ket{0}_e,\ket{1}_e}$ with an operator, such that the state (\ref{Acting on the cluster state with multiplication of Zb' when b' neighbor of a', multiple by Xa'}) transforms to  (\ref{Cluster state after measuring a,b for general unitary transformation}), we need that one of $(A,B)$ to be zero and also one of $(C,D)$. If $A=C=0$ or $B=D=0$, then the state is a product state, so these options are irrelevant, but if $A=D=0$ or $B=C=0$, then acting with $Z_e$ yields the required. This yields the regular cluster stabilizer (\ref{Clusters eigenvalue equation}) for $a'$, which commutes with all the other regular stabilizers, and also with the new stabilizer of $e$ in (\ref{New stabilizer for e}), as follows from the same argument as in proof \ref{Proof appendix - Lemma - Conditions for stabilizer in e and the stabilizer} in proof appendix.
    
\subsection{Proof of Theorem \ref{Theorem - when it's graph state}}
    \label{Proof appendix - Theorem - when it's graph state}
We begin with the case where $|n(b)|=1$, and let $d$ be the qubit in $n(b)$. Let us put all the qubits in the $\ket{+}$ state, and then act on every connected qubit with $CZ$, except for $d,e$. Thus, we get two clusters - one containing $e$, and the other one containing $d$. The total wave function is:
    \begin{gather}
    (\ket{0}_e\ket{\phi_L}_{V_L\setminus \{L\}}+\ket{1}_e \prod_{c\in n(L)} Z_c \ket{\phi_L}_{V_L\setminus \{L\}})
    \ket{\phi_b}_{V_b\setminus \{b\}} \ .
    \label{Graph state wave function before acting with the gate between d,e}
    \end{gather}
We act with a two-qubit gate on $(d,e)$, and get the state (\ref{Cluster state after measuring a,b for general unitary transformation}). We require this gate to commute with all the $CZ$ gates, thus it has to be diagonal in the standard basis $\ket{0},\ket{1}$. This leaves only one option - the operator $T_{e,n(b)}$, defined in (\ref{The 2-qubits gate between e and n(b) for graph state}). In order for this operator to be unitary (up to normalization) we have to require  $|A\pm B|=|C\pm D|$.
From this we get:
    \begin{gather}
        1=|A|^2+|B|^2+|C|^2+|D|^2=\nonumber \\
        \frac{1}{2}(|A+B|^2+|A-B|^2
        +|C+D|^2+|C-D|^2)
        \nonumber\\
        \Longrightarrow \frac{1}{\sqrt{2}}=|A\pm B|^2=|C \pm D|^2 \nonumber \\
        = |A|^2+|B|^2\pm 2\Re{AB^*}
        =|C|^2+|D|^2\pm 2\Re{CD^*} \ ,
    \end{gather}
    and the conditions in (\ref{Conditions on A,B,C,D for graph state}) follow.

    These conditions can also be written as the parameterization in (\ref{Parameterization of A,B,C,D in the case of graph state}). In this parameterization, the gate $T_{e,n(b)}$ reads:
 \begin{gather}
        T_{e,n(b)}=e^{i(\theta_1+\phi_1)}\ket{0_e,0_d}\bra{0_e,0_e}+e^{i(\theta_1-\phi_1)}\ket{0_e,1_d}\bra{0_e,1_e}\nonumber \\
        +e^{i(\theta_2+\phi_2)}\ket{1_e,0_d}\bra{1_e,0_e}+e^{i(\theta_2-\phi_2+\pi)}\ket{1_e,1_d}\bra{1_e,1_e} \ .
    \end{gather}
Denote $\chi_{1,2}=\theta_1\pm\phi_1, \chi_3=\theta_2+\phi_2, \chi_4=\theta_2-\phi_2+\pi$.
Then, assuming we apply $Z$-rotations and a global phase rotation, we can first perform a global rotation that takes $\chi_1$ to zero and all the other $\chi_i$ to $\chi_i'=\chi_i-\chi_1$. Then, by $Z$-rotations and global phase rotations we act with a phase shift gate on the qubit $e$ with a shift $\chi_1-\chi_3$, which takes $\chi_3'$ to $\chi_3''=\chi_3'+\chi_1-\chi_3=0$ and $\chi_4'$ to $\chi_4''=\chi_4'+\chi_1-\chi_3=\chi_4-\chi_3$. By the same method we can perform a phase shift gate on qubit $d$ with shift $\chi_1-\chi_2$, which takes $\chi_2'$ to $\chi_2''=\chi_2'+\chi_1-\chi_2=0$, and we are left with $\chi_4''=\chi_4'+\chi_1-\chi_2=\chi_1+\chi_4-\chi_2-\chi_3$. This is the required $\chi$, and by substituting the values of $\chi_i$ and performing modulo $2\pi$, we get (\ref{chi by phi1,2}) as required. 

    If these conditions are not satisfied, we can still perform a unitary transformation on the qubits $(d,e)$, that takes the state in (\ref{Graph state wave function before acting with the gate between d,e}) to the state (\ref{Cluster state after measuring a,b for general unitary transformation}), and in fact there are many such transformations. But, none of those will commute with the $CZ$ gates, so we must use this gate only after all the $CZ$ gates.
Consider next the case $|n(b)|=2$ (or greater than 2), and denote by $d_1,d_2$ the qubits in $n(b)$. Assume we again prepared the state in (\ref{Graph state wave function before acting with the gate between d,e}) and we wish to get the state (\ref{Cluster state after measuring a,b for general unitary transformation}) by applying 2-qubits gates on the couples $(e,d_1)$ and $(e,d_2)$. 
    
    In this case, one of $A,B$ has to be zero, because if not, then we must apply on the couple $e,d_1$ 2-qubits gate that contains the terms $\ket{0}_e\bra{0}_e$, $\ket{0}_e\bra{0}_eZ_{d_1}$, which will give us part of the total wave function which is $(A\ket{0}_e\ket{\phi_L}_{V_L\setminus \{L\}}\ket{\phi_{d_1}}+B\ket{0}_e\ket{\phi_L}_{V_L\setminus \{L\}}Z_{d_1}\ket{\phi_{d_1}})\ket{\phi_{d_2}}$. Then we need to apply some 2-qubits gate on $e,d_2$ that will contain the terms $\ket{0}_e\ket{\phi_L}_{V_L\setminus \{L\}}\ket{\phi_{d_1}}\ket{\phi_{d_2}}$ and $\ket{0}_e\ket{\phi_L}_{V_L\setminus \{L\}}Z_{d_1}\ket{\phi_{d_1}}Z_{d_2}\ket{\phi_{d_2}}$, but this will also generate the terms $\ket{0}_e\ket{\phi_L}_{V_L\setminus \{L\}}Z_{d_1}\ket{\phi_{d_1}}\ket{\phi_{d_2}}$ and $\ket{0}_e\ket{\phi_L}_{V_L\setminus \{L\}}\ket{\phi_{d_1}}Z_{d_2}\ket{\phi_{d_2}}$ which makes our mission impossible.
Following the same argument, one of $C,D$ must be zero. In that case, we can apply the gate in (\ref{The 2-qubits gate between e and n(b) for graph state}) on $e,d_1$ and then on $e,d_2$ to get the wanted state in (\ref{Cluster state after measuring a,b for general unitary transformation}). If $A=C=0$ or $B=D=0$ then the state is a product state. If $A=D=0$ or $B=C=0$, then from the conditions in (\ref{Conditions on A,B,C,D for graph state}) it follows that $|A|=|D|$ or $|B|=|C|$, respectively. We, thus, get exactly the conditions of theorem \ref{Theorem - when it's cluster state up to 1-qubit gate operation} for a stabilizer state.
    
\subsection{Proof of Theorem \ref{Theorem - when it can be made cluster state by operating with 2 and 3 qubits gates}}
    \label{Proof appendix - Theorem - when it can be made cluster state by operating with 2 and 3 qubits gates}
    Back to the more convenient way of the $f$'s, we have $f_2=\left(\prod_{c\in n(L)} Z_c\right) f_1$ and require $g_2=\left(\prod_{c\in n(L)} Z_c\right) g_1$, and similarly for $3,4$ with $\left(\prod_{d\in n(b)} Z_d\right)$. We can write the transformation of $f_1,f_2$ as in (\ref{Hard transformations}):
\begin{gather}
        \begin{bmatrix}
            g_1 \\
            g_2
        \end{bmatrix}
        =
        \begin{bmatrix}
            T_{11}&T_{12} \\
            T_{21}&T_{22}
        \end{bmatrix}
        \begin{bmatrix}
            f_1 \\
            f_2
        \end{bmatrix} \ .
    \end{gather}  
Note, that $\left(\prod_{c\in n(L)} Z_c\right)^2=I$ and the same for $d$, so we have the condition:
     \begin{gather}
        T_{21}f_1+T_{22}f_2=g_2=\left(\prod_{c\in n(L)} Z_c\right)g_1
        \nonumber \\ =\left(\prod_{c\in n(L)} Z_c\right)(T_{11}f_1+T_{12}f_2)
        =T_{11}f_2+T_{12}f_1
        \nonumber \\ \Longrightarrow T_{11}=T_{22}, \hspace{0.1cm} T_{12}=T_{21} \ .
    \end{gather}
Since $T$ is unitary, we get:
    \begin{gather}
        |T_{11}|^2+|T_{12}|^2=1, 
        \nonumber\\
        0=T_{11}T_{21}^*+T_{12}T_{22}^*=T_{11}T_{12}^*+T_{11}^*T_{12}
        \nonumber\\
        \Longrightarrow Re(T_{11}T_{12}^*)=0 \ .
    \end{gather}
Writing $T_{11}$ in a polar representation as $cos(\phi_1) e^{i\theta_1}$, the matrix $T$ reads:
    \begin{gather}
        T=e^{i\theta_1}\begin{bmatrix}
            \cos{\phi_1} &i \sin{\phi_1}  
            \\
            i \sin{\phi_1}&\cos{\phi_1}
        \end{bmatrix}
        =e^{i(\theta_1I+\phi_1\sigma_x)} \ .
    \end{gather}
Similarly, we represent the transformation of $f_3,f_4$ using $\theta_2, \phi_2$. Thus, the state (\ref{Schmidt decomposition f's language}) is (note that we required $\alpha=\beta=\frac{1}{\sqrt{2}}$):
     \begin{gather}
        \alpha g_1 g_3+\beta g_2 g_4 =
        \begin{bmatrix}
            f_1 & f_2
        \end{bmatrix}
        e^{i(\theta_1I+\phi_1\sigma_x)}
        \frac{1}{\sqrt{2}}
        e^{i(\theta_2I+\phi_2\sigma_x)}
        \begin{bmatrix}
            f_3 \\
            f_4
        \end{bmatrix}
        \nonumber \\
        =\frac{1}{\sqrt{2}}
        \begin{bmatrix}
            f_1 & f_2
        \end{bmatrix}
        e^{i((\theta_1+\theta_2)I+(\phi_1+\phi_2)\sigma_x)}
        \begin{bmatrix}
            f_3 \\
            f_4
        \end{bmatrix} \ .
    \end{gather}
Denoting $\phi=\phi_1+\phi_2$ and $\theta=\theta_1+\theta_2$ we get 
    \begin{gather}
        \frac{e^{i\theta}}{\sqrt{2}}(\cos{\phi}f_1f_3+i\sin{\phi}f_1f_4+i\sin{\phi}f_2f_3+\cos{\phi}f_2f_4) \ .
    \end{gather}
Finally, we can absorb a phase between $\alpha$ and $\beta$ in $g_3,g_4$. Another way to see this is, that we can make 1 qubit phase shift gate on $e$. This means, that we can multiply $C$ and $D$ by some phase, which gives us the more general state:
\begin{gather}
        \frac{e^{i\theta_1}}{\sqrt{2}}(\cos{\phi}f_1f_3+i\sin{\phi}f_1f_4)+\frac{e^{i\theta_2}}{\sqrt{2}}(i\sin{\phi}f_2f_3+\cos{\phi}f_2f_4) \ .
    \end{gather}
This state has the same structure as a weighted graph state, with $\phi_1=\phi_2=\phi$ (\ref{Parameterization of A,B,C,D in the case of graph state}) and $\chi=0$ (\ref{chi by phi1,2}). That means that we can get this state by placing all the qubits in a $\ket{+}$ state, operating on every couple of qubits that are connected in the original clusters with $CZ$ gates, and operating on $\{e\}\cup n(b)$ with a gate of the form:
\begin{gather}
        T_{e,n(b)}=A\ket{0}_e\bra{0}_e+B\ket{0}_e\bra{0}_e\prod_{d\in n(b)} Z_d
        \nonumber \\+C\ket{1}_e\bra{1}_e+D\ket{1}_e\bra{1}_e\prod_{d\in n(b)} Z_d
        \label{The 3-qubits gate between e and n(b) for cluster state up to 2-local rotations} \ .
    \end{gather}
We apply $Z$-rotation on $e$, and a two-qubit gate on $d_1,d_2\in n(b)$, which is analogous to the $Z$-rotation on $d\in n(b)$ when $|n(b)|=1$ for the weighted graph state (with the replacement $\ket{0}_d\rightarrow \ket{0}_{d_1}\ket{0}_{d_2}$ and $\ket{1}_d\rightarrow \ket{1}_{d_1}\ket{1}_{d_2}$). This implies a gate of the form (\ref{The 2-qubits gate between e and n(b) for graph state after z-rotations}) (again with the replacement $\ket{0}_d\rightarrow \ket{0}_{d_1}\ket{0}_{d_2}$ and $\ket{1}_d\rightarrow \ket{1}_{d_1}\ket{1}_{d_2}$) with $\chi=0$. This means that what we get is the application of $CZ$ on $b$ and $d_1$, and on $b$ and $d_2$, which is the required cluster state. Thus, we can get the required cluster state after a two-local rotation (two-qubit gate on $n(b)$ and single-qubit gate on $e$). When $|n(b)|=1$, this becomes a single-qubit rotation.
    
    \subsection{Proof of Equation (\ref{m,n,t,k relations})} 
    \label{Proof appendix - m,n,t,k relations}
    Since $U$ (\ref{Unitary transformation}) is unitary we have:
\begin{gather}
        m_1+m_2+m_3+m_4=\sum_{i=1}^{i=4} (|U_{1i}|^2+|U_{2i}|^2)
        \nonumber\\=\sum_{i=1}^{i=4} |U_{1i}|^2+\sum_{i=1}^{i=4} |U_{2i}|^2=1+1=2 \ ,
    \label{Sum of m's}
    \end{gather}
    and therefore
    \begin{gather}
        \sum_{i=1}^{i=4} n_i=\sum_{i=1}^{i=4} (\frac{1}{2}-m_i)=2-\sum_{i=1}^{i=4} m_i=0 \ .
    \label{Sum of n's}
    \end{gather}
Since $U$ is unitary we also have:
    \begin{gather}
       \sum_{i=1}^{i=4} t_i =\sum_{i=1}^{i=4} U_{1i}U_{2i}^*=0 \ ,
    \label{Sum of t's}
    \end{gather}
and 
\begin{gather}
        \sum_{i=1}^{i=4} k_i = \sum_{i=1}^{i=4} (|U_{1i}|^2-|U_{2i}|^2)
        \nonumber\\=\sum_{i=1}^{i=4} |U_{1i}|^2-\sum_{i=1}^{i=4} |U_{2i}|^2=1-1=0  \ .
        \label{Sum of k's}
    \end{gather}

\subsection{Proof of Lemma \ref{Lemma - the wave function}}
    \label{Proof appendix - Lemma - the wave function}
        Denote:
\begin{gather}
        x_1=c_H, x_2=c_V, x_3=d_H, x_4=d_V \ .
    \label{Denoting c,d by x} 
    \end{gather}
Before the measurement, the wave function after operating with $U$ reads:
\begin{gather}
        \ket{\phi}=\hat{C_1}\hat{C_2}=\frac{1}{2} \left( f_1 a^\dag_H + f_2 a^\dag_V \right) \left( f_3 b^\dag_H + f_4 b^\dag_V \right)\ket{vac}
        \nonumber\\
        \rightarrow \frac{1}{2} [\sum_{1\leq i<j\leq 4}(a_{ij}f_{1}f_{3}+b_{ij}f_{1}f_{4}+c_{ij}f_{2}f_{3}+d_{ij}f_{2}f_{4})x^\dag_i x^\dag _j 
        \nonumber\\
        +\sum_{1\leq i\leq 4}(a_{ii}f_{1}f_{3}+b_{ii}f_{1}f_{4} +c_{ii}f_{2}f_{3}+d_{ii}f_{2}f_{4})(x^\dag _i)^2]\ket{vac}
        \nonumber\\
        =\frac{1}{2}\sum_{1\leq i<j\leq 4}(a_{ij}f_{1}f_{3}+b_{ij}f_{1}f_{4}+c_{ij}f_{2}f_{3}+d_{ij}f_{2}f_{4})\ket{1_i1_j}
        \nonumber\\
        +\frac{\sqrt{2}}{2}\sum_{1\leq i\leq 4}(a_{ii}f_{1}f_{3}+b_{ii}f_{1}f_{4}  +c_{ii}f_{2}f_{3}+d_{ii}f_{2}f_{4})\ket{2_i} \ ,
    \label{Wave function before measuring}
    \end{gather}
where
 \begin{gather}
        i\ne j:   a_{ij}=U_{1i}U_{3j}+U_{1j}U_{3i}
    \hspace{0.5cm}
    b_{ij}=U_{1i}U_{4j}+U_{1j}U_{4i} \nonumber\\
    c_{ij}=U_{2i}U_{3j}+U_{2j}U_{3i}
    \hspace{0.5cm}
    d_{ij}=U_{2i}U_{4j}+U_{2j}U_{4i} \nonumber\\
    a_{ii}=U_{1i}U_{3i}\hspace{0.5cm}b_{ii}=U_{1i}U_{4i}\hspace{0.5cm}c_{ii}=U_{2i}U_{3i}\hspace{0.5cm}d_{ii}=U_{2i}U_{4i} \ .
    \label{The coefficients of the wave function proofs appendix}
    \end{gather}

After measuring one photon in $i$, and one photon in $j$ (when $i=j$ we measure two photons in the $i$ state), the wave function reads:
\begin{gather}
       \frac{1}{N_{ij}}(a_{ij}f_{1}f_{3}+b_{ij}f_{1}f_{4}+c_{ij}f_{2}f_{3}+d_{ij}f_{2}f_{4})\ket{vac} \ ,
   \label{Wave function after measuring proofs appendix}
   \end{gather}
where $N_{ij}^2=|a_{ij}|^2+|b_{ij}|^2+|c_{ij}|^2+|d_{ij}|^2$ is a normalization factor.

    \subsection{Proof of Lemma \ref{Lemma - the states probabilities}}
    \label{Proof appendix - Lemma - the states probabilities}
    For a non-relevant state:
\begin{gather}
        p_{ii}=\frac{1}{2}(|a_{ii}|^2+|b_{ii}|^2+|c_{ii}|^2+|d_{ii}|^2)
        \nonumber\\
        =\frac{1}{2}(|U_{1i}|^2+|U_{2i}|^2)(|U_{3i}|^2+|U_{4i}|^2) \ .
        =\frac{1}{2}m_i(1-m_i)
    \label{Probability of non-relevant state proofs appendix}
    \end{gather}
By substituting $m_i=0.5-n_i$ and $m_j=0.5-n_j$, we get $p_{ii}=\frac{1}{8}-\frac{1}{2}n_i^2$.
For a relevant state $i\ne j$:
\begin{gather}
4p_{ij}=N_{ij}^2=|a_{ij}|^2+|b_{ij}|^2+|c_{ij}|^2+|d_{ij}|^2=\nonumber\\|U_{1i}U_{3j}+U_{1j}U_{3i}|^2+|U_{1i}U_{4j}+U_{1j}U_{4i}|^2+\nonumber\\|U_{2i}U_{3j}+U_{2j}U_{3i}|^2+|U_{2i}U_{4j}+U_{2j}U_{4i}|^2=\nonumber\\
        =|U_{1i}|^2|U_{3j}|^2+|U_{1j}|^2|U_{3i}|^2+U_{1i}U_{3j}U_{1j}^*U_{3i}^*+h.c.\nonumber\\+|U_{1i}|^2|U_{4j}|^2+|U_{1j}|^2|U_{4i}|^2+U_{1i}U_{4j}U_{1j}^*U_{4i}^*+h.c.\nonumber\\+|U_{2i}|^2|U_{3j}|^2+|U_{2j}|^2|U_{3i}|^2+U_{2i}U_{3j}U_{2j}^*U_{3i}^*+h.c.\nonumber\\+|U_{2i}|^2|U_{4j}|^2+|U_{2j}|^2|U_{4i}|^2+U_{2i}U_{4j}U_{2j}^*U_{4i}^*+h.c.\nonumber\\=(|U_{1i}|^2+|U_{2i}|^2)(|U_{3j}|^2+|U_{4j}|^2)\nonumber\\+(|U_{1j}|^2+|U_{2j}|^2)(|U_{3i}|^2+|U_{4i}|^2)\nonumber\\+(U_{1i}U_{1j}^*+U_{2i}U_{2j}^*)(U_{3j}U_{3i}^*+U_{4j}U_{4i}^*)+h.c.\nonumber\\=m_i(1-m_j)+m_j(1-m_i)\nonumber\\+(U_{1i}U_{1j}^*+U_{2i}U_{2j}^*)(-U_{1j}U_{1i}^*-U_{2j}U_{2i}^*)+h.c.\nonumber\\=m_i(1-m_j)+m_j(1-m_i)\nonumber\\-(U_{1i}U_{1j}^*+U_{2i}U_{2j}^*)(U_{1i}U_{1j}^*+U_{2i}U_{2j}^*)^*+h.c.\nonumber\\=m_i(1-m_j)+m_j(1-m_i)-2|U_{1i}U_{1j}^*+U_{2i}U_{2j}^*|^2 \ .
    \label{Probability of relevant state proofs appendix}
    \end{gather}
Thus, indeed $p_{ij}=\frac{1}{4}(m_{i}(1-m_{j})+m_{j}(1-m_{i})) - \frac{1}{2} |U_{1i}U_{1j}^{*}+U_{2i}U_{2j}^{*}|^2$.
By substituting $m_i=0.5-n_i$ and $m_j=0.5-n_j$ we get:
    \begin{gather}
      p_{ij}=\frac{1}{8}-\frac{1}{2}n_in_j - \frac{1}{2} |U_{1i}U_{1j}^{*}+U_{2i}U_{2j}^{*}|^2\nonumber\\ \leq \frac{1}{8}-\frac{1}{2}\frac{1}{2}(-\frac{1}{2})-0=\frac{1}{4}   \ . 
    \label{Probability of relevant state by n proofs appendix}
    \end{gather}

\subsection{Proof of Lemma \ref{Lemma - the determinant satisfies}}
    \label{Proof appendix - Lemma - the determinant satisfies}
    Assuming $i\ne j$ and using equation (\ref{Density matrix}):
\begin{gather}
        N_{ij}^4 det(\rho_{ij})=(|a_{ij}|^2+|b_{ij}|^2)(|c_{ij}|^2+|d_{ij}|^2)
         \nonumber\\
        -(a_{ij}^*c_{ij}+b_{ij}^*d_{ij})(a_{ij}^*c_{ij}+b_{ij}^*d_{ij})^*=|a_{ij}|^2|c_{ij}|^2+|a_{ij}|^2|d_{ij}|^2\nonumber\\
        +|b_{ij}|^2|c_{ij}|^2+|b_{ij}|^2|d_{ij}|^2-|a_{ij}|^2|c_{ij}|^2-|b_{ij}|^2|d_{ij}|^2
        \nonumber\\
        -a_{ij}^*c_{ij}b_{ij}d_{ij}^*-h.c.=|a_{ij}d_{ij}|^2+|b_{ij}c_{ij}|^2-(a_{ij}d_{ij})^*(b_{ij}c_{ij}) \nonumber\\
        -h.c.
        =|a_{ij}d_{ij}-b_{ij}c_{ij}|^2 \ .
    \label{Determinant identities part 1 proofs appendix}
    \end{gather}
Dividing by $N_{ij}^4$ we get $det(\rho_{ij})=\frac{|a_{ij}d_{ij}-b_{ij}c_{ij}|^2}{N_{ij}^4}$, which is what we aimed to prove. 

We can now write $a,b,c,d$ in terms of the components of $U$ using (\ref{The coefficients of the wave function}):
 \begin{gather}
        a_{ij}d_{ij}-b_{ij}c_{ij}=(U_{1i}U_{3j}+U_{1j}U_{3i})(U_{2i}U_{4j}+U_{2j}U_{4i})
        \nonumber\\
        -(U_{1i}U_{4j}+U_{1j}U_{4i})(U_{2i}U_{3j}+U_{2j}U_{3i})
        \nonumber\\
        =(U_{1i}U_{2j}-U_{2i}U_{1j})(U_{3j}U_{4i}-U_{3i}U_{4j}) \ .
    \label{Determinant identities part 2 proofs appendix}
    \end{gather}
Using this, and the fact that $N_{ij}=\sqrt{4p_{ij}}$, we get that $det(\rho_{ij})=\left|\frac{(U_{1i}U_{2j}-U_{2i}U_{1j})(U_{3j}U_{4i}-U_{3i}U_{4j})}{4p_{ij}}\right|^2$, as required.
    
\subsection{Proof of the general form of a maximally entangled two-qubit state}
\label{Proof appendix - General form of 2-qubits maximally entangled state}
From equation (\ref{Density matrix}) of the reduced density matrix, we must have $|A|^2+|B|^2=|C|^2+|D|^2=\frac{1}{2}$, $AC^*+BD^*=0$. Using the first condition we represent $A,B,C,D$ as:
\begin{gather}
    A=\frac{1}{\sqrt{2}}e^{i\theta_a}\cos{\phi_1}\hspace{0.5cm}
    B=\frac{1}{\sqrt{2}}e^{i\theta_b}\sin{\phi_1}
    \nonumber \\
    C=\frac{1}{\sqrt{2}}e^{i\theta_c}\sin{\phi_2}\hspace{0.5cm}
    D=-\frac{1}{\sqrt{2}}e^{i\theta_d}\cos{\phi_2}
    \label{General form of states with |A|^2+|B|^2=|C|^2+|D|^2} \ .
\end{gather}
The second condition implies:
\begin{gather}
    \frac{1}{2}(e^{i(\theta_a-\theta_c)}\cos{\phi_1}\sin{\phi_2}+e^{i(\theta_b-\theta_d)}\sin{\phi_1}\cos{\phi_2})=0 \nonumber \\
    \longrightarrow e^{i(\theta_a-\theta_c)}\cos{\phi_1}\sin{\phi_2}=-e^{i(\theta_b-\theta_d)}\sin{\phi_1}\cos{\phi_2} \ .
\end{gather}
Taking absolute value of both sides gives $|\cos{\phi_1}\sin{\phi_2}|=|\cos{\phi_1}\sin{\phi_2}|$. If $\cos{\phi_1}=0$, then it follows that $\sin{\phi_1}=\pm 1$ so $\cos{\phi_1}=0$. Then $A=D=0, |B|=|C|=\frac{1}{\sqrt{2}}$, and we can choose $\phi_1=\phi_2=\frac{\pi}{2}$ and the appropriate values of $\theta_a,\theta_b,\theta_d$, such that we get the state in (\ref{General form of 2-qubits maximally entangled state}). We can apply the same argument when $\cos{\phi_2}=0$, and we can assume that $\cos{\phi_1},\cos{\phi_2}\ne 0$. Dividing by them gives $|\tan{\phi_1}|=|\tan{\phi_2}|$. By absorbing $-1$ into the phases $\theta_c,\theta_d$ if needed, we can assume that the vectors $(\cos{\phi_1},\sin{\phi_1}),(\cos{\phi_2},\sin{\phi_2})$ lie at the same quarter of the plane, from which it follows that $\phi_1=\phi_2$. We denote $\phi=\phi_1=\phi_2$. 
If $\sin{\phi}=0$, then we get $B=C=0$ and $|A|=|D|=\frac{1}{\sqrt{2}}$, which can be dealt similarly. Else, our condition turns into:
\begin{gather}
    e^{i(\theta_a-\theta_c)}=-e^{i(\theta_b-\theta_d)}\Longleftrightarrow e^{i\theta_c}=-e^{\theta_a+\theta_d-\theta_b} \ .
\end{gather}
Plugging it in (\ref{General form of states with |A|^2+|B|^2=|C|^2+|D|^2}), together with $\phi_1=\phi_2=\phi$ gives us the general form in (\ref{General form of 2-qubits maximally entangled state}).

\subsection{Proof of condition 1 of Lemma \ref{Lemma - Conditions for Bell state}}
    \label{Proof appendix - Lemma - Conditions for Bell state - condition 1}
     For $i\ne j$ by using (\ref{m,n,t,k definitions}), (\ref{The coefficients of the wave function}) and the unitary of $U$ (\ref{Unitary transformation}):
\begin{gather}
        a_{ij}c_{ij}^*+b_{ij}d_{ij}^*=(U_{1i}U_{3j}+U_{1j}U_{3i})(U_{2i}U_{3j}+U_{2j}U_{3i})^*\nonumber\\+(U_{1i}U_{4j}+U_{1j}U_{4i})(U_{2i}U_{4j}+U_{2j}U_{4i})^*\nonumber\\=U_{1i}U_{2i}^*|U_{3j}|^2+U_{1j}U_{2j}^*|U_{3i}|^2+U_{1i}U_{2j}^*U_{3j}U_{3i}^*\nonumber\\+U_{1j}U_{2i}^*U_{3i}U_{3j}^*+U_{1i}U_{2i}^*|U_{4j}|^2+U_{1j}U_{2j}^*|U_{4i}|^2\nonumber\\+U_{1i}U_{2j}^*U_{4j}U_{4i}^*+U_{1j}U_{2i}^*U_{4i}U_{4j}^*\nonumber\\=U_{1i}U_{2i}^*(|U_{3j}|^2+|U_{4j}|^2)+U_{1j}U_{2j}^*(|U_{3i}|^2+|U_{4i}|^2)\nonumber\\+U_{1i}U_{2j}^*(U_{3i}^*U_{3j}+U_{4i}^*U_{4j})+U_{1j}U_{2i}^*(U_{3i}U_{3j}^*+U_{4i}U_{4j}^*)\nonumber\\=U_{1i}U_{2i}^*(1-|U_{1j}|^2-|U_{2j}|^2)+U_{1j}U_{2j}^*(1-|U_{1i}|^2-|U_{2i}|^2)\nonumber\\-U_{1i}U_{2j}^*(U_{1i}^*U_{1j}+U_{2i}^*U_{2j})-U_{1j}U_{2i}^*(U_{1i}U_{1j}^*+U_{2i}U_{2j}^*)\nonumber\\=U_{1i}U_{2i}^*(1-2|U_{1j}|^2-2|U_{2j}|^2)+U_{1j}U_{2j}^*(1-2|U_{1i}|^2-2|U_{2i}|^2)\nonumber\\=2n_it_j+2n_jt_i \ .
    \label{Condition 1 proofs appendix}
    \end{gather}
    
\subsection{Proof of condition 2 of Lemma \ref{Lemma - Conditions for Bell state}}
    \label{Proof appendix - Lemma - Conditions for Bell state - condition 2}
    For $i\ne j$ by using (\ref{m,n,t,k definitions}), (\ref{The coefficients of the wave function}) and the unitary of $U$ (\ref{Unitary transformation}):
\begin{gather}
        |a_{ij}|^2+|b_{ij}|^2-|c_{ij}|^2-|d_{ij}|^2\nonumber\\=|U_{1i}U_{3j}+U_{1j}U_{3i}|^2+|U_{1i}U_{4j}+U_{1j}U_{4i}|^2\nonumber\\-|U_{2i}U_{3j}+U_{2j}U_{3i}|^2-|U_{2i}U_{4j}+U_{2j}U_{4i}|^2\nonumber\\=|U_{1i}|^2|U_{3j}|^2+|U_{1j}|^2|U_{3i}|^2+U_{1i}U_{3j}U_{1j}^*U_{3i}^*+h.c.\nonumber\\+|U_{1i}|^2|U_{4j}|^2+|U_{1j}|^2|U_{4i}|^2+U_{1i}U_{4j}U_{1j}^*U_{4i}^*+h.c.\nonumber\\-|U_{2i}|^2|U_{3j}|^2-|U_{2j}|^2|U_{3i}|^2-U_{2i}U_{3j}U_{2j}^*U_{3i}^*-h.c.\nonumber\\-|U_{2i}|^2|U_{4j}|^2-|U_{2j}|^2|U_{4i}|^2-U_{2i}U_{4j}U_{2j}^*U_{4i}^*-h.c.\nonumber\\=(|U_{1i}|^2-|U_{2i}|^2)(|U_{3j}|^2+|U_{4j}|^2)\nonumber\\+(|U_{1j}|^2-|U_{2j}|^2)(|U_{3i}|^2+|U_{4i}|^2)\nonumber\\+(U_{1i}U_{1j}^*-U_{2i}U_{2j}^*)(U_{3i}^*U_{3j}+U_{4i}^*U_{4j})+h.c.\nonumber\\=k_i(1-m_j)+k_j(1-m_i)\nonumber\\-(U_{1i}U_{1j}^*-U_{2i}U_{2j}^*)(U_{1i}^*U_{1j}+U_{2i}^*U_{2j})-h.c.\nonumber\\=k_i(1-m_j)+k_j(1-m_i)\nonumber\\-2Re(|U_{1i}|^2|U_{1j}|^2-|U_{2i}|^2|U_{2j}|^2+U_{1i}U_{1j}^*U_{2i}^*U_{2j}-h.c.)\nonumber\\=k_i(1-m_j)+k_j(1-m_i)-2(|U_{1i}|^2|U_{1j}|^2-|U_{2i}|^2|U_{2j}|^2)\nonumber\\=k_i(1-m_j)+k_j(1-m_i)-((|U_{1i}|^2+|U_{2i}|^2)(|U_{1j}|^2-|U_{2j}|^2)\nonumber\\+(|U_{1j}|^2+|U_{2j}|^2)(|U_{1i}|^2-|U_{2i}|^2))=k_i(1-m_j)+k_j(1-m_i)\nonumber\\-(m_ik_j+m_jk_i)=k_i(1-2m_j)+k_j(1-2m_i)\nonumber\\
        =2n_ik_j+2n_jk_i \ .
    \label{Condition 2 part 1 proofs appendix}
    \end{gather}
We can also write it as:
 \begin{gather}
        =2n_i(2|U_{1j}|^2-(|U_{1j}|^2+|U_{2j}|^2))\nonumber\\+2n_j(2|U_{1i}|^2-(|U_{1i}|^2+|U_{2i}|^2))\nonumber\\=2n_i(2|U_{1j}|^2-(\frac{1}{2}-n_j))+2n_j(2|U_{1i}|^2-(\frac{1}{2}-n_i))\nonumber\\=4n_i|U_{1j}|^2+4n_j|U_{1i}|^2+4n_in_j-n_i-n_j \ .
    \label{Condition 2 part 2 proofs appendix}
    \end{gather}
    
\subsection{Proof of Lemma \ref{Lemma - probability for specific Bell state}}
    \label{Proof appendix - Lemma - probability for specific Bell state}
     We will prove, that given a relevant state $(i,j)$ which is a maximally entangled state, then the probability to obtain this state (\ref{Probability of relevant state}) is less than or equal to $\frac{1}{8}$, and if equality holds then necessarily $n_i=n_j=0$.
First, when $n_i=0$ or $n_j=0$, then by (\ref{Probability of relevant state}) $p_{ij}=\frac{1}{8}-\frac{1}{2}|U_{1i}U_{1j}^*+U_{2i}U_{2j}^*|^2\leq\frac{1}{8}$. Also,  when $n_i=n_j$ then $p_{ij}=\frac{1}{8}-\frac{1}{2}n_i^2-\frac{1}{2}|U_{1i}U_{1j}^*+U_{2i}U_{2j}^*|^2\leq \frac{1}{8}$. Thus, we shall assume in the following that $n_i,n_j,0$ differ from each other. 
    
    Using both conditions (\ref{condition 1 for Bell state},\ref{condition 2 for Bell state}) proven in Lemma \ref{Lemma - Conditions for Bell state} we get that:
    \begin{gather}
        n_i^2m_j^2=n_i^2(4|t_j|^2+k_j^2)=4|n_it_j|^2+|n_ik_j|^2\nonumber\\=4|n_jt_i|^2+|n_jk_i|^2=n_j^2(4|t_i|^2+k_i^2)=n_j^2m_i^2 \ .
    \label{n_i^2*m_j^2=n_j^2*m_i^2}
    \end{gather}
    We can write this as:
    \begin{gather}
        0=n_i^2m_j^2-n_j^2m_i^2=\frac{(n_i-n_j)(n_i+n_j-4n_in_j)}{4}
    \label{0=n_i^2*m_j^2-n_j^2*m_i^2} \ .
    \end{gather}
Because $n_i\neq n_j$ we conclude that $n_i+n_j=4n_in_j$. Thus, 
    \begin{gather}
        0=2n_ik_j+2n_jk_i=4n_i|U_{1j}|^2+4n_j|U_{1i}|^2\nonumber\\+4n_in_j-n_i-n_j=4(n_i|U_{1j}|^2+n_j|U_{1i}|^2) \ .
    \label{0=2n_i*k_j+2n_j*k_i}
    \end{gather}
Hence, $n_i|U_{1j}|^2=-n_j|U_{1i}|^2$. Because $n_i$ and $n_j$ are different from zero, if $U_{1i}=0$ then also $U_{1j}=0$ and vice versa, but then by Remark \ref{Remark - when two elements of U are zero} the state is a product state. Thus, also $U_{1i}$ and $U_{1j}$ are different from zero, and we can write the equation as $\frac{|U_{1i}|^2}{|U_{1j}|^2}=-\frac{n_i}{n_j}$. Similarly we get that:
    \begin{gather}
        \frac{|U_{1i}|^2}{|U_{1j}|^2}=\frac{|U_{2i}|^2}{|U_{2j}|^2}=-\frac{n_i}{n_j}=1-4n_i=4m_i-1 \ .
    \label{=4*m_i-1}
    \end{gather}
Thus, $4m_i-1>0$. We denote:
    \begin{gather}
        U_{1i}=\sqrt{m_i}\cos{\theta}e^{i\phi_{1i}}
        \hspace{1cm}
        \nonumber\\
        U_{2i}=\sqrt{m_i}\sin{\theta}e^{i\phi_{2i}}
        \nonumber\\
        U_{1j}=\sqrt{\frac{m_i}{4m_i-1}}\cos{\theta}e^{i\phi_{1j}}
        \hspace{1cm}
        \nonumber\\
        U_{2j}=\sqrt{\frac{m_i}{4m_i-1}}\sin{\theta}e^{i\phi_{2j}}
    \label{U_1i,U_2i,U_1j,U_2j by m_i}
    \end{gather}
    Plugging it gives:
    \begin{gather}
        4m_i-1=-\frac{n_i}{n_j}=\frac{t_i}{t_j}=(4m_i-1)e^{i(\phi_{1i}-\phi_{2i}-\phi_{1j}+\phi_{2j})} \ , 
    \label{Phases condition}
    \end{gather}
    thus,  $\phi_{1i}-\phi_{2i}-\phi_{1j}+\phi_{2j}=0$ up to modulo $2\pi$.
    
    Using the fact that $U$ is unitary, we have:
    \begin{gather}
        U_{3i}U_{3j}^*+U_{4i}U_{4j}^*=-U_{1i}U_{1j}^*-U_{2i}U_{2j}^*\nonumber\\=-\frac{m_i}{\sqrt{4m_i-1}}(\cos^{2}\theta e^{i(\phi_{1i}-\phi_{1j})}+\sin^{2}\theta e^{i(\phi_{2i}-\phi_{2j})})\nonumber\\=-\frac{m_i}{\sqrt{4m_i-1}} e^{i(\phi_{1i}-\phi_{1j})} \ .
    \label{Finding (U_3i)(U_3j*)+(U_4i)(U_4j*)}
    \end{gather}
Using this and the Cauchy–Schwarz inequality we have:
    \begin{gather}
        \frac{m_i^2}{4m_i-1}=|U_{3i}U_{3j}^*+U_{4i}U_{4j}^*|^2
        \nonumber\\
        \leq(|U_{3i}|^2+|U_{4i}|^2)(|U_{3j}|^2+|U_{4j}|^2)=(1-m_i)(1-m_j)
        \nonumber\\
        =\frac{-3m_i^2+4m_i-1}{4m_i-1} \ .
    \label{Using Cauchy-Schwarz inequality}
    \end{gather}
$4m_i-1>0$ so we have:
     $0\geq m_i^2-(-3m_i^2+4m_i-1)=(2m_i-1)^2$ \ .
We get $m_i=\frac{1}{2}$ and $n_i=0$, so again $p_{ij}\leq \frac{1}{8}$.
    
    In the case of equality, we have $n_i=n_j$ or $n_i=0$. If $n_i=n_j$ then $p_{ij}=\frac{1}{8} - \frac{1}{2} n_{i}^2 - \frac{1}{2} |U_{1i}U_{1j}^{*}+U_{2i}U_{2j}^{*}|^2 \leq \frac{1}{8}$, and for equality we must have $n_i=0$. 
    If $n_i=0$, let us assume that $n_j\neq 0$. Then, from $n_it_j+n_jt_i=0$ (\ref{condition 1 for Bell state}) and $n_ik_j+n_jk_i=0$ (\ref{condition 2 for Bell state}) we get that $t_i=k_i=0$. Now, $m_i^2=4|t_i|^2+k_i^2=0$ so $m_i=0$, in contradiction with $n_i=0$.

\subsection{Completing the proof of Lemma \ref{Lemma - six Bell states}}
    \label{Proof appendix - Lemma - six Bell states - completing the proof}
Summing over $j \neq i$ in equation (\ref{condition 1 for Bell state}), and using the fact that the sum of the $n$'s and the $t$'s is zero (\ref{m,n,t,k relations}), we get that: $t_i(0-n_i)+n_i(0-t_i)=0$ and therefore $n_i=0$ or $t_i=0$ for every $1\leq i \leq 4$.
There are three cases: (1) Three of the $n$'s are zero, which implies that all the $n$'s are zero because their sum is zero.
   (2) Three of the $t$'s are zero, which implies that all the $t$'s are zero because their sum is zero.
   (3) Two of the $n$'s are zero and two of the $t$'s are zero. Assume that $n_1=n_2=t_3=t_4=0$, then $t_2n_3=-t_3n_2=0$ and therefore $n_3=0$ or $t_2=0$. In both cases we get that three of the $n$'s are zero or three of the $t$'s are zero, thus we are back to case (1) or case (2).
   
\subsection{Proof of Lemma \ref{Lemma - five Bell states}}
    \label{Proof appendix - Lemma - five Bell states}
    When five of the relevant states are maximally entangled states, we have the $n_it_j+t_in_j=0$ conditions (\ref{condition 1 for Bell state}), which at least five of the six $(i,j)$ satisfy.  If the sixth satisfies as well, then by proof \ref{Proof appendix - Lemma - six Bell states - completing the proof} we have one of two options:
    (1) all the $n$'s are zero, thus by Lemma \ref{Lemma - six Bell states} all the states are maximally entangled states, contradiction;
    (2) all the t's are zero, so for every $1\leq i\leq 4$ one of $U_{1i},U_{2i}$ is zero. For every relevant state $(i,j)$ for which $U_{1i}=U_{1j}=0$ or $U_{2i}=U_{2j}=0$, the state is a product state by Remark \ref{Remark - when two elements of U are zero}, and we can find two different relevant states like that. 
    
    If for three of the indices $U_{1i}$ is zero, then we can choose any two of them (three options), and the same holds if for three indices $U_{2i}$ is zero. If for two indices $U_{1i}=0$ and for the other two $U_{2i}=0$, then we still have two couples. This is a contradiction for five of the relevant states being maximally entangled states.
     In both cases we get a contradiction to the assumption. Hence, $t_in_j=-t_jn_i$ for all $i\neq j$ except for one, $n_3t_4\neq t_3n_4$. So, by Lemma \ref{Lemma - Conditions for Bell state}, the state $(3,4)$ is not a maximally entangled state and every relevant state different from $(3,4)$ state must be maximally entangled state.
    We cannot  have three of the $n$'s being zero, or three of the $t$'s being zero, because then also the fourth $n$ or $t$ is zero, and we arrive at the same contradiction as before.
    
    For $i=1$ or $i=2$ we can still take the equation $n_it_j+n_jt_i=0$ (\ref{condition 1 for Bell state}), and sum on all $j\neq i$ as in proof \ref{Proof appendix - Lemma - six Bell states - completing the proof}, and get that $n_i=0$ or $t_i=0$ for $i=1,2$. We again have three cases: (1) If $t_1=t_2=0$,  then $t_3,t_4\ne 0$ because we cannot have three $t$'s that are zero, and from $n_1t_3=-n_3t_1=0$ and $n_1t_4=-n_4t_1=0$ we get $n_1=n_2=0$. (2) If $n_1=n_2=0$, then by the same argument of (1) (by exchanging the $t$'s and the $n$'s in the proof, we get that $t_1=t_2=0$. (3) If $n_1=0$ and $t_2=0$, then $n_2t_1=-n_1t_2=0$ and therefore $n_2=0$ or $t_1=0$. We end up with case (1) or (2). The same holdsfor $n_2=0$ and $t_1=0$.
    
In all the cases we get that $n_1=n_2=t_1=t_2=0$ and $n_3,n_4,t_3,t_4\ne 0$. Now, we can use the $n_ik_j+n_jk_i=0$ conditions (\ref{condition 2 for Bell state}). From $n_3k_1=-n_1k_3=0$ we get that $k_1=0$, and as in the proof \ref{Proof appendix - Lemma - probability for specific Bell state}, $m_1^2=4|t_1|^2+k_1^2=0$ so $m_1=0$, a contradiction with $n_1=0$.
    Thus, we cannot have exactly five maximally entangled states.

\subsection{Proof of Lemma \ref{Lemma - subsets of (1,2,3,4)}}
    \label{Proof appendix - Lemma - subsets of (1,2,3,4)}
    We aim to show, that if we choose four couples of two different indices from the set $\{1,2,3,4\}$, then we can choose two couples $(i,j)$ and $(k,l)$, such that $\{i,j,k,l\}=\{1,2,3,4\}$. In other words, the indices $i,j,k,l$ are all different from each other. Assume 
    that there exists a counter example. Say, one of our couples is $(1,2)$. This means, that every other couple has $1$ or $2$. If all the other couples have the same number between $1$ and $2$, say $1$, then we can only add $(1,3)$ and $(1,4)$ and we have only three couples at most. Thus, we must have a couple with $1$, say $(1,3)$, and also a couple with $2$. In order for the couple with $2$ to share one index with $(1,3)$, this couple must be $(2,3)$. Now, we have the couples $(1,2),(1,3),(2,3)$, and we also have a couple that can have at most one index from $(1,2,3)$, say $1$. But, then this couple does not share any index with $(2,3)$, hence a contradiction.

\section{Proof of the equivalence between the stabilizer definition and the recursive definition of the graph state}
\label{sec:Proof of equivalence}
In order to show the equivalence of
(\ref{Graph state eigenvalue equation}) and (\ref{Graph state recursive definition}), we substitute the wave function (\ref{Graph state recursive definition}) in (\ref{Graph state eigenvalue equation}), where we assume that $\ket{\phi_a}_{V_a\setminus \{a\}}$ satisfies (\ref{Graph state eigenvalue equation}). 
When $a'=a$ we have:
\begin{gather}
    K_{a}\ket{\phi}=X_{a} \otimes 
    \prod _{b\in n(a)} Z_{b}  \ (\ket{0}_a\ket{\phi_a}_{V_a\setminus \{a\}}
    \nonumber\\
    +\ket{1}_a \prod_{b\in n(a)} Z_b \ket{\phi_a}_{V_a\setminus \{a\}})=\ket{1}_a\prod _{b\in n(a)} Z_{b} \ket{\phi_a}_{V_a\setminus \{a\}} 
    \nonumber \\
    +\ket{0}_a\ket{\phi_a}_{V_a\setminus \{a\}}=\ket{\phi} \ .
    \label{The recursive and eigenvalue definitions are equivalent, a'=a case, Graph states}
\end{gather}

When $a'\in n(a)$ then, using $X_{a'}Z_{a'}=-Z_{a'}X_{a'}$:
\begin{gather}
   K_{a'}\ket{\phi}=Z_{a}X_{a'} \otimes 
    \prod _{b'\in n(a')\setminus \{a\}} Z_{b'} (\ket{0}_a\ket{\phi_a}_{V_a\setminus \{a\}}
    \nonumber\\
    +\ket{1}_a Z_{a'}\prod_{b\in n(a)\setminus\{a'\}} Z_b \ket{\phi_a}_{V_a\setminus \{a\}})=\ket{0}_a \ket{\phi_a}_{V_a\setminus \{a\}}
    \nonumber\\
    +(-\ket{1}_a)(-Z_{a'}\prod_{b\in n(a)\setminus\{a'\}} Z_bX_{a'}\prod _{b'\in n(a')\setminus \{a\}} Z_{b'}\ket{\phi_a}_{V_a\setminus \{a\}})
    \nonumber\\
    =\ket{0}_a\ket{\phi_a}_{V_a\setminus \{a\}}
    +\ket{1}_a\prod _{b\in n(a)} Z_{b}  \ket{\phi_a}_{V_a\setminus \{a\}}
    =\ket{\phi} \ .
    \label{The recursive and eigenvalue definitions are equivalent, a' in n(a) case, Graph states}
\end{gather}
When $a'$ is not $a$ and not in $n(a)$, then $K_{a'}\ket{\phi}=\ket{\phi}$, simply because $\ket{\phi_a}_{V_a\setminus \{a\}}$ satisfies (\ref{Graph state eigenvalue equation}).

\section{Haar Measure}
\label{sec:Haar measure}
We sample the fusion matrices (\ref{Unitary transformation}) from the Haar measure distribution on on $SU\left(4\right)$. 
The Haar measure $\mu$ is defined on a locally compact topological group $G$, such that for any set $S$ in the Borel set of $G$ and for every $a\in G$, $\mu(aS)=\mu(S)$.
We realize the distribution by the Gauss–Gram–Schmidt construction (see \cite{HaarMeasureBook} chapter 1.2 for more details). In this construction, we first construct a $4\times 4$ complex matrix $M$, by drawing $32$ random real numbers from
standard normal distribution (normal distribution with $\mu =0$ and $\sigma =1$), and using them
as the real and imaginary parts of the entries of the matrix. We perform a $QR$-decomposition of $M$:
\begin{gather}
    M=QR \ ,
\end{gather}
where $Q$ is a unitary matrix, and $R$ is an upper triangular matrix with real nonzero numbers on the diagonal. This decomposition is not unique, and in order to make it unique we require that all the numbers on the diagonal of $R$ are positive. We obtain this unique decomposition by applying a Gram-Schmidt algorithm to the columns of $M$, and in this case the matrix $Q$ is distributed according to the Haar measure. The Gram-Schmidt algorithm is not numerically stable, thus most of the algorithms that we use are different from the Gram-Schmidt one, hence gives a $QR$-decomposition, where $Q$ does not obey Haar measure distribution.

This can be fixed by changing the matrices $Q,R$: for every index $i$ such that $R_{ii}<0$, we change the signs of the $i$'th row of $R$ and the $i$'th column of $Q$, and we get the required
$QR$-decomposition with $R_{ii}>0$ for every $i$. Since we are only interested in $Q$, we define the matrix:
\begin{gather}
    E_{ij}=sign(R_{ii})\delta_{ij} \ ,
\end{gather}
which is a diagonal matrix with $1$'s when the diagonal element of $R$ is positive, and $-1$'s when the diagonal element of $R$ is negative. Then, the matrix:
\begin{gather}
    U=QE \ ,
\end{gather}
is the required  "Q" in the unique QR-decomposition and follows the Haar measure distribution. 

\section{Summary of Section \ref{sec:Generalized Type II Fusion and the resulting state}}
\label{sec:Summary of Generalized Type II Fusion}

\begin{figure}
    \centering
    \includegraphics[width=1.0\linewidth]{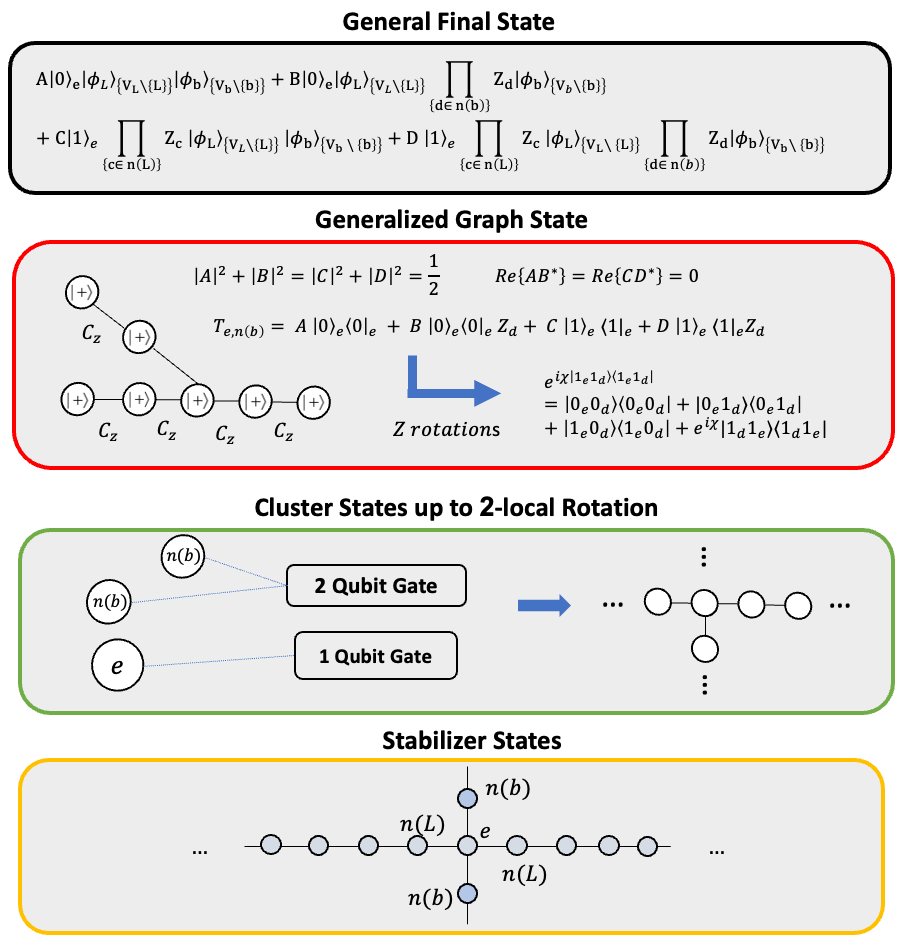}
    \caption{Summary of the results of section \ref{sec:Generalized Type II Fusion and the resulting state}. The black rectangle shows the general form of the final state. The red rectangle shows the results about weighted graph states. The green rectangle illustrates the states that are cluster states up to 2-local rotation and the appropriate rotations. The yellow rectangle shows the stabilizer states.}   
    \label{fig:SetsHeirarchy}
\end{figure}
We summarize in Figure \ref{fig:SetsHeirarchy} the resulting final states, that were discussed in section \ref{sec:Generalized Type II Fusion and the resulting state}. 
The general final state is written in the black rectangle (as in equation (\ref{Cluster state after measuring a,b for general unitary transformation})), where its parameters $A,B,C,D$ depend on the elements of the fusion matrix $U$, which represents the transformation of the creation operators of the qubits $a,b$ as in equation (\ref{Unitary transformation}), and the result of the measurement of qubits $c,d$. The precise dependence is determined in equation (\ref{The coefficients of the wave function}). The wave function $\ket{\phi_L}_{V_L\setminus \{L\}}$ describes the total wave function of the one-dimensional cluster of the Logical qubit, without this qubit. The total wave function of this cluster is as in equation (\ref{Graph state recursive definition}), and the wave function $\ket{\phi_b}_{V_b\setminus \{b\}}$ is defined similarly. 

The yellow rectangle describes the final states, that are stabilizer states. In theorem \ref{Theorem - when it's cluster state up to 1-qubit gate operation} we proved that the appropriate stabilizers (the generators of the group $S$ as in the definition of the stabilizer states) consist of the regular stabilizers $K_a'$ as in equation (\ref{Clusters eigenvalue equation}) for $a'\ne e$, where for $e$ one has a new stabilizer as in (\ref{New stabilizer for e}). We identified these final states in theorem \ref{Theorem - when it's cluster state up to 1-qubit gate operation}, as the states for which $A=D=0$ and $|B|=|C|$ or $B=C=0$ and $|A|=|D|$, and proved that these states become
cluster states after operating with a single-qubit gate on $e$. 

In theorem \ref{Theorem - when it's graph state} we obtained the conditions on $A,B,C,D$, such that if $|n(b)|=1$ the final state is a weighted graph state (if $|n(b)|=2$ then by the theorem this states are simply the stabilizer states). In the red rectangle there is the general form of the weighted graph state. In this weighted graph state, the qubits are put in the $\ket{+}$ state, and the edges are  realized by $CZ$ gates between the proper qubits, as in the regular cluster state, except for the edge between $e$ and the qubit in $b$, for which the gate is (\ref{The 2-qubits gate between e and n(b) for graph state}), which also appears in the red rectangle.

The green rectangle shows the states, that are cluster states up to two-local rotations: two-qubit gate applied to the two qubits in $n(b)$, and a single-qubit gate on $e$. After the application of these gates, the resulting state is the required cluster state. In theorem \ref{Theorem - when it can be made cluster state by operating with 2 and 3 qubits gates} we proved that if $|n(b)|=1$, then these states (that are cluster states up to single-qubit rotations - one on $e$ and one on the single qubit of $n(b)$), are weighted graph states. Furthermore, we proved in theorem \ref{Theorem - maximally entangled graph state}, that these states are exactly the weighted graph states with maximal entanglement entropy (still in the case $|n(b)|=1$).

\section{Summary of the results from sections \ref{sec:prob_def} and \ref{sec:anal_bound} that are not our main theorems}
\label{sec:Summary of usefull computations}


Table \ref{tab:ComputationsSummarized} summarizes results from sections \ref{sec:prob_def} and \ref{sec:anal_bound}, that were used to prove the main theorems in these sections, deduce the boundary values of $P(S)$ (\ref{eq:P_S_target}), and perform the numerical simulations in section \ref{sec:numerical_res}.

\begin{table*}[!t]
    \centering
    \begin{tabular}{|p{9cm}|p{8cm}|}
    \hline
         $m_i=\left|U_{1 i}\right|^2+\left|U_{2 i}\right|^2 \quad m_1+m_2+m_3+m_4=2$ \newline $n_i=\frac{1}{2}-m_i \quad n_1+n_2+n_3+n_4=0$ \newline $t_i=U_{1 i} U_{2 i}^* \quad t_1+t_2+t_3+t_4=0$ \newline $k_i=\left|U_{1 i}\right|^2-\left|U_{2 i}\right|^2 \quad k_1+k_2+k_3+k_4=0$ & {The constants $m_i,n_i,t_i,k_i$ that we defined (\ref{m,n,t,k definitions}) and their relations (\ref{m,n,t,k relations})}.\\ \hline
         $|\phi\rangle_{i j}=\frac{1}{N_{i j}}\left(a_{i j} f_1 f_3+b_{i j} f_1 f_4+c_{i j} f_2 f_3+d_{i j} f_2 f_4\right)$ & The wave function $\ket{\phi}_{ij}$ (\ref{Wave function of relevant state}) of the relevant state $(i,j)$ (so $i\ne j$).\\ \hline
         $a_{i j}=U_{1 i} U_{3 j}+U_{1 j} U_{3 i} \quad b_{i j}=U_{1 i} U_{4 j}+U_{1 j} U_{4 i}$ \newline $c_{i j}=U_{2 i} U_{3 j}+U_{2 j} U_{3 i} \quad d_{i j}=U_{2 i} U_{4 j}+U_{2 j} U_{4 i}$ \newline $N_{i j}=\sqrt{\left|a_{i j}\right|^2+\left|b_{i j}\right|^2+\left|c_{i j}\right|^2+\left|d_{i j}\right|^2}=\sqrt{4 p_{i j}}$ & The coefficients $a_{ij}$,$b_{ij}$,$c_{ij}$,$d_{ij}$ (\ref{The coefficients of the wave function}) and the normalization factor $N_{ij}$ (\ref{The coefficients of the wave function}) of the wave function (\ref{Wave function of relevant state}) of the relevant state $(i,j)$.\\ \hline
         $p_{i i}=\frac{1}{2} m_i\left(1-m_i\right)=\frac{1}{8}-\frac{1}{2} n_i^2$ \newline $p_{\text {non-relevant }}=\frac{1}{2}\left(1-n_1^2-n_2^2-n_3^2-n_4^2\right) \leq \frac{1}{2}$ & The probability $p_{ii}$ to get to the non-relevant state $(i,i)$ (\ref{Probability of non-relevant state}) and the total probability for all the non-relevant states $p_{non-relevant}$ (\ref{Probability of all non-relevant states}).\\ \hline
         $p_{i j}=\frac{1}{8}-\frac{1}{2} n_i n_j-\frac{1}{2}\left|U_{1 i} U_{1 j}^*+U_{2 i} U_{2 j}^*\right|^2 \leq \frac{1}{4}$ & The probability $p_{ij}$ to get to the relevant state $(i,j)$ (\ref{Probability of relevant state}).\\ \hline
         $\rho_{i j}=\frac{1}{N_{i j}^2}
        \begin{bmatrix}
                |a_{ij}|^2+|b_{ij}|^2 & a_{ij}^*c_{ij}+b_{ij}^*d_{ij} \\ a_{ij}c_{ij}^*+b_{ij}d_{ij}^* & |c_{ij}|^2+|d_{ij}|^2
        \end{bmatrix}$
         & The reduced density matrix $\rho_{ij}$ (\ref{Density matrix}).\\ \hline
         $\operatorname{det}\left(\rho_{i j}\right)=\frac{\left|a_{i j} d_{i j}-b_{i j} c_{i j}\right|^2}{N_{i j}^4}=\left|\frac{\left(U_{1 i} U_{2 j}-U_{1 j} U_{2 i}\right)\left(U_{3 j} U_{4 i}-U_{3 i} U_{4 j}\right)}{4 p_{i j}}\right|^2$ & The determinant $\det(\rho_{ij})$ (\ref{Determinant}) of the reduced density matrix (\ref{Density matrix}), which is monotonic in the entanglement entropy $S_{ij}$ (\ref{Entropy}).\\ \hline
         $\lambda_{i j}, 1-\lambda_{i j}=\frac{1 \pm \sqrt{1-4 \operatorname{det}\left(\rho_{i j}\right)}}{2}$ \newline $S_{i j}=-\lambda_{i j} \log _2 \lambda_{i j}-\left(1-\lambda_{i j}\right) \log _2\left(1-\lambda_{i j}\right)$ & The computing of the eigenvalues $\lambda_{ij}$,$1-\lambda_{ij}$ (\ref{eigenvalues of the reduced density matrix}) of the reduced density matrix (\ref{Density matrix}) and the entanglement entropy $S_{ij}$ (\ref{Entropy}) out of $\det(\rho_{ij})$ (\ref{Determinant}).\\ \hline
         $S_{i j}=1 \Leftrightarrow$
        $ 0=a_{i j} c_{i j}^*+b_{i j} d_{i j}^*=2 t_i n_j+2 t_j n_i $
        , \newline
        $0=\left|a_{i j}\right|^2+\left|b_{i j}\right|^2-\left|c_{i j}\right|^2-\left|d_{i j}\right|^2=2 n_i k_j+2 n_j k_i$
         & The conditions for the relevant state $(i,j)$ to be maximally entangled (\ref{Condition 1 with details}),(\ref{eq:Condition 2 with details}).\\ \hline
         If $S_{i j}=1$ then $p_{i j} \leq \frac{1}{8}$ and $p_{i j}=\frac{1}{8} \Rightarrow n_i=n_j=0$ & The upper bound on $p_{ij}$ for maximally entangled relevant state as proven in Lemma \ref{Lemma - probability for specific Bell state}.\\ \hline
    \end{tabular}
    \caption{Summary of the results from sections \ref{sec:prob_def} and \ref{sec:anal_bound}, that are needed to prove the main theorems.}
    \label{tab:ComputationsSummarized}
\end{table*}



\end{document}